\documentclass[]{jfm}

\usepackage{CJKutf8}

\usepackage{graphicx}
\usepackage{newtxtext}
\usepackage{newtxmath}
\usepackage{ulem}
\usepackage{natbib}
\usepackage{hyperref}
\hypersetup{
    colorlinks = true,
    urlcolor   = blue,
    citecolor  = black,
}

\newcommand{\RomanNumeralCaps}[1]
\linenumbers

\usepackage{xcolor}
\usepackage{array}
\usepackage{tikz} 

\definecolor{mygreen}{rgb}{0, 0.5, 0}

\newcommand{\mylab}[3]{\raisebox{#2}[0mm][0mm]{\makebox[0mm][l]{\hspace*{#1}{#3}}}}

\newcommand{\Ret}[0]{Re_\tau}
\newcommand{\ut}[0]{u_\tau}

\renewcommand{\vec}[1]{\boldsymbol{#1}}

\newcommand{\sijsij}[0]{s_{ij}s_{ij}}

\newcommand{\err}[0]{\varepsilon}
\newcommand{\lcell}[0]{l_{\mbox{\scriptsize\it cell}}}
\newcommand{\ycell}[0]{y_{\mbox{\scriptsize\it cell}}}

\newcommand{\erru}[0]{\err_{\vec{u}}}

\newcommand{\errw}[0]{\err_{\vec{\omega}}}

\newcommand{\sig}[0]{\sigma}
\newcommand{\fsig}[0]{\phi}
\newcommand{\tsig}[0]{t_{\mbox{\scriptsize\it sig}}}
\newcommand{\tatt}[0]{t_{att}}
\newcommand{\sigu}[0]{\sig_{\vec{u}}}
\newcommand{\sigur}[0]{\sig_{\vec{u}r}}

\newcommand{\Perr}[0]{P_{\err u}}
\newcommand{\Cerr}[0]{C_{\err u}}
\newcommand{\Derr}[0]{D_{\err u}}
\newcommand{\dif}[1]{\widehat{#1}}
\newcommand{\ave}[1]{\overline{#1}}
\newcommand{\cave}[1]{\langle{#1}\rangle_{c}} 
\newcommand{\cstd}[1]{{#1}^{\prime c}} 
\newcommand{\sfw}[0]{\langle\partial_y u\rangle_{\mbox{\scriptsize\it fw}}} 
\newcommand{\figpath}{./}

\def\dd{{\, \rm{d}}}
\def\dr{{\rm{d}}}

\def\bra{\langle}
\def\ket{\rangle}
\def\p{\partial}
\def\beq{\begin{equation}}
\def\eeq{\end{equation}}
\def\la{\label}

\def\r#1{(\ref{#1})}
\def\sref{{\mbox{\scriptsize\it ref}}}
\def\smod{{\mbox{\scriptsize\it mod}}}

\newlength{\unitl}
\title{Causal features in turbulent channel flow}

\author{Kosuke Osawa \corresp{\email{kosuke.osawa@upm.es}} \and Javier Jim{\'e}nez}

\affiliation{School of Aeronautics, Universidad Polit{\'e}cnica de Madrid, 28040 Madrid, Spain. }

\begin{document}

\maketitle

\begin{abstract}
 
The causal relevance of local flow conditions in wall-bounded turbulence is analysed using ensembles
of interventional experiments in which the effect of perturbing the flow within a small cell is
monitored at some future time. When this is done using the relative amplification of the
perturbation energy, causality depends on the flow conditions within the cell before it is perturbed,
and can be used as a probe of the flow dynamics. The key scaling parameter is the ambient shear,
which is also the dominant diagnostic variable for wall-attached perturbations.
Away from the wall, the relevant variables are the streamwise and wall-normal velocities. Causally significant cells are associated with sweeps that carry the perturbation towards the
stronger shear near the wall, whereas irrelevant ones are associated with ejections that carry it
towards the weaker shear in the outer layers.
Causally significant and irrelevant cells are themselves organised into structures that share many
characteristics with classical sweeps and ejections, such as forming spanwise pairs whose
dimensions and geometry are similar to those of classical quadrants.
At the wall, this is consistent with causally significant configurations in which a high-speed
streak overtakes a low-speed one, and causally irrelevant ones in which the two streaks
pull apart from each other. It is argued that this is probably associated with streak meandering.

\end{abstract}


\section{Introduction}\la{sec_intro}

Although turbulence is a high-dimensional chaotic system, it is often modelled
as a collection of compact and approximately autonomous coherent structures.
These are typically intermittent, emerging and vanishing with a lifetime and frequency that depend
on their nature and size, and are characterised both by evolving relatively independently from
their flow environment, and by having a measurable influence on the rest of the flow \citep{jim18}. As
such, it is important to clarify not only how they behave individually, but how are they
connected among themselves in space and in time.

Such causal connections would help us understand how turbulence works, both from the fundamental point
view and in practical applications connected with flow control and prediction. For example, it is
important to avoid introducing in the initial conditions of numerical weather forecasting spurious
perturbations that would later amplify significantly \citep{rodwell23}, and identifying such highly
influential events would help us improve prediction accuracy. Another example is flow control, which
intrinsically tries to modify the future of the flow by altering its present state. Understanding which
structures are causally important  and which ones have no significant effect in the evolution of the flow would
clearly help in optimising this process.

Conversely, elucidating the connections between different flow regions, not necessarily initially
identified as coherent, may lead to the discovery of novel coherent structures that describe turbulence
better than the known ones, or to previously overlooked connections between known
structures that can be incorporated into better flow models \citep{JJ20,jimPF23}.

For example, quasi-streamwise rollers, streamwise-velocity streaks and wall-normal velocity bursts
are believed to be essential for maintaining wall-bounded turbulence. The most common hypothesis is
that there is a self-sustaining process (SSP) in which at least two of these structures mutually
induce each other \citep{jmoin,Waleffe95,Waleffe97}, but the details are incomplete. For example,
recent evidence suggests that bursts are able to sustain a cycle by themselves \citep{jim18}, while
streaks are byproducts rather than actors in the SSP \citep{jjnostreak22}. Even apparently
straightforward connections, such as the generation of the streaks by bursts \citep{kkr} are only
incompletely understood, because the two phenomena have very different length scales \citep{jim18}.
Establishing the causality relations between these different structures would throw light on whether
they are indeed connected, on the sequence in which they are linked and on whether some component
is missing from the model.

With the goal of minimising bias, our strategy is to exclusively characterise flow regions in terms
of their influence on the future of the flow, without necessarily relating them to previously known
coherent structures. Only once a particular flow template has been identified as highly
causal or as especially irrelevant will we try to classify it within existing theories, or to
recognise it as something new.

There two general approaches to causality. The first one is observational and non-intrusive, and is
often the only option when the system is hard to replicate (e.g. astrophysics), difficult to
experiment with (e.g. some social sciences), or simply too large to easily simulate. Unfortunately,
it is generally believed that observation is not enough to unambiguously establish cause and effect,
because correlation does not imply causation \citep{granger:69,Pearl:09} but, even in those cases, a
careful consideration of the temporal evolution of the system may lead to the identification of
causal histories when they cross neighbourhoods of particular interest, typically extreme events
\citep{AngEtal:96}. A related approach is the operator representation of turbulence time series,
examples of which are
\cite{Froy:09,KaiserNoack_JFM14,schmid:iop18,Brunt:ARFM20,FernexNoack:21,Taira22,jimPF23,Souza23}, among
others. Another example is the analysis of data series from wall-bounded turbulence by
\citet{LozBaeEnc20} using tools of transfer entropy, or the improved version in which their
applicability to subgrid modelling and flow control was demonstrated by \cite{LozArra22}.

The alternative is interventional causality, in which the system is directly modified and the
consequences observed. This offers more control over what is being analysed, and safer inferences
\citep{Pearl:09}, but presumes a sufficiently cheap way of modifying the system. Essentially, in
dynamical system notation, non-interventional methods provide information about the behaviour of the
system while it moves within its attractor, while interventional ones give additional information about
the system by observing what happens outside it.

Turbulence, which is expensive to simulate and hard to modify experimentally, was for a long
time considered to be in the group of phenomena that could only be observed, but the increased speed of
computers, as well as better experimental techniques, slowly eroded that difficulty
\citep{jmoin,jpin}. More recently, fast GPUs have speeded the numerical simulation of 
realistic turbulent flows to the point of allowing the practical simulation of artificially modified
flow ensembles that can be considered interventional \citep{Vela:21}. They have opened the
possibility of Monte Carlo studies in which the consequences of `randomly' modified flows are
examined.

Examples of this approach are \cite{jimploff18,JJ20}, who introduced localised perturbations in
two-dimensional turbulence in order to determine which parts of the flow result in significant perturbation growth or
decay after a certain time. This allowed the identification of causally significant and irrelevant
flow structures, including the relatively unexpected relevance of vortex dipoles rather than individual
vortices, and eventually led to new models for the two-dimensional energy cascade \citep{jim2DT:21}.
\citet{encinar23} extended the technique to three-dimensional homogeneous isotropic turbulence and
demonstrated that causal events are in that case characterised by either high kinetic energy or high dissipation rate,
depending on the spacial scale of the initial perturbation, and that strong strain, rather than
high vorticity, is the main prerequisite for perturbation growth. In these two cases, it is interesting that 
some of the significant structures were not the classically expected ones, underlining the ability
of Monte Carlo interventional experiments to mitigate the bias of conventional wisdom.

In this study, we adopt the interventional approach, following the basic methodology in
\citet{jimploff18}. Spatially localized perturbations are imposed on a fully developed turbulent
channel flow, and their influence is measured by their ability to alter the future evolution of the
flow.

Numerical experiments that track the development of perturbation ensembles in wall turbulence are
not a new, probably starting with the computation by \cite{keef92} of the Lyapunov
spectrum in a low-Reynolds number channel. On a similar subject, \citet{Nikitin08,Nikitin18}
investigated the Reynolds number scaling of the leading Lyapunov exponent of a turbulent channel.
Lyapunov analyses do not typically control the form of the initial perturbation, leaving the system
to choose the most unstable direction in state space, and taking precautions to avoid
nonlinearities, but \cite{Cherubini10} and \cite{Cherubini17} turned the problem around by searching
for weakly or fully nonlinear perturbations that optimally grow in energy after a given target time.
They work on an initially stationary flow with a turbulent profile and, when they constrain the
initial energy of the perturbation, they obtain optimals that are localised in physical space. More
recently \cite{Ciola23} extended the analysis to snapshots of real turbulence, finding that, for
properly chosen target times, the optimal precursor is an early stage of an Orr burst. However, due
to the difficulty of convergence over long times, the result only applied to short delays of the
order of a few tens of viscous units. Moreover, the solution is only optimal for the snapshot at
which it is applied, making it difficult to generalise the result.

The choice of the size of the initial perturbations is important, and data assimilation experiments
have been conducted to estimate the minimum size below which perturbations are enslaved to their
environment. The first was probably \citet{Yoshida05} in isotropic three-dimensional
turbulence, who showed that randomised scales smaller than 30 \cite{kol41} viscous lengths are
regenerated if continuously assimilated to larger structures. \citet{Wang22} conducted similar
experiments in channel turbulence. They replace some layers with white noise and showed how they
synchronised with the original flow when assimilated through their boundaries. The maximum
synchronisation thickness is approximately 30 viscous lengths (approximately 12 Kolmogorov units)
for layers attached to the wall, and twice the Taylor microscale for layers away from it. However,
since \citet{encinar23} found that freely evolving perturbations grow even below the assimilation
limit, the result in wall-bounded turbulence remains uncertain.

The present study targets the nonlinear evolution of localised perturbations applied to
instantaneous snapshots of turbulent channels at a moderate but non-trivial Reynolds number, over
times of the order of an eddy turnover. A Monte Carlo search is used to apply the analysis across
snapshots, and across as many combinations of perturbation location, size and target time as
practicable. The basic assumption is that causality depends of the local state of the neighbourhood
at which the perturbation is applied, and the details of this dependence are extracted from the
database of numerical experiments using standard methods of data analysis.

The organisation of the paper is as follows. The numerical setup and the definition of the initial perturbations
are described in \S\ref{sec_setup}. How their evolution can be used to determine causality is discussed in \S\ref{sec_signi}--\ref{sec_classify}, and the relation between causal structures and the surrounding flow field is discussed in \S\ref{sec_template}. Conclusions are offered in \S\ref{sec_conc}.   

\section{Numerical setup}\la{sec_setup}

\newcommand{\colwd}{2mm}
\begin{table}
\begin{center}
\def~{\hphantom{0}}
    \begin{tabular}{>{\hspace{\colwd}}c>{\hspace{\colwd}}c>{\hspace{\colwd}}c>{\hspace{\colwd}}c>{\hspace{\colwd}}c>{\hspace{\colwd}}c>{\hspace{\colwd}}c>{\hspace{\colwd}}c<{\hspace{\colwd}}}
$U_b h/\nu$ & $Re_\tau$ & $L_x \times L_y\times L_z$    & $N_x \times N_y \times N_z$  & $\Delta x^+$
& $\Delta y^+_{\min}$ &  $\Delta y^+_{\max}$  & $\Delta z^+$ \\[3pt]
       11180   & 600.9  & $\pi h \times h \times \pi h$ & $128 \times 192 \times 256$ & 14.7 & 0.46 & 6.51 & 7.4\\
    \end{tabular}
\caption{Computational parameters. $L_{i}$ is the domain size along the $i$-th direction, $h$ is
the `half-channel' height, equivalent to the domain height in open channels, and $U_b$ is
the bulk velocity. The grid dimensions, $N_i$, and effective resolutions, $\Delta{x_i}$, are expressed in terms of Fourier modes.}
    \label{table_para}
    \end{center}
\end{table}

\begin{figure}
\centerline{%
\includegraphics[width=0.44\textwidth]{\figpath 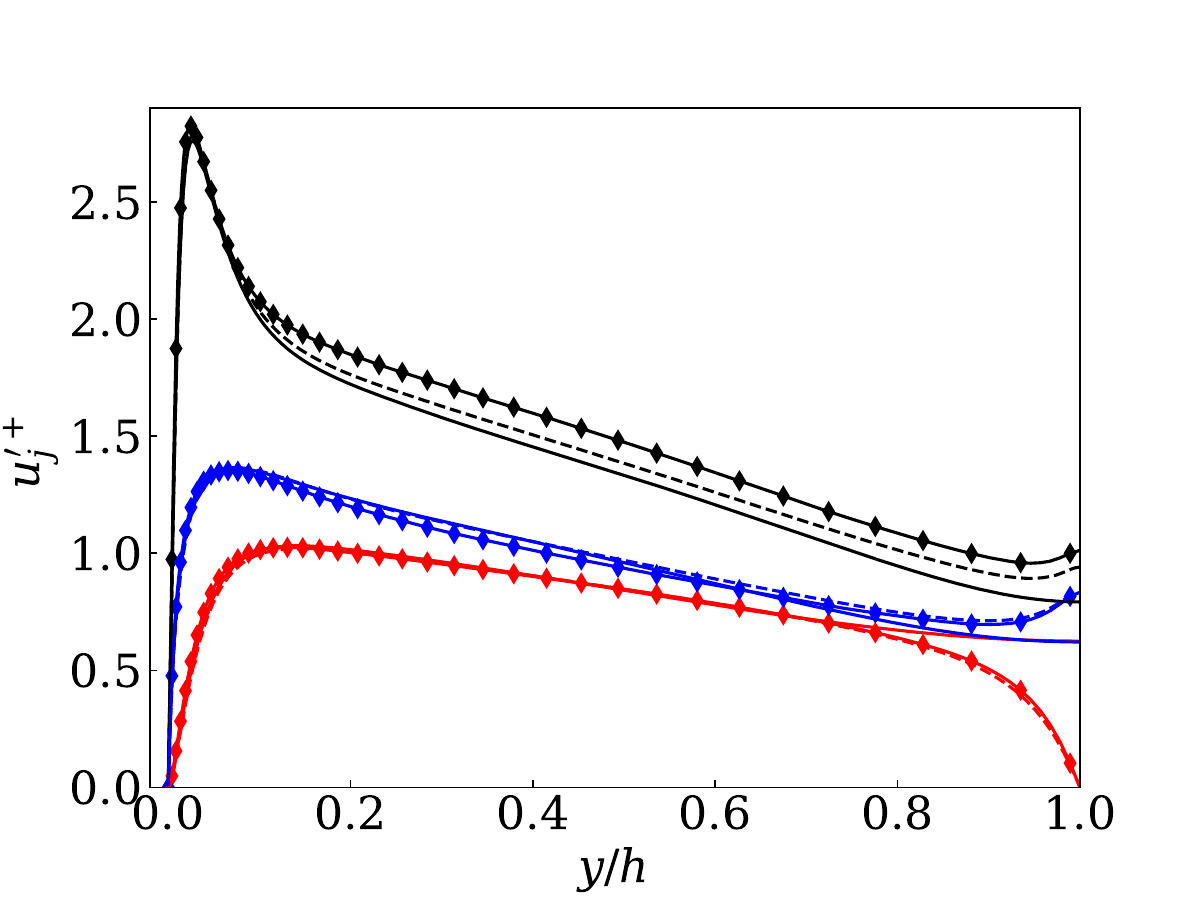}%
\mylab{-0.09\textwidth}{0.26\textwidth}{(a)}%
\hspace*{2mm}%
\includegraphics[width=0.40\textwidth]{\figpath 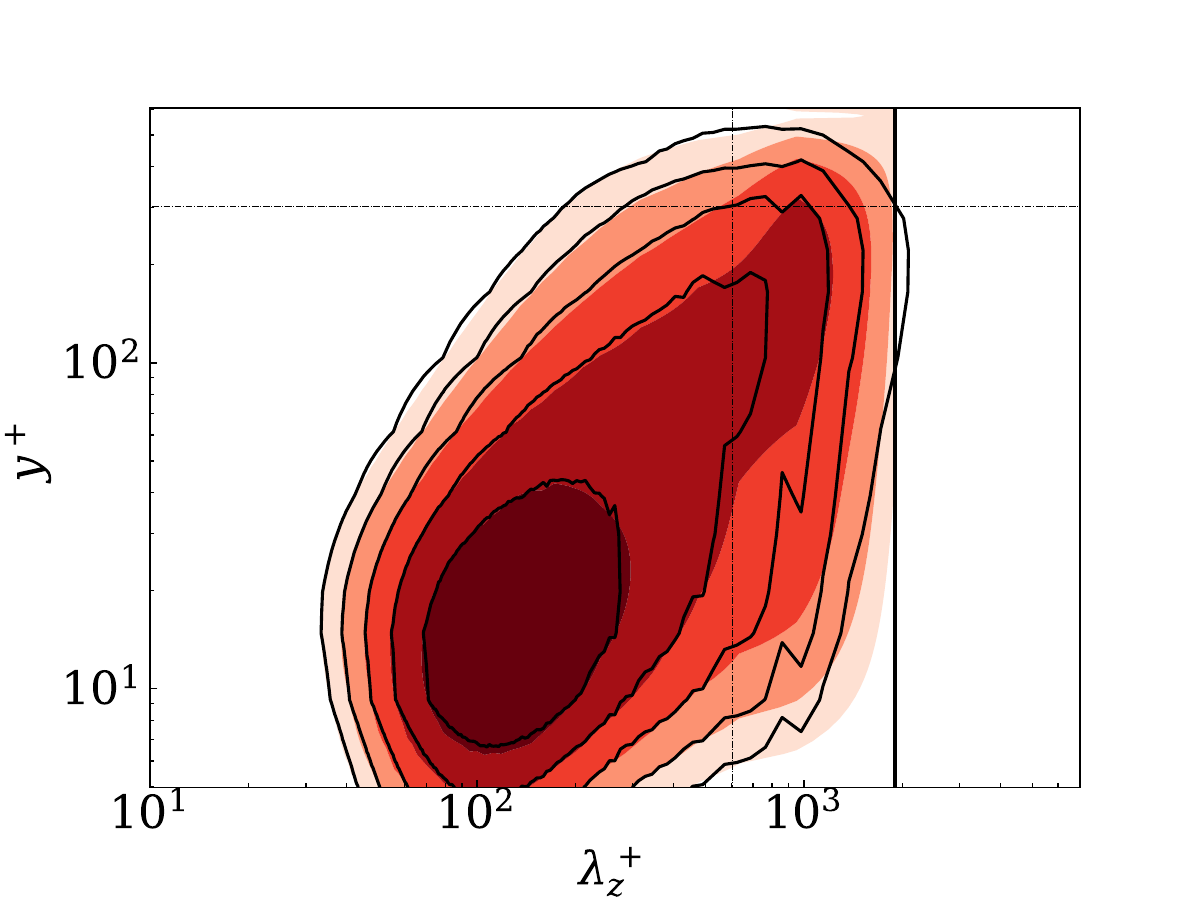}%
\mylab{-0.05\textwidth}{0.26\textwidth}{(b)}%
}
\caption{%
(a) Velocity fluctuation intensities. Symbols, present open channel at $Re_\tau=601$; dashed, open
channel at $Re_\tau=541$ \citep{Pirozz:23}; solid, full channel at $Re_\tau=547$ \citep{juanc03}.
Black: $u'$; red, $v'$; blue, $w'$.
(b) Premultiplied spanwise spectrum of the turbulent kinetic energy, normalised with $\ut^2$.
Contours are logarithmically equispaced from $k_zE_{KK}/\ut^2=0.56$ to 2.8. The vertical solid line
is the present computational box. The vertical thinner line is $\lambda_z=2\pi/k_z=h$, and the
horizontal line is $y^+=300$. Filled contours are the present simulation; lines are \cite{juanc03}.
}%
\label{fig:openchann}
\end{figure}

To save computational resources, we analyse simulations of a pressure-driven turbulent open channel
flow in a doubly periodic domain, between a no-slip wall at $y=0$ and an impermeable free-slip one
at $y=h$. The streamwise, wall-normal and spanwise directions are $x,y$ and $z$, respectively, and
the corresponding velocities are $u, v$ and $w$, although position and velocities are occasionally
denoted by their components $\vec{x}=\{x_j\}, j=1\ldots 3$. The domain size is $L_x\times L_y\times
L_z=\pi h \times h \times \pi h$, and the Reynolds number is $\Ret=\ut h/\nu=600.9$. The `+'
superscript denotes wall units, normalised with the friction velocity $\ut$ and with the kinematic
viscosity $\nu$. Capital letters, as in $U(y)$, denote variables averaged over the simulation
ensemble and over wall-parallel planes, lower case ones are fluctuations with respect to this
average, and primes are root-mean-squared fluctuation intensities. Repeated indices, including
squares, imply summations unless otherwise noted. The simulation code is standard dealiased Fourier
spectral along $x$ and $z$, as in \cite{kmm}, but uses seven-points-stencil compact finite differences for
the wall-normal derivatives, as in \cite{hoyas06}. Time marching is semi-implicit third-order
Runge--Kutta \citep{spalart91}, and the mass flux is kept constant. The numerical $y$-grid is
stretched at the no-slip wall with a hyperbolic tangent. See table \ref{table_para} for other
numerical parameters.

Figure \ref{fig:openchann}(a) compares the resulting fluctuation profiles with existing data from
regular and open channels. It was shown by \cite{lozano14} that a computational box with $L_z/h=\pi$
reproduces well the statistics of regular channels, and figure \ref{fig:openchann} shows that the
same is true for open ones. In particular, figure \ref{fig:openchann}(b) shows that the spanwise
kinetic energy spectrum fits well within the computational box. On the other hand, the figure 
shows that open and full channels only agree below $y/h\approx 0.5$, above which the effect of
`splatting' at the top wall is particularly visible in the fluctuations of the cross-flow velocities
\citep{perot:95}. We will only use the range $y^+\lesssim 300$ for the rest of the paper. It is also
clear from the figure that the energy at long wavelengths above $y^+=100$ is higher than in
\cite{juanc03}. This is due to the short computational box, which inhibits the
instability of the streaks \citep{toh2018MFU}, and results in two pairs of large streamwise streaks
and rollers that dominate the flow.
 
The original code was ported to CUDA by \cite{VelaEtal:21} for the efficient simulation of
high-Reynolds number channel turbulence in GPU clusters. It has been adapted to a single GPU
for the present experiments, but the original reference should be consulted for full details.

\subsection{The initial perturbations}\la{sec_perturb}
%

\renewcommand{\colwd}{3mm}
\begin{table}
  \begin{center}
\def~{\hphantom{0}}
  \begin{tabular}{>{\hspace{\colwd}}l>{\hspace{\colwd}}c>{\hspace{\colwd}}c<{\hspace{\colwd}}c<{\hspace{\colwd}}}
      $\lcell^+$  & $\ycell^+$ (set 1)   &   $\ycell^+$ (set 2) & Symbol \\[5pt]
       25   & 0, 12.5, 37.6, 62.6, 87.6, 113, 138, 163, 188, 213, 238, 288  & 0, 138 & \full \\
       50   & 0, 25.0, 50.0, 75.0, 100, 125, 150, 175, 200, 225, 275          & NA & \fulltriangle\\
       75   & 0, 12.6, 37.6, 62.6, 87.6, 113, 138, 163, 188, 213, 263         & 0, 38, 113, 188, 263 & \fullsquare\\
       100  & 0, 25.0, 50.0, 75.0, 100, 125, 150, 175, 200, 250                & NA & \fulldiamond \\
       150  & 0, 25.0, 50.0, 75.0, 100, 125, 150, 175, 225                       & 0,75,225& \fullcirc \\       
  \end{tabular}
  \caption{Parameters of the perturbation cells. See text for details.}
  \label{tab_cell}
  \end{center}
\end{table}
 
\begin{figure}
\centerline{\includegraphics[width=0.70\textwidth]{\figpath 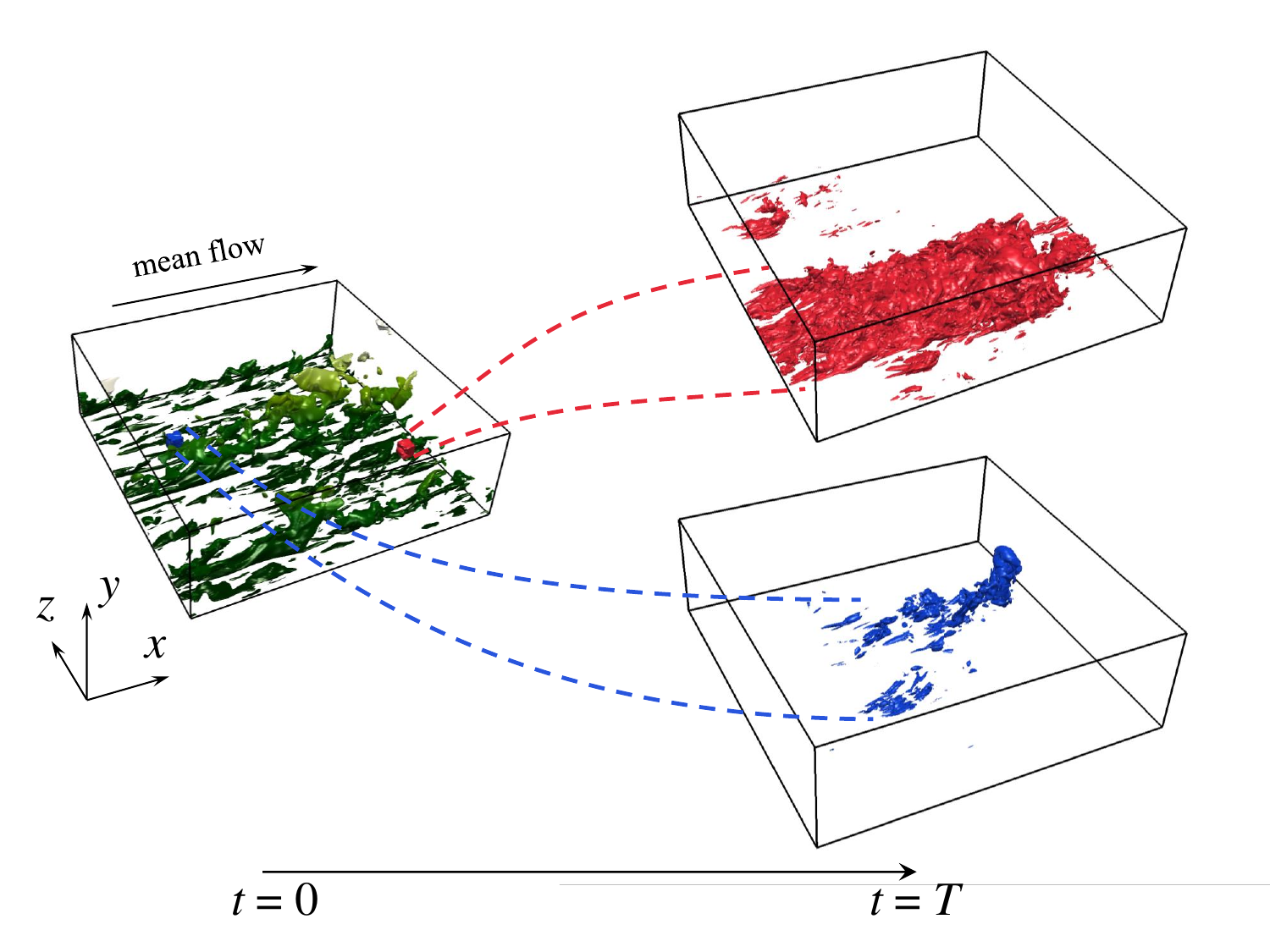}}%
\caption{ Schematic of the numerical experiment. Green, isosurface of the turbulent kinetic energy
for the reference flow at $t=0$, $|\vec{u}_\sref|^+=4.5$. Colour intensity encode the distance from
the wall; red, perturbation kinetic energy at some later time,
$|\vec{u}_\sref(T)-\vec{u}_\smod(T)|^+=0.17$, for a causally significant perturbation ; blue, same
for a causally irrelevant perturbation.
\label{fig_schem}
}
\end{figure}

As mentioned in the introduction, the interventional identification of causality follows
\cite{jimploff18,JJ20}. The idea is to apply a spatially localized initial perturbation to existing
turbulence, after which the flow is allowed to develop naturally (see figure \ref{fig_schem}). The
effect is measured after some time. Unlike the sensitivity analysis of the mean velocity profile in
\cite{Cherubini17}, each causality experiment is the response to a particular perturbation on a
particular location of a given flow snapshot, and the numerical experiment has to be repeated many
times for different snapshots and perturbations. The goal is to create a data base of responses from
which to extract the characteristics that make a particular flow location influential for the future
behaviour of turbulence (i.e., causally significant). To ensure independence, the 40 initial
reference snapshots used for our experiments are separated by at least 1.4 turnovers (defined as
$h/\ut$).

Perturbations modify the flow within a cubical cell of side $\lcell$ centred at $\vec{x}_c$.
Although there are countless choices for the form of the disturbance, and even if experience shows
that the manner in which the flow is disturbed influences the outcome of the experiment \citep{JJ20,
encinar23}, cost considerations limit us to a single perturbation scheme. Specifically, the flow is
modified by removing the velocity fluctuations within the cell, overwriting the velocity field with
its $y$-dependent cell average. Defining the cell average of a variable $f$ as
\begin{equation}
    \cave{f}(y)=\lcell^{-2} \int^{x_c+\lcell/2}_{x_c-\lcell/2}\int^{z_c+\lcell/2}_{z_c-\lcell/2}{f(x,y,z) \dd x \dd z},
    \la{fig:cellav}
\end{equation}
the perturbed velocity $\vec{u}_\smod$ is 
\begin{equation}
\vec{u}_\smod =
\begin{cases}
    \cave{\vec{u}_\sref} &\mbox{when}\; |x_j-x_{cj}|\leq\lcell/2 ,\\
    \vec{u} _\sref   &\mbox{otherwise.}
\end{cases}
\la{eq:perturb2}
\end{equation}
where $\vec{u}_\sref\/$ is the unperturbed flow, and an extra pressure step is applied after
\r{eq:perturb2} to restore continuity at the edges of the cell. The experiment is repeated as many
times as possible, applying it to different reference flow fields while changing the location and
size of the perturbation cell.

Table \ref{tab_cell} summarises the parameters of the experimental cells. They are expressed in
terms of the distance from the wall to the bottom of the cell, $\ycell= y_c-\lcell/2$, which was
found to collapse some results better than the cell centre, and are separated in two sets. In the
first one, involving the 40 reference snapshots, perturbations are applied to a $6\times 6$ grid of
cells evenly spaced in $x$ and $z$, such that their centres are separated by $\pi h/6$ in each
direction (approximately 315 wall units). In the $y$-direction, perturbations are applied at the
heights detailed in the second column of table \ref{tab_cell}, ranging from cells touching the wall
to those centred at the middle of the computational domain, $y_c^+\approx 300$. Each of them is run
for 0.65 turnovers, and consumes approximately 6 minutes in an Nvidia A100 GPU, so that the
approximately 76000 experiments in this set spent 318 GPU-days.

While these experiments test a wide range of sizes at sparsely spaced locations across
the flow, the ones in the last column of table \ref{tab_cell} aim at building heat maps
that explore possible large-scale causality distributions not limited to a single cubical cell. Each
reference snapshot is divided into a $30 \times 30$ grid in the $x-z$ plane, approximately spaced by
75 wall units, and perturbations are centred at each point of that grid. For cells with
$\lcell^+\ge 75$, this procedure uniformly samples the whole plane but, due to its cost, it
was limited to 20 initial snapshots and five different heights, each of which only ran for
0.49 turnovers. The resulting 180000 tests spent 565 GPU-days. 

In both sets of experiments, the temporal evolution of the perturbation is measured by the energy 
of the perturbation velocity integrated over the whole computational domain,
\beq
\erru (t)     = V^{-1}\, \int {|\vec{u}_\smod - \vec{u}_\sref|^2 dV},
\la{eq:pertu1}
\eeq
which evolves from some initial $\erru (0)$ at the moment at which the perturbation is applied, to
$\erru(\infty)=2 K\equiv 2V^{-1} \int |\vec{u}|^2 \dd V$ when the reference and perturbed flow fields
decorrelate after a sufficiently long time. For a chaotic system such as turbulence, $\erru(\infty)
\gg \erru(0)$ and, even if the evolution of the perturbation is far from linear over times of the
order of a turnover, the perturbation energy typically grows almost exponentially for a while
before levelling at $\erru(\infty)$. These considerations lead to two definitions of causal
significance: an absolute one that disregards the initial perturbation magnitude and vanishes as
$t\to \infty$,
\beq
\sigu (t)     = \log_{10} \erru(t) / \erru(\infty)=\log_{10} \erru(t) / 2K,  
\label{def_sigu}
\eeq
and a relative one,
\beq
\sigur (t)     = \log_{10} \erru(t) /\erru(0),
\label{def_sigur}
\eeq
which measures relative growth and vanishes at $t=0$. Both definitions typically grow with time, but
we will be interested in cases in which the growth is particularly fast or slow, as defined by the
top and bottom $\fsig$ percentile of the significance distribution. For most of the paper,
experiments within the top $\fsig=10\%$ of the significance distribution will be defined as
`causally significant', and those in the bottom 10\%, as `causally irrelevant'. This fraction is
broadly compatible with the percolation analysis often used to defined thresholds. For example, the
optimal percolation threshold in three-dimensional wall-bounded turbulence fills volume fractions of
order of 5\%--10\% \citep{jim18}, while in two-dimensional vorticity fields, which are more directly
comparable with the present application to individual planes, the covered area is closer to
20\%--30\% \citep{jimdipoles:20}. Tests using $\fsig=5\%$ or 15\% showed few differences in the
present results.

\section{Temporal evolution of the significance}\la{sec_signi}

\begin{figure}
\vspace{4mm}%
\centerline{%
\mylab{0.23\textwidth}{0.31\textwidth}{(a)}%
\includegraphics[width=0.43\textwidth]{\figpath 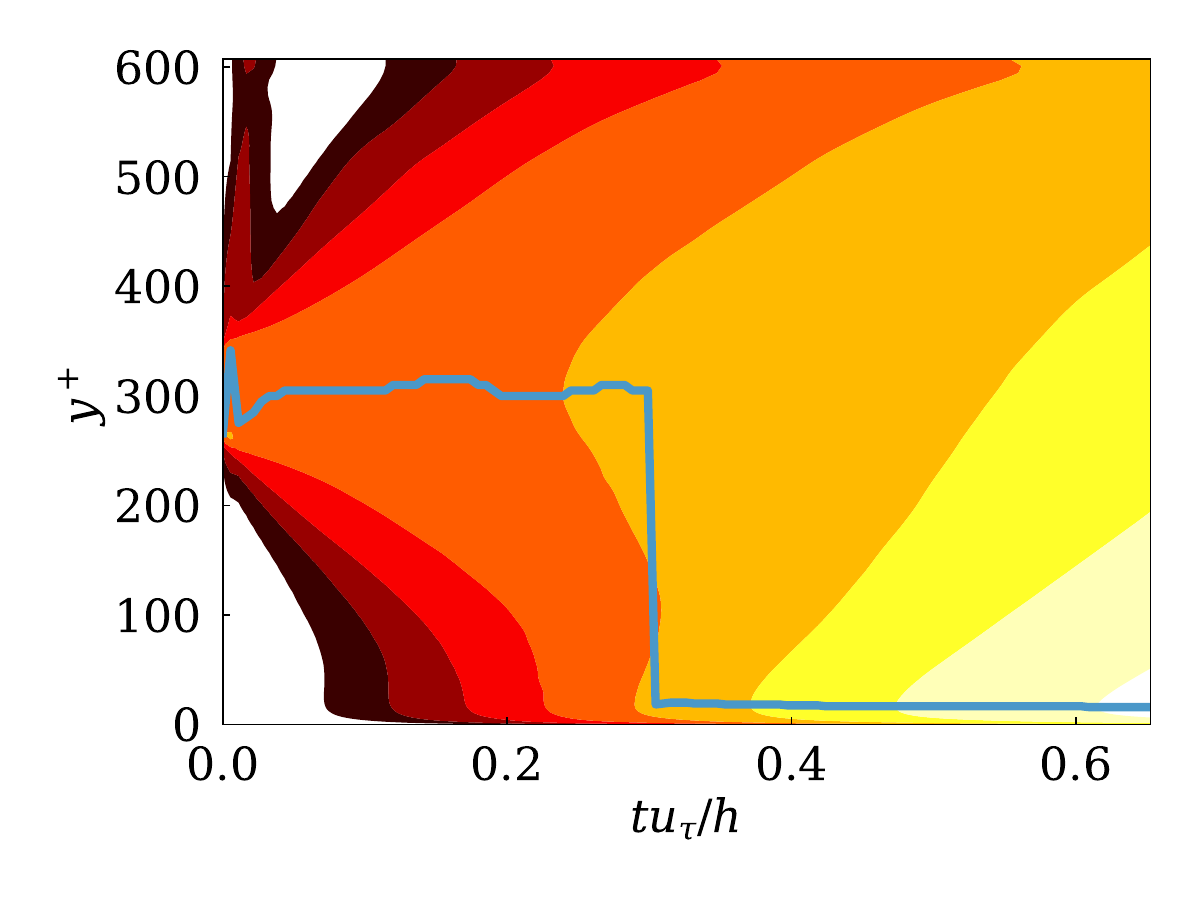}%
\hspace*{3mm}%
\mylab{0.23\textwidth}{0.31\textwidth}{(b)}%
\includegraphics[width=0.43\textwidth]{\figpath 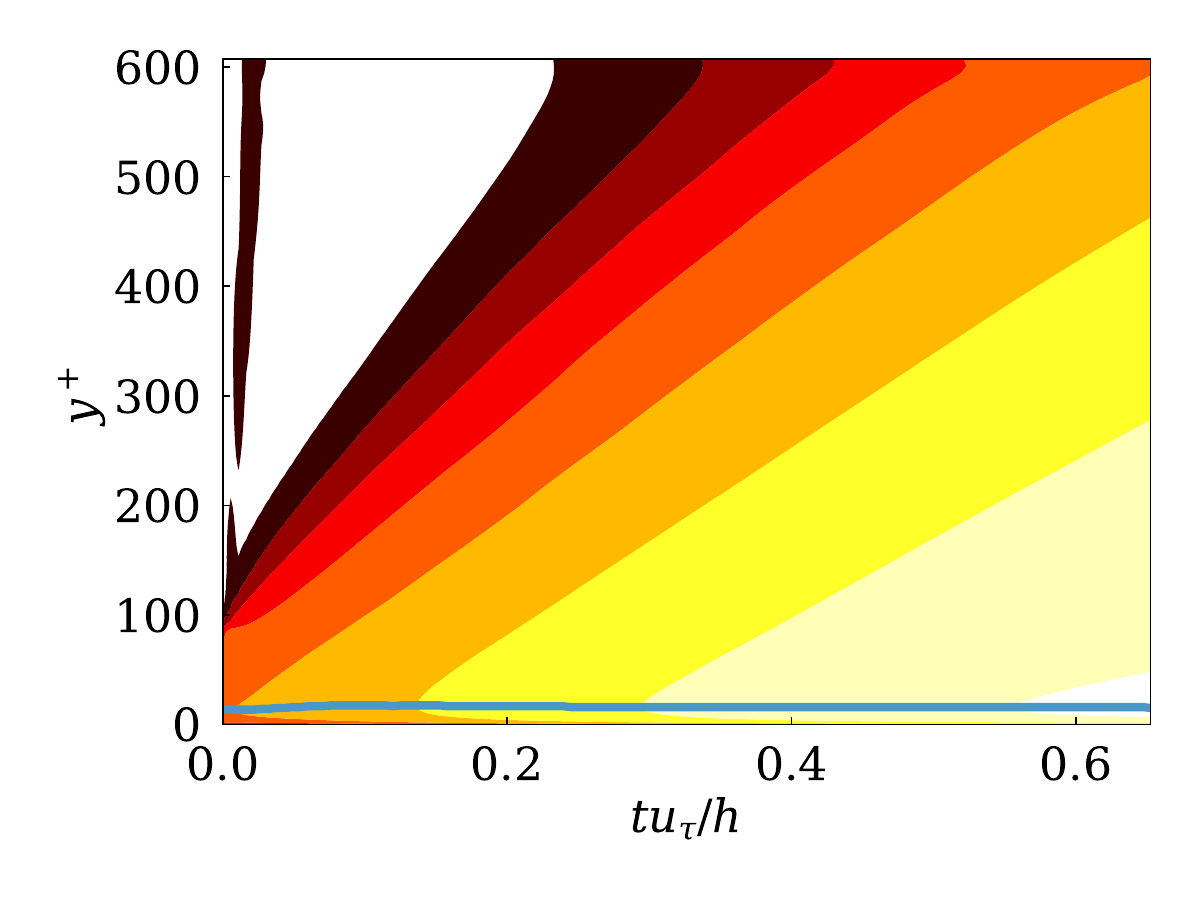}%
}%
\caption{Plane-averaged perturbation magnitude \r{eq:avexz}, unconditionally averaged over all
perturbations with $\lcell^+=75$ introduced at a given height, normalized with its maximum at $t=0$.
The light blue line is the instantaneous position of the perturbation maximum. Contours are
${\bra\erru\ket}(y,t)/\max_y\bra\erru\ket(y,0)= 10^{-4}(\times 10)10^3$. (a) $y_c^+=300$. (b)
$\ycell=0$.
\label{fig_err_y_t}}
\end{figure}

To further study the growth of the perturbations we use the $y$-dependent averaged  intensity, 
\beq
{\erru}(y,t)=(L_xL_z)^{-1}\, \iint |\vec{u}_\smod - \vec{u}_\sref|^2 \dd x \dd z,
\la{eq:avexz}
\eeq
equivalent to \r{eq:pertu1} but integrated over wall-parallel planes instead of over the whole
domain, together with corresponding definitions for the significances. To minimise notational
clutter, we use for them the same symbols as in (\ref{eq:pertu1}--\ref{def_sigur}), with the
inclusion of $y$ as a parameter. Figure \ref{fig_err_y_t}(a) shows the growth of $\erru(y)$,
unconditionally averaged over all the perturbations introduced at a particular size and distance
from the wall and normalised with its maximum at $t=0$. The light blue line is the
position at which the perturbation is maximum. It initially stays at the height at which the
perturbation is introduced, but a new peak grows near the wall and becomes dominant after
$t\ut/h\approx 0.3$. In figure \ref{fig_err_y_t}(b), where the perturbation is initially attached to
the wall, the peak is always attached. During the very early stage of evolution $(t\ut/h\lesssim
0.05)$, a low-intensity perturbation spanning the whole channel appears in both cases. This is
almost surely due to the pressure pulse that enforces continuity at the edges of the perturbation
cell, but it quickly dissipates and does not seem to influence later development.

\begin{figure}
\centerline{%
\includegraphics[height=0.35\textwidth]{\figpath 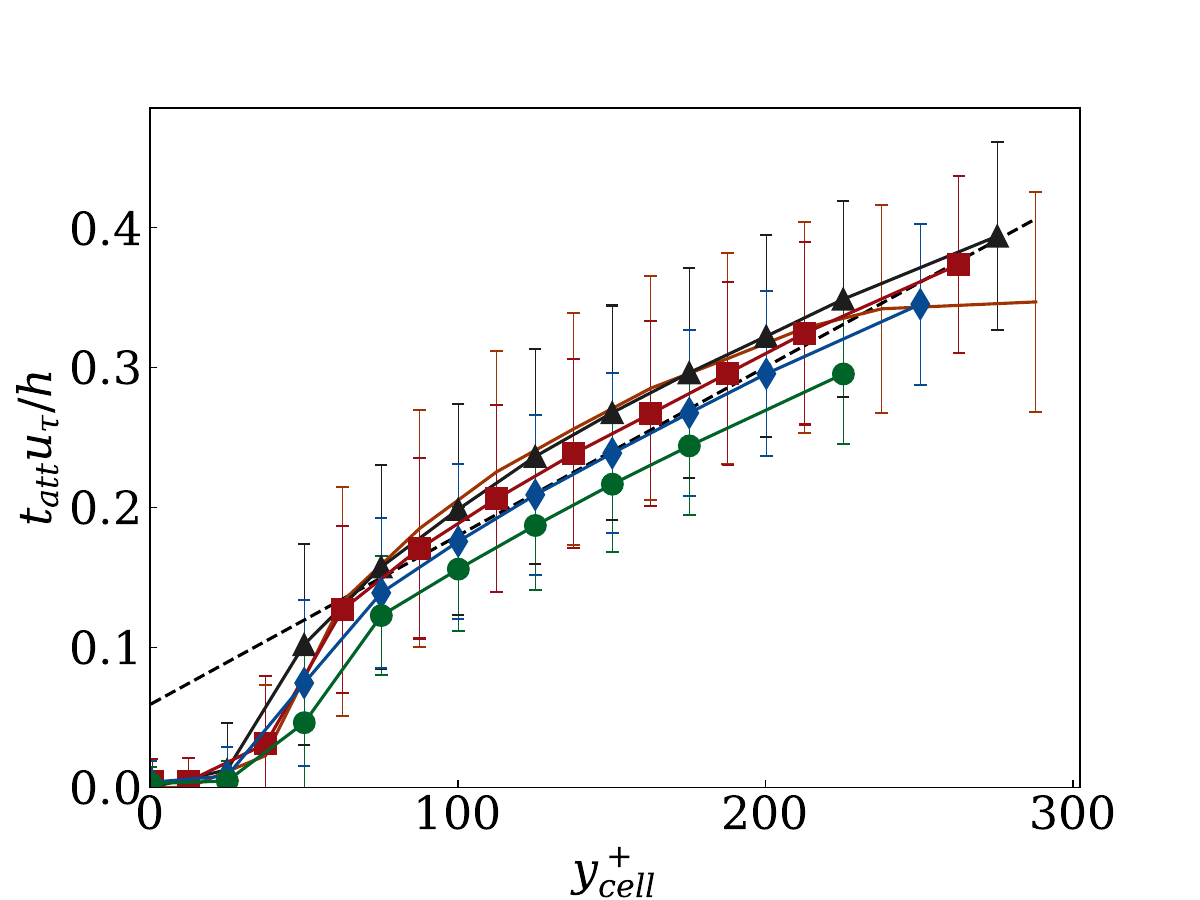}%
}%
\caption{Attachment time as a function of $\ycell$, for different $\lcell$. The dashed line is a
least-square fit to the curves with $\ycell>0$, with slope $1.37\ut$. Times are computed for
individual tests, and symbols and bars are their average and standard deviation. Symbols
as in table \ref{tab_cell}.
\label{fig_tatt}}
\end{figure}

Figure \ref{fig_tatt} shows that the attachment time, $\tatt$, defined for individual tests as the
moment when the perturbation maximum falls below $y^+=50$, and later averaged over the experimental
ensemble. It is approximately proportional to $\ycell$, at least for $\ycell^+\gtrsim 50$, with a
propagation velocity $\dr \ycell/\dr t=1.37\ut$. This is faster than the observed vertical advection
velocity of coherent features in channels, $\dr y/\dr t\approx \pm \ut$ \citep{lozano_time}, and
suggests that the perturbation is not simply advected by the flow, but actively amplified by it. In
fact, the production term in the evolution equation for the perturbation energy is proportional to
the mean shear (see Appendix \ref{sec_evoleq}), and the most likely interpretation of figure
\ref{fig_tatt} is that, while all perturbations are advected to and from the wall by the background
turbulence, those approaching the wall, where the shear is most intense, grow faster than those
moving away from it, resulting in a mean downwards migration of the perturbation maximum. Notice,
for example, the different slopes of downwards and upwards contours in figure \ref{fig_err_y_t}. It
is also relevant that $\tatt$ scales better with the distance from the wall to the bottom of the
cell, $\ycell$, than with its centre, $y_c$ (not shown), because it is the bottom that predominantly
feels the stronger shear near the wall.

\begin{figure}
\centerline{%
\includegraphics[height=0.35\textwidth]{\figpath 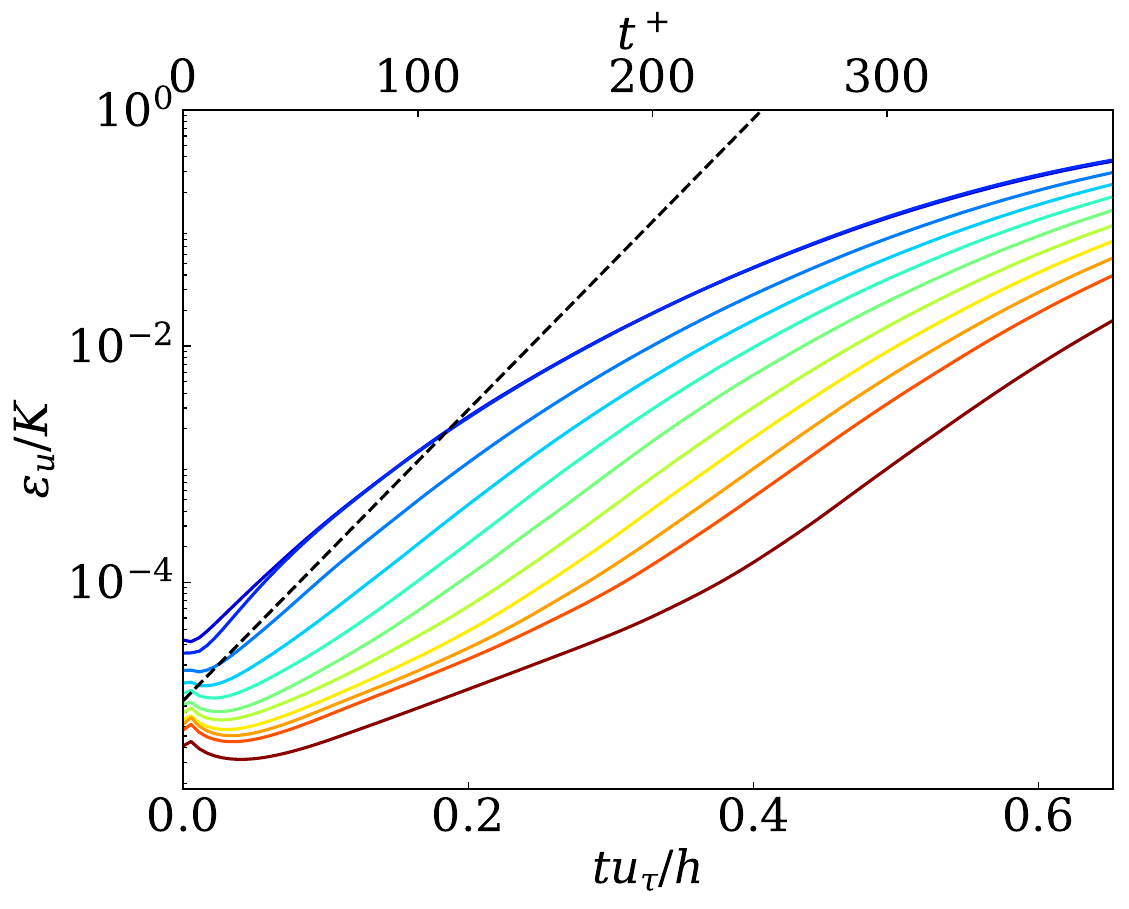}%
\hspace{3mm}%
 \includegraphics[height=0.35\textwidth]{\figpath 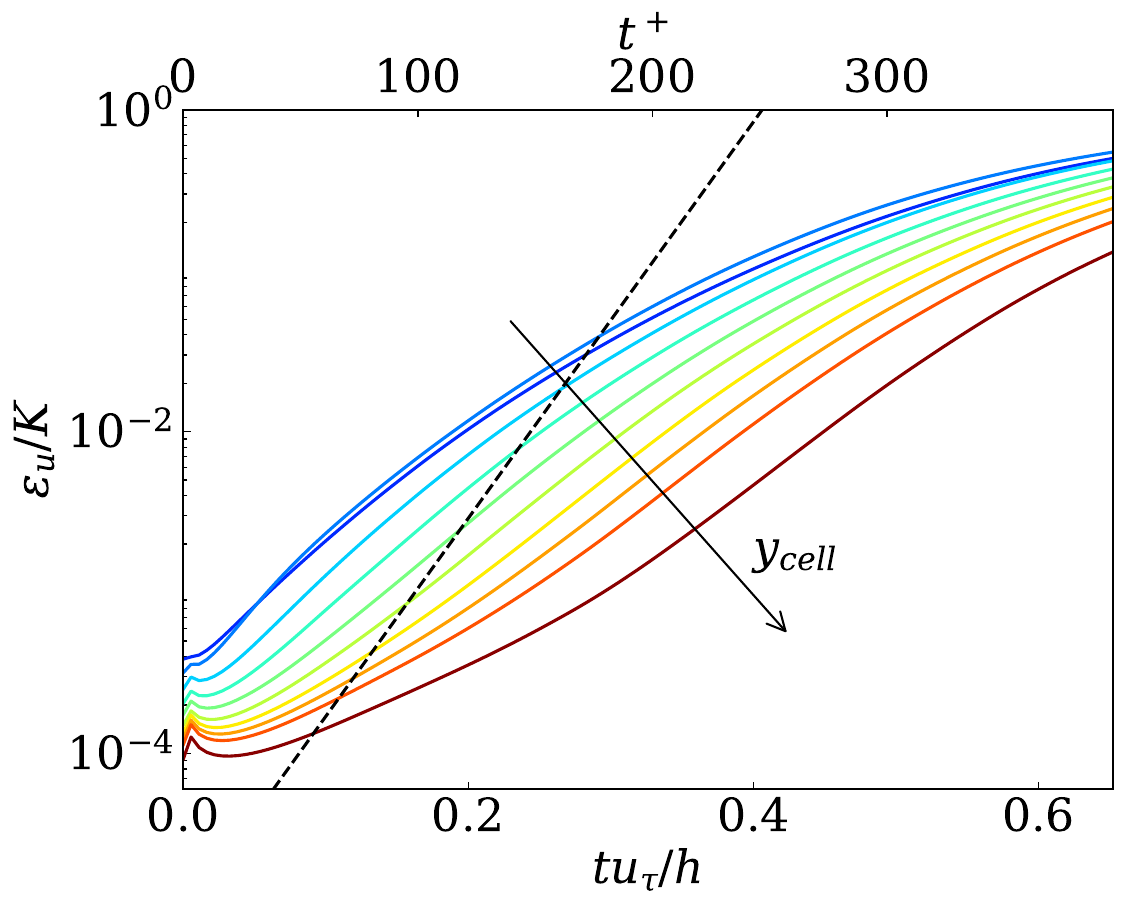}%
\mylab{-0.78\textwidth}{.27\textwidth}{(a)}%
\mylab{-0.35\textwidth}{.27\textwidth}{(b)}%
}%
\caption{Temporal development of the unconditionally averaged domain-integrated perturbation.
The cell distance from the wall increases from blue to red, as in table \ref{tab_cell}, and the
diagonal dashed lines are the exponential Lyapunov growth rate from \cite{Nikitin18}. 
(a) $\lcell^+=50$. (b) $\lcell^+=100$.
\label{fig_errt}}
\end{figure}

Fig \ref{fig_errt} shows two examples of the temporal evolution of the domain-integrated perturbation
$\erru$. The two panels in the figure are different cell sizes, and the line colour is the cell
height. The figure shows that $\erru$ is higher for larger cells, which is to be expected since it
is an integrated quantity, and also for lower $\ycell$, also expected for a perturbation that
removes velocity fluctuations, which are stronger near the wall. More interesting is that cells near the
wall grow faster than those away from it, which may be understood as supporting the model in which
their growth rate is controlled by the ambient shear.

The dashed straight lines in figure \ref{fig_errt} are the exponential growth from the Lyapunov
analysis by \cite{Nikitin18}, who reports a Lyapunov time for $\erru$ (the inverse of the leading
exponent) of $T_L^+\approx 19\, (\ut T_L/h=0.032)$ in a turbulent channel at $\Ret=586$. The
leading Lyapunov vector is concentrated in the buffer layer, and the exponent scales in wall units,
again consistent with a model in which the growth is controlled by the near-wall shear.

It is clear that our analysis shares many characteristics with the classical Lyapunov analysis,
albeit with important differences. The most obvious is that the classical Lyapunov exponent assumes
that the perturbation behaves linearly for an infinitely long time, while figure \ref{fig_errt} shows
that our experiments saturate for times that, even if much longer than $T_L$, remain of interest for
the flow evolution. A second important difference is that our initial perturbations, which are
intended to probe the local structure of the flow rather than its mean properties, are compact with
predetermined shapes, while those in Lyapunov analysis are allowed to spread across the flow field
to their optimal structure. It may be relevant in this respect that there is an initial transient in
which perturbations decay in most of our tests, $\ut t/h\lesssim0.1$, and that this period is
shorter for cells near the wall. This is reminiscent of the similar transient in Lyapunov
calculations, during which perturbations align themselves to the most unstable direction. Our
limited range of initial conditions is probably partly compensated by the substitution of the
temporal averaging of classical analysis by averaging over tests, and it is interesting that the
short-time growth rate of the smallest perturbations in figure \ref{fig_errt}(a), which mostly
sample the buffer layer, approximately agree with \cite{Nikitin18}. Larger or higher perturbations,
which sample weaker shears, grow more slowly. We will provide in \S\ref{sec_shear} further support
for the relevance of local shear to perturbation growth. 

\begin{figure}
\centerline{%
    \includegraphics[height=0.32\textwidth]{\figpath 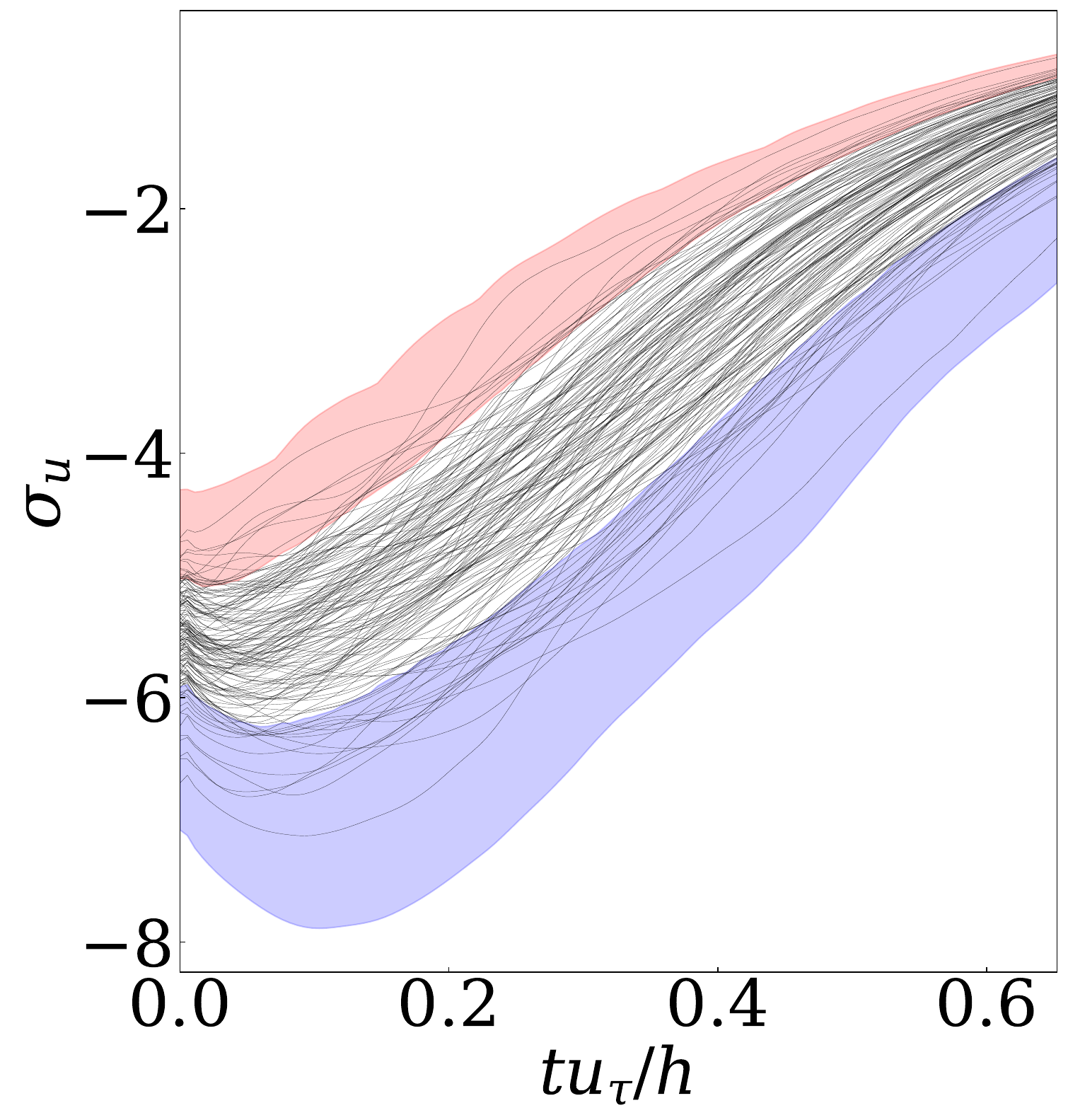}%
    \mylab{-0.23\textwidth}{0.29\textwidth}{(a)}
    \hspace{2mm}%
    \includegraphics[height=0.33\textwidth]{\figpath  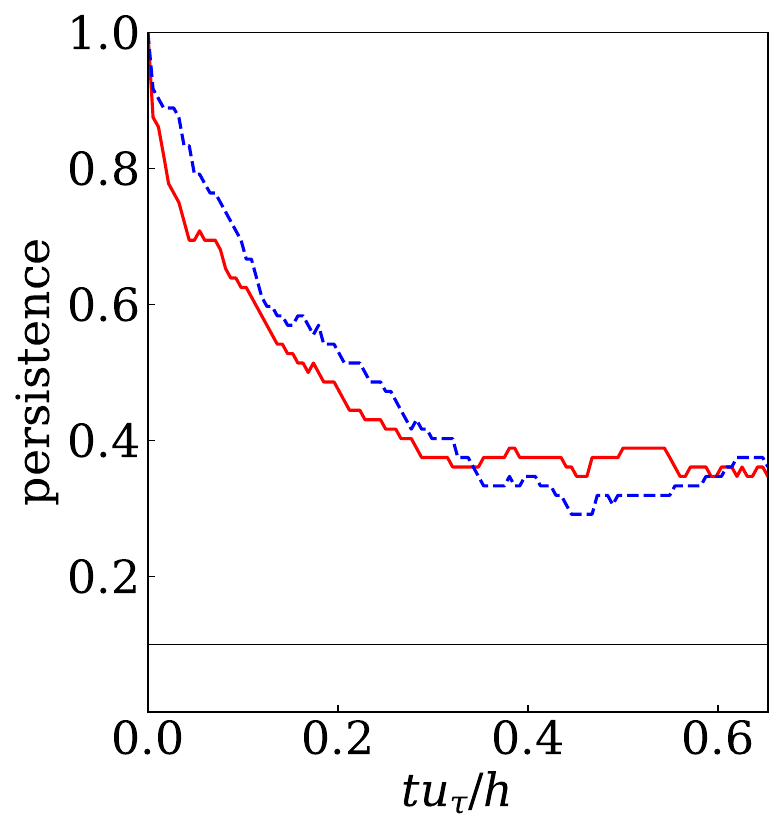}%
    \mylab{-0.05\textwidth}{0.28\textwidth}{(b)}%
    \hspace{2mm}%
    \includegraphics[height=0.32\textwidth]{\figpath  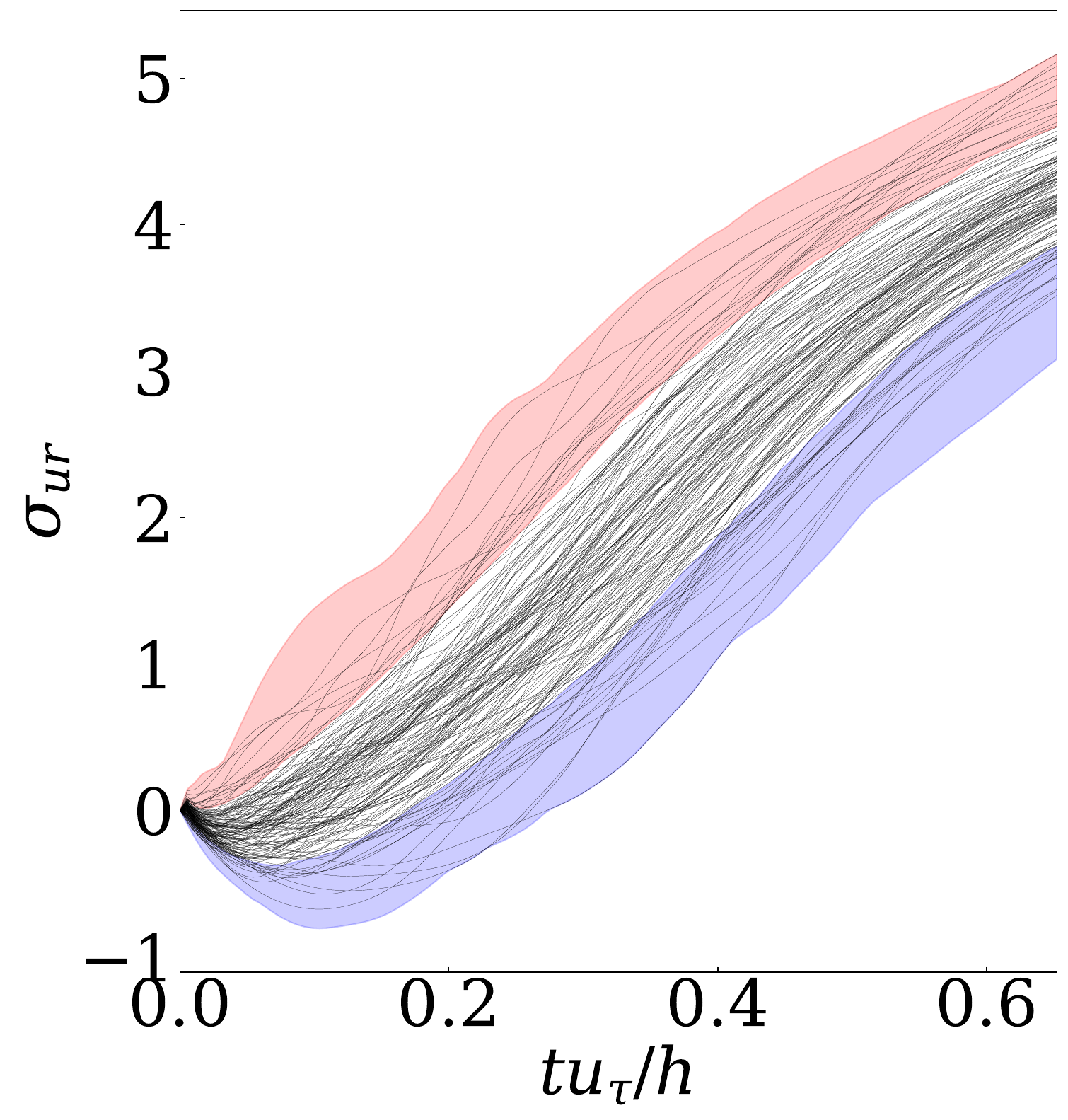}%
    \mylab{-0.23\textwidth}{0.29\textwidth}{(c)}%
}
\caption{(a) Lines are the absolute significance \r{def_sigu} of individual experiments as a function of time; only
20\% of the total are included. The red and blue bands are envelopes that respectively contain the 10\%
significant and irrelevant samples. 
(b) Fraction of experiments that continue to be classified as significant or irrelevant in terms of
$\sigu$ at different times, after being so classified at $t=0$. Red: significants; blue:
irrelevants. The black horizontal line is the probability threshold, $\fsig=10\%$.
(c) As in (a), for the relative significance \r{def_sigur}.
In all cases, $\lcell^+=50, \ycell^+=125$.
\label{fig_sigt}}
\end{figure}
\vspace{2mm}%
\begin{figure}
    \centering
    \includegraphics[height=0.32\textwidth]{\figpath 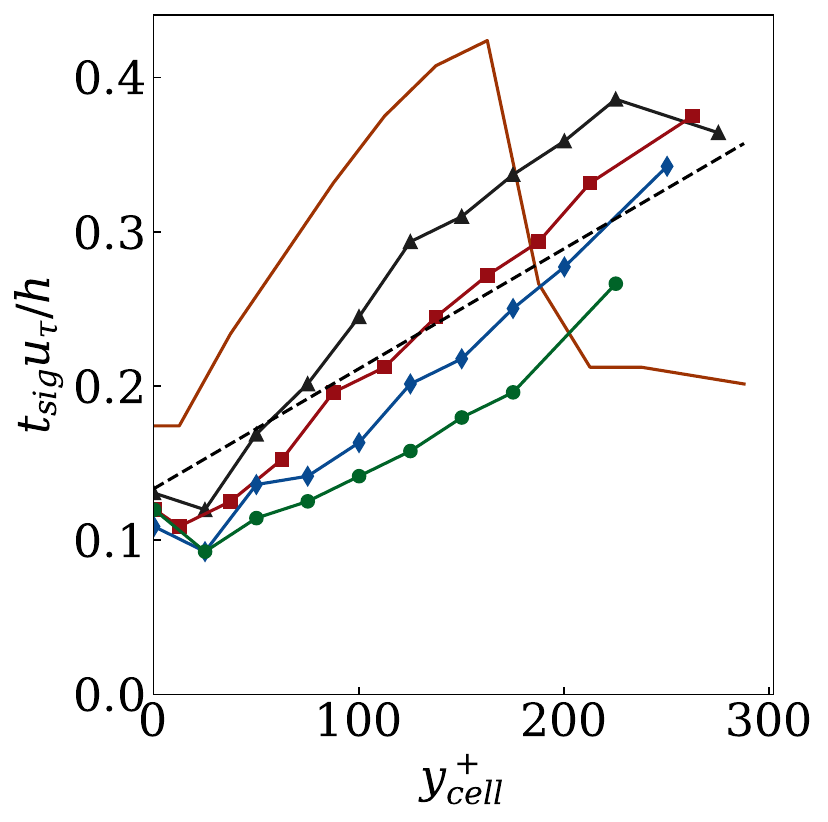}%
\mylab{-0.25\textwidth}{.27\textwidth}{(a)}%
    \includegraphics[height=0.32\textwidth]{\figpath 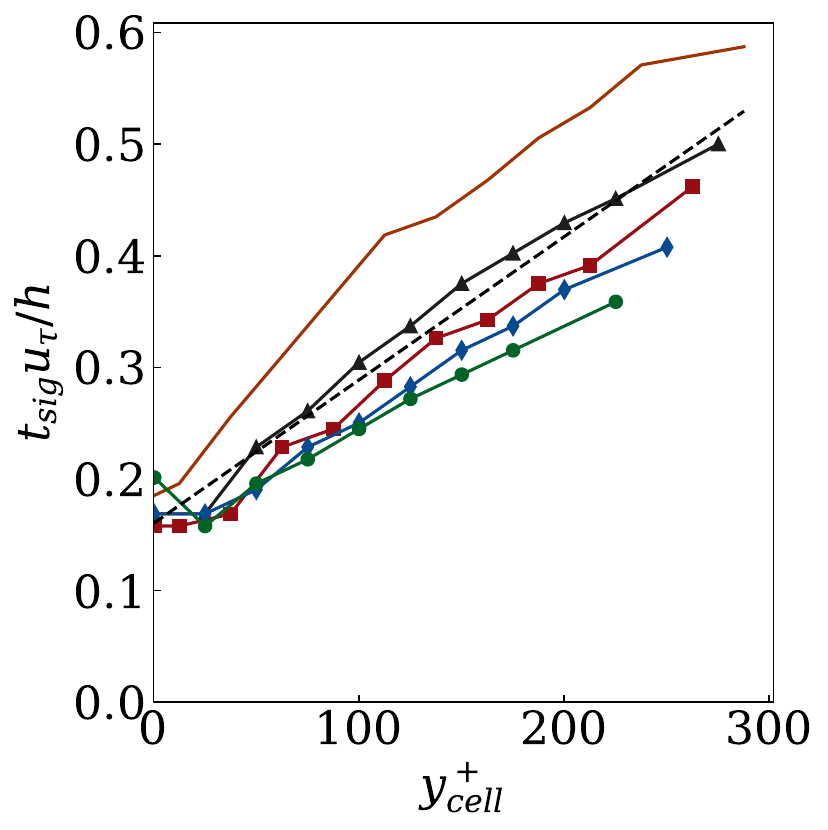}%
\mylab{-0.25\textwidth}{.27\textwidth}{(b)}%
    \includegraphics[height=0.32\textwidth]{\figpath 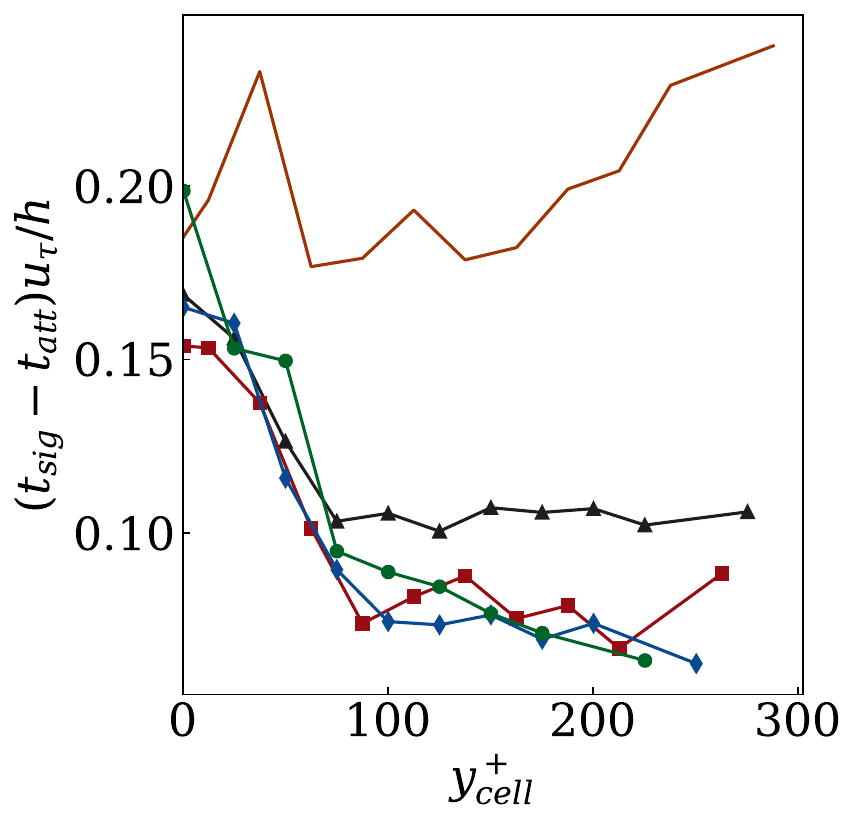}%
\mylab{-0.055\textwidth}{.27\textwidth}{(c)}%
\caption{Classification time $\tsig$ as a function of $\ycell$. Symbols as in table \ref{tab_cell}.
The dashed lines in (a,b) are least-square linear fits, whose slope is $2.12 u_\tau$ in (a) and
$1.28 u_\tau$ in (b).
(a) Using $\sigu$. (b) Using $\sigur$. (c) Offset between the $\sigur$ classification time and the
attachment time.
\label{fig_tsig}}
\end{figure}

Figure \ref{fig_sigt}(a) shows a typical evolution of the absolute significance, $\sigu$, for 
perturbations with a given $\lcell$ and $\ycell$. Each of the grey lines is the result of a different
experiment, and the red and blue regions are the envelopes of the significant and irrelevant
samples, individually classified according to their intensity at each moment of their evolution. The
figure shows that the perturbations approximately maintain the ordering of their initial intensity.
Initially stronger perturbations tend to remain strong for long times, although it follows from
its definition that $\sigu$ vanishes on average as $t\to\infty$.
Figure \ref{fig_sigt}(b) displays the persistence of the causality classification based of $\sigu$,
defined as the fraction of samples identified as significant or irrelevant at $t=0$ that remain
significant or irrelevant when classified at subsequent times. In the case illustrated in the figure, 34\% of the
initially significant samples and 29\% of the initially irrelevant ones remain at the end of our
experiments in the same class in which they were classified at $t=0$. This fraction is at least 20\% in
all the experiments  in this paper, which is substantially higher than the
10\% expected from a random selection.
 
Figure \ref{fig_sigt}(c) shows the evolution of the relative significance. Unlike the absolute
significance, $\sigur$ vanishes at $t=0$ but does not reach the same long-time limit in all cases.
In fact, $\sigur(\infty) = \log_{10} (2K) - \log_{10} \erru(0)$, and $\sigur(\infty)$ is essentially
equivalent to the initial perturbation magnitude.

These considerations show that both $\sigu$ and $\sigur$ characterise the evolution of the
perturbations at short and intermediate times. The former mostly reveals that the perturbation
intensity stays approximately proportional to its initial value for some time, while the latter,
which compensates for this effect, describes its intrinsic growth. When $\erru\to 2K$ at longer
times, the system forgets its initial conditions and neither measure of significance is very useful.

Although figure \ref{fig_sigt}(b) shows that the significance classification of a given experiment
is not a completely random variable, the fact that the persistence is not unity implies that the
time at which the classification is performed is important. Consider, for example, the mean
significance of the set $\{I\}$ of tests classified at time $t_c$ as irrelevant, $\sig_I(t_c) =N_{\{I\}}^{-1}
\sum_{j\in\{I\}} \sig_j$, and define a similar $\sig_S (t_c)$ average for significant perturbations. The
difference $\sig_S-\sig_I$ typically increases initially and reaches a maximum before decaying at
long times. The time, $\tsig$, at which this difference is maximum is also when the classification
is less ambiguous, and we will preferentially use it from now on to define our significance classes.

Figure \ref{fig_tsig}(a,b) shows how $\tsig$ changes as a function of $\lcell$ and $\ycell$, using
either $\sigu$ or $\sigur$ as a causality measure. Disregarding the case $\lcell^+=25$, which is
well within the dissipative range of scales and tends to behave differently from larger cells,
$\tsig$ is mainly explained by $\ycell$, and it is clear that $\sigur$ is a better indicator for
this purpose than $\sigu$. We will mostly use it from now on. It is interesting that $\tsig$ is very
close to, and generally slightly larger than, the attachment time in figure \ref{fig_tatt}, $\tatt$,
as shown by the difference of the two values in figure \ref{fig_tsig}(c), suggesting again that the
arrival of the disturbances to the wall is an important factor in determining causality.

We have seen above that the initial intensity of the perturbations has an effect on their subsequent
evolution. This is also true of our significance classification, and can be quantified by the
correlation of $\sigur(t_c)$ with $\erru(0)$ (not shown). This correlation tends to $-1$ at long
times, as explained above, but remains moderately positive for $t_c\lesssim \tsig$, confirming that the
initial relative growth rate for strong perturbations is faster than for weak ones. In most cases,
$\tsig$ approximately coincides with the moment at which the correlation changes sign and is close
to zero, making the classification relatively independent of the initial perturbation intensity. At
this moment, the energy of the perturbation is still a small fraction of the total energy of the
flow. Taking as example the topmost curve in figure \ref{fig_errt}(b) $(\lcell^+=100,\,
\ycell=0)$, the average energy of the initial perturbations is approximately $3\times 10^{-4} K$,
and grows to $0.6 K$ at the end of the experimental runs, but it is still $8\times 10^{-3} K$ at the
optimal classification time $\tsig\approx 0.17 h/\ut$. This does not mean that the perturbation can
be linearised up to that time. The intensity of the perturbation is always $O(K)$, and the growth of
its integrated energy is mostly due to its geometric spreading(see figure \ref{fig_schem}).

%

\section{Diagnostic properties for causal significance}\label{sec_classify}

Having described how the significance of an initial condition can be characterised, we recover our
original task of determining which properties of the perturbed cells are responsible for their
causality. The basic assumption is that the characterisation of causality can be reduced to a
single observable of the cell at the perturbation time, $t=0$, such as its average vorticity, rather
than requiring several conditions to be simultaneously satisfied, or even some property of the
extended environment of the cell, or of its history. As mentioned in the introduction, the strategy is to perform many
experiments modifying individual cells, to label them according to their significance at some later
time, and to test which cell observables at $t=0$ can be used to separate the
classes thus labelled.

\begin{table}
\renewcommand{\colwd}{3mm}
\centering
\def~{\hphantom{0}}
    \begin{tabular}{>{\hspace{\colwd}}l>{\hspace{\colwd}}l<{\hspace{\colwd}}}
     $\cave{u_i}$, $\cave{\omega_i}$ &  Mean velocities and vorticities.\\
     $\cave{u_i^2}$  & Kinetic energy. \\
     $\cave{uv}$, $\cave{(uv)^2}$ & Mean and mean-squared Reynolds product.\\    
     $\cave{u_i^2}$, $\cave{\omega_i^2}$ (no sum over $i$) & Mean-squared components.\\
      $\cave{\mbox{\it Prod}}=-\cave{uv\partial_yU}$ & Mean energy production.  \\
     $\cave{\mbox{\it Tdif}}=-\cave{\partial_y v u_i^2}/2$  & Mean energy transport. \\ 
     $\cave{\partial_y u}$   & Mean shear. \\
     $\sfw$  & Mean shear using only $y^+>5$. \\
     $\cave{\partial_i u_i}$ (no summation)  & Mass conservation.  \\
\end{tabular}
\caption{Cell observables. All averages are taken over cells.}%
\la{tab_obscell}
\end{table}
\begin{table}
\centering
\def~{\hphantom{0}}
    \begin{tabular}{>{\hspace{\colwd}}l>{\hspace{\colwd}}l<{\hspace{\colwd}}}
     $\erru$, $\errw$  &  Mean squared velocity and vorticity fluctuations.\\
     $\cave{\omega_i^2}$, $\cave{\sijsij}$  & Enstrophy, strain. \\
     $\cstd{u_i}$, $\cstd{\omega_i}$ (no sum over $i$) & In-cell standard deviations. \\
     $\cave{\Perr}$, $\cave{\Cerr}$, $\cave{\Derr}$  & Production, transport and dissipation of $\erru$
   (Appendix\ref{sec_evoleq}).  \\
\end{tabular}
\caption{Perturbation and small-scale observables}
\la{tab_obspert}
\end{table}

Following \cite{JJ20} and \cite{encinar23}, the ranking of observables uses a linear kernel Support
Vector Machine \citep[SVM,][]{SVM00}, implemented in the \texttt{\small scikit-learn} Python library
\citep{scikit-learn}, which determines an optimal separating hyperplane between two pre-labelled
data classes. In our case, we look for the optimal separation of significant or irrelevant
experiments in terms of a single quantity, and the SVM hyperplane reduces to a threshold. For each
combination of $\lcell$ and $\ycell$ in the second column of table \ref{tab_cell}, and for each
classification time $t_c$, two-thirds of the initial conditions are collected into a training set,
with the remaining third reserved for testing. The 10\% most significant experiments of the training
set are labelled as significant, and the bottom 10\%, as irrelevant. The remaining 80\% are not used
for classification purposes. An optimum partition threshold is computed for each of the observables
detailed below, and a SVM classification score is assigned to each observable using the test set.
The score measures the fraction of data allocated to their correct class by the SVM threshold, and
ranges from unity for perfect separability to 0.5 for cases in which the two classes are fully
mixed. The procedure is repeated three times after randomly separating the data into training and
test sets, and the diagnostic score for the observable is defined as the average of the three
results.

The whole process can be automated and is reasonably fast. The experimental description in
\S\ref{sec_perturb} shows that each SVM run is only requested to classify two sets of 96 points each, and to
test the classification on two sets of 48 points. This allows us to minimise preexisting biases by
testing many possible observables.

The observables can be physically classified into average cell properties, such as its kinetic
energy, and perturbation or small-scale properties, such as the kinetic energy of the velocity
fluctuations with respect to the cell mean. The former are summarised in table \ref{tab_obscell},
and the latter in table \ref{tab_obspert}. In both cases, properties that are statistically
symmetric with respect to reflections on $z$ are used as absolute values, and positive definite
quantities, such as mean squares, are used as logarithms. Otherwise, all observables are processed
in the same way.

\begin{figure}
    \centering
    \includegraphics[width=0.95\textwidth]{\figpath 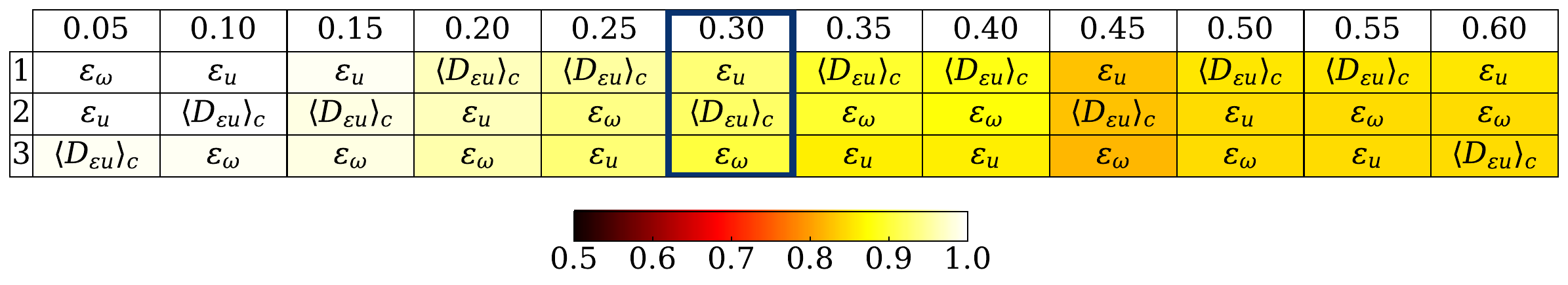}
\caption{Classification score of the three best observables for a classification based on
absolute significance, $\sigu$. Row: rank; Column: evaluation time in turnovers. Colour:
classification score. $\lcell^+=50$, $\ycell^+=125$.
\label{fig_class_yt_erru}}
\end{figure}

The diagnostic score of an initial condition depends on the cell height, on its size, and on the
moment at which it is classified. Figure \ref{fig_class_yt_erru} shows a typical table of the three
best observables identified by the absolute significance $\sigu$, as functions of the classification
time. Cells are coloured by the classification score. In all cases, the best observable is the
initial perturbation amplitude $\erru$, a related quantity such as $\errw$, or the viscous
dissipation of the perturbation intensity $\cave{\Derr}$. Although not included in the table, the
next best observable is usually also a small-scale quantity closely correlated with the intra-cell
velocity fluctuations, such as $\sijsij$, $\omega_i^2$ or the in-cell standard deviation $\cstd{v}$.
Table \ref{fig_class_yt_erru} is thus equivalent to the fluctuation persistence in figure
\ref{fig_sigt}(a,b).

\begin{figure}
\centerline{\includegraphics[width=0.95\textwidth]{\figpath 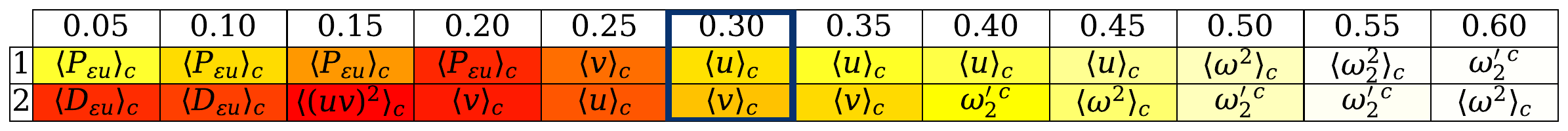}}%
\centerline{\includegraphics[width=0.95\textwidth]{\figpath 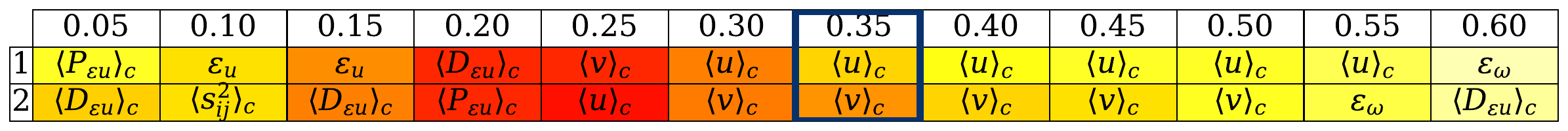}}%
\centerline{\includegraphics[width=0.95\textwidth]{\figpath 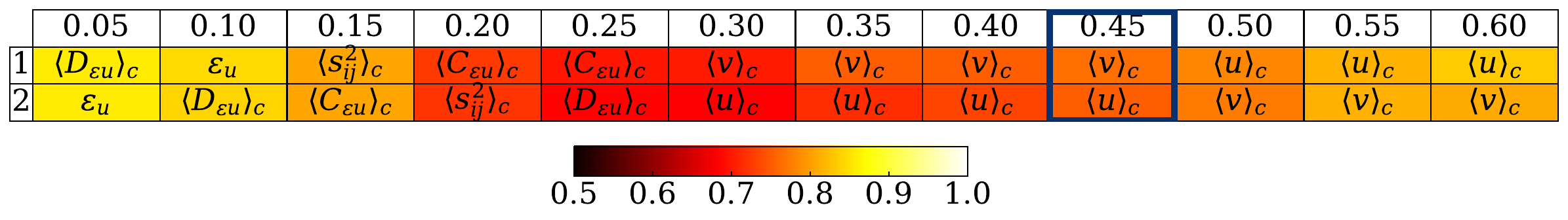}}%
\mylab{-0.01\textwidth}{0.26\textwidth}{(a)}%
\mylab{-0.01\textwidth}{0.185\textwidth}{(b)}
\mylab{-0.01\textwidth}{0.105\textwidth}{(c)}%
    \caption{ Classification score of  the two best observables for a classification based on
relative significance, $\sigur$.
    (a) $\lcell^+=150,\,\ycell=150$, (b) $\lcell^+=75,\,\ycell=138$, (c) $\lcell^+=25,\,\ycell=138$.
    Row: rank; Column: evaluation time in turnovers. The highlighted columns are $\tsig$.
    Colour: classification score.
    $\ycell^+\approx 150$.
\label{fig_class_yt_errur_lcell}}
\vspace{3mm}%
\centerline{\includegraphics[width=0.95\textwidth]{\figpath 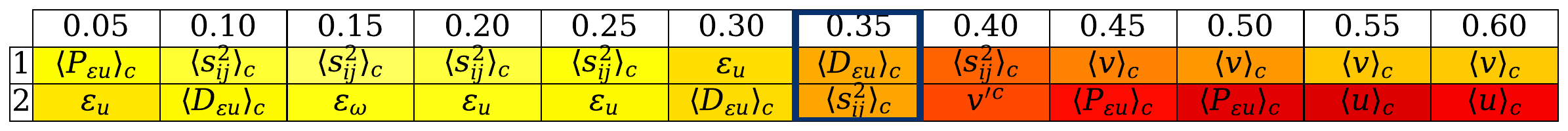}}%
\centerline{\includegraphics[width=0.95\textwidth]{\figpath 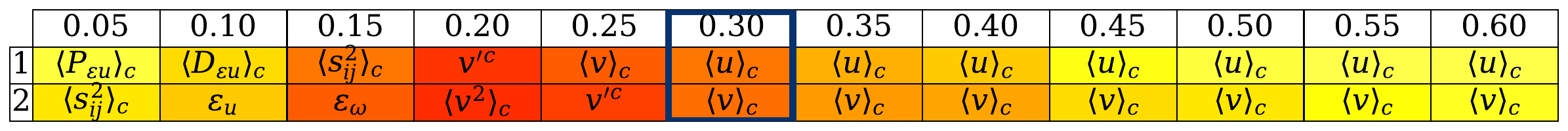}}%
\centerline{\includegraphics[width=0.95\textwidth]{\figpath 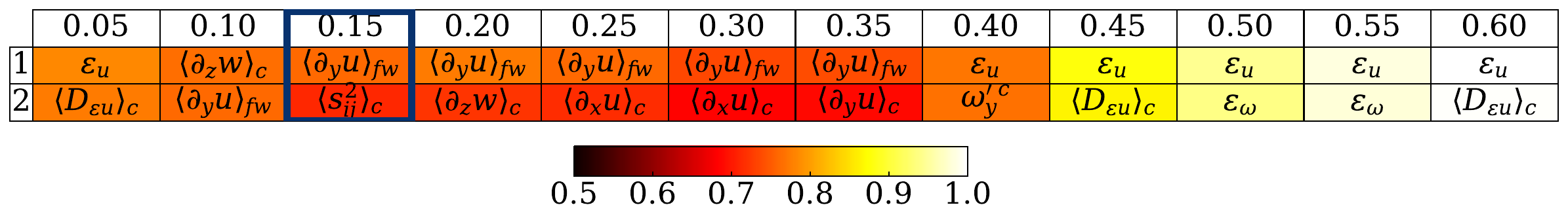}}%
%
\mylab{-0.01\textwidth}{0.26\textwidth}{(a)}%
\mylab{-0.01\textwidth}{0.185\textwidth}{(b)}
\mylab{-0.01\textwidth}{0.105\textwidth}{(c)}%
\caption{Classification score of the two best observables for a classification based on
relative significance, $\sigur$. (a) $\ycell^+=275$, (b) $\ycell^+=125$, (c) $\ycell=0$. Row: rank;
Column: evaluation time in $h/\ut$. The highlighted columns are $\tsig$. Colour: classification score.
$\lcell^+=50$.
\label{fig_class_yt_errur_ycell}}
\end{figure}

\begin{figure}
    \centerline{%
    \includegraphics[width=0.405\textwidth]{\figpath 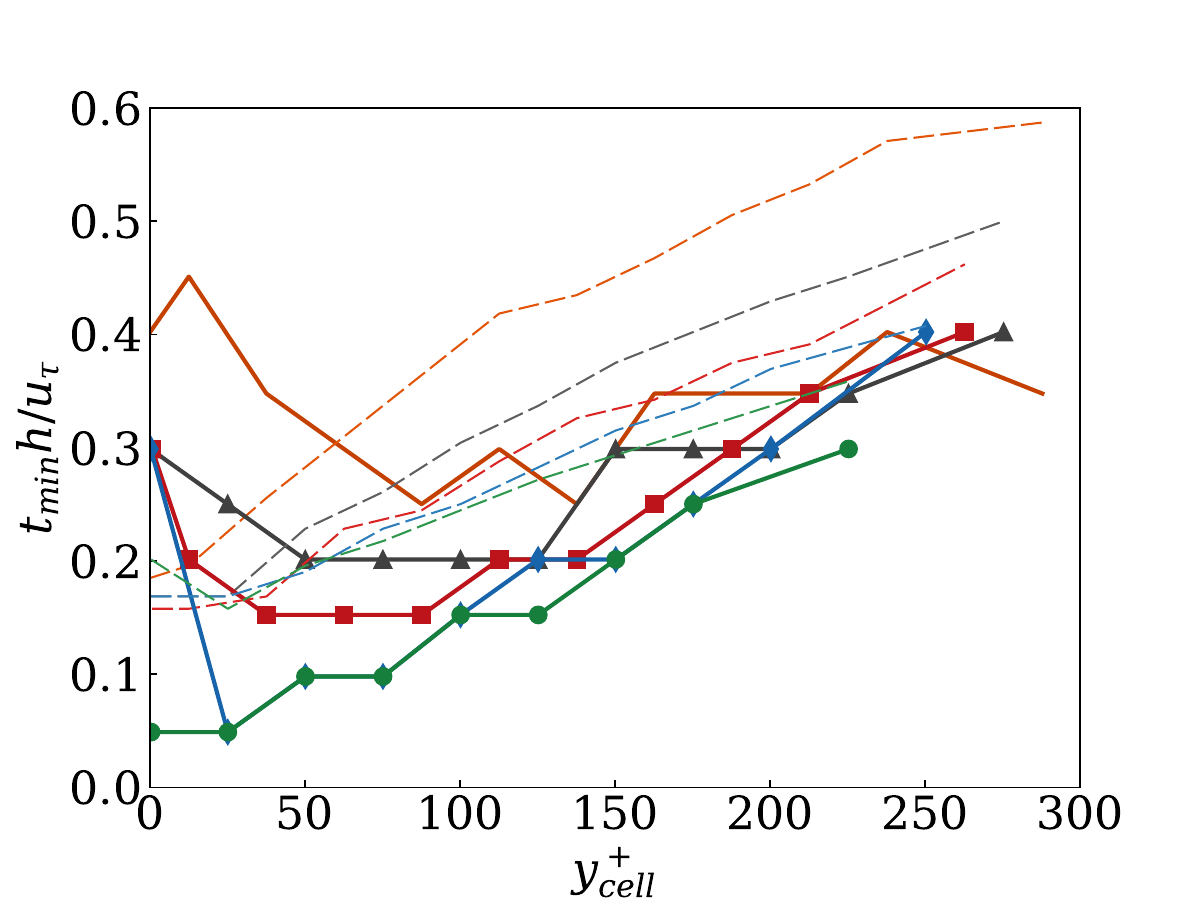}%
    \mylab{-0.33\textwidth}{0.24\textwidth}{(a)}%
\hspace{2mm}%
    \includegraphics[width=0.4\textwidth]%
    {\figpath 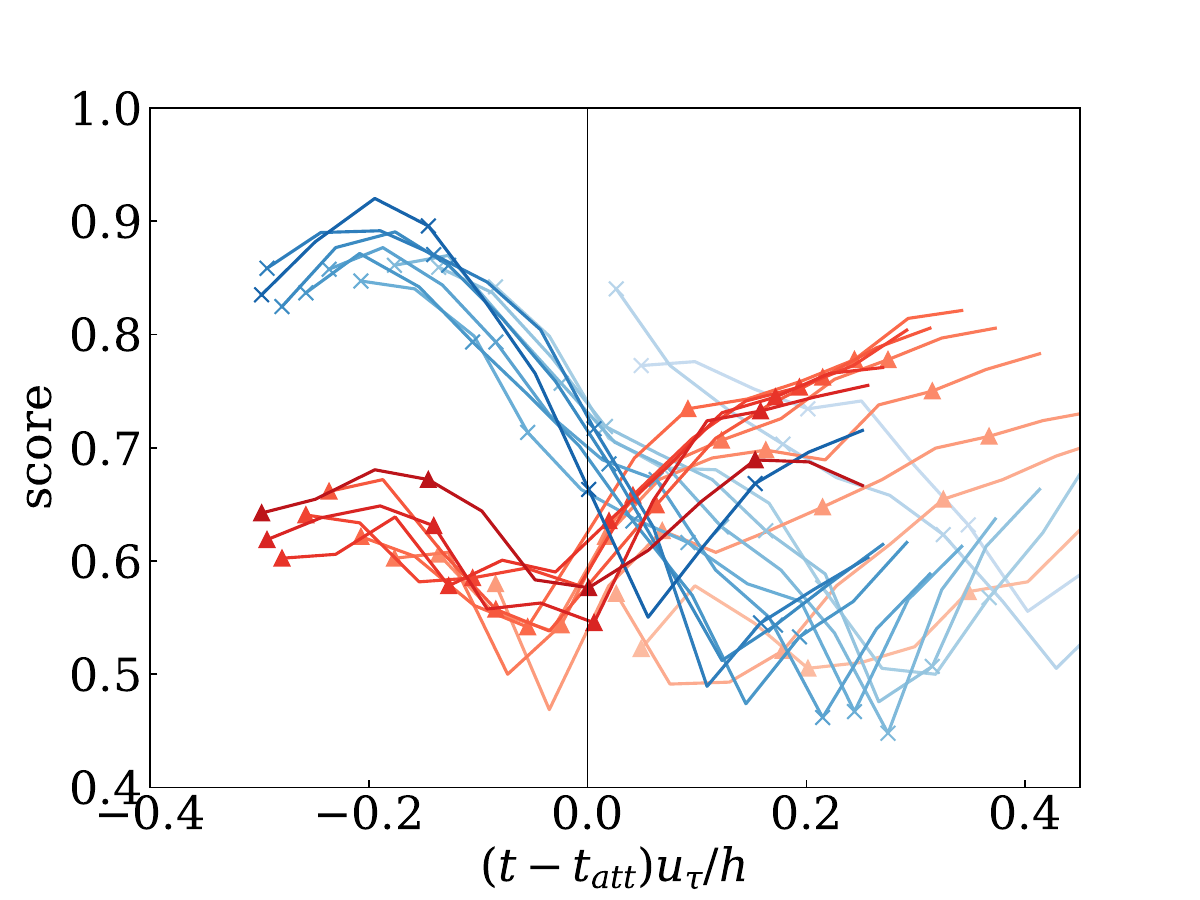}%
    \mylab{-0.33\textwidth}{0.24\textwidth}{(b)}%
}
\vspace{3mm}%
\centerline{%
    \includegraphics[width=0.4\textwidth]%
    {\figpath 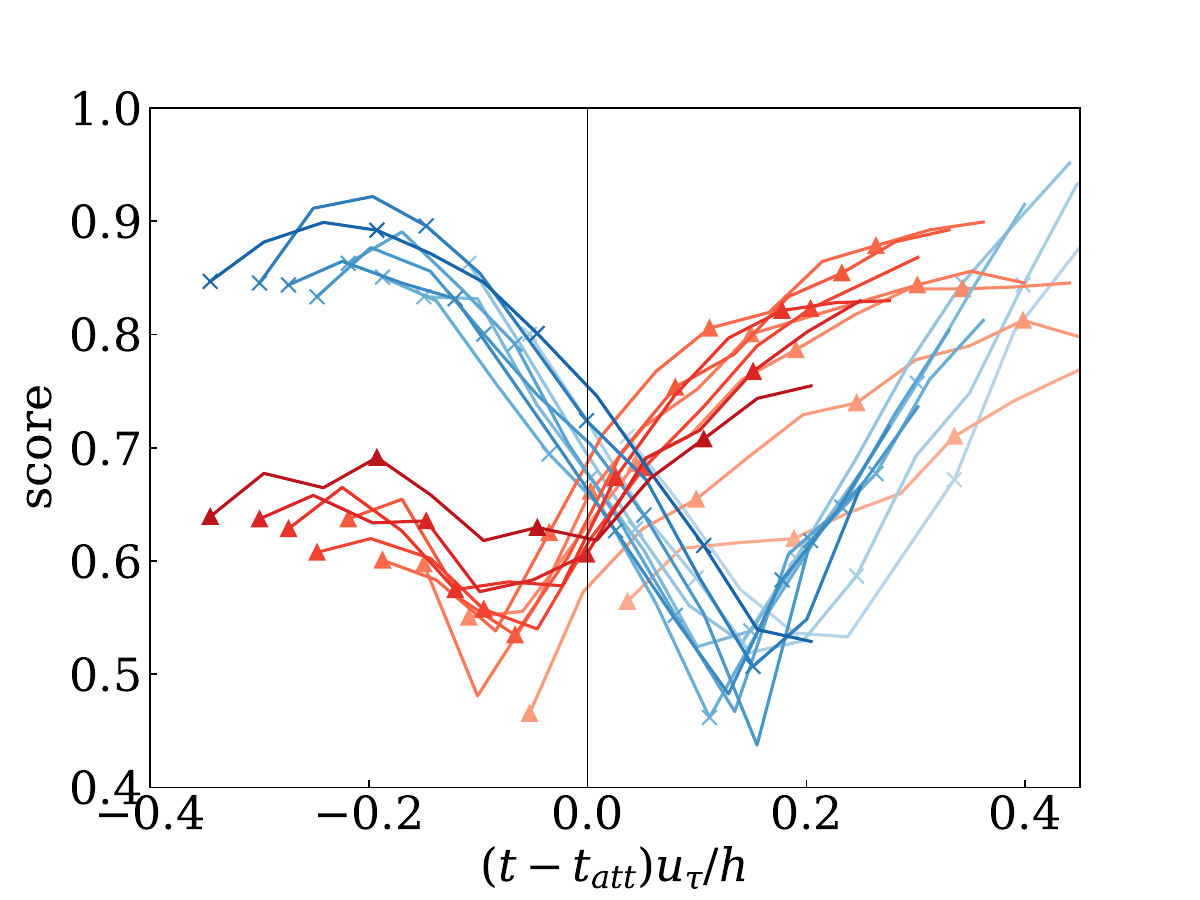}%
    \mylab{-0.33\textwidth}{0.24\textwidth}{(c)}%
\hspace{3mm}%
    \includegraphics[width=0.4\textwidth]%
    {\figpath 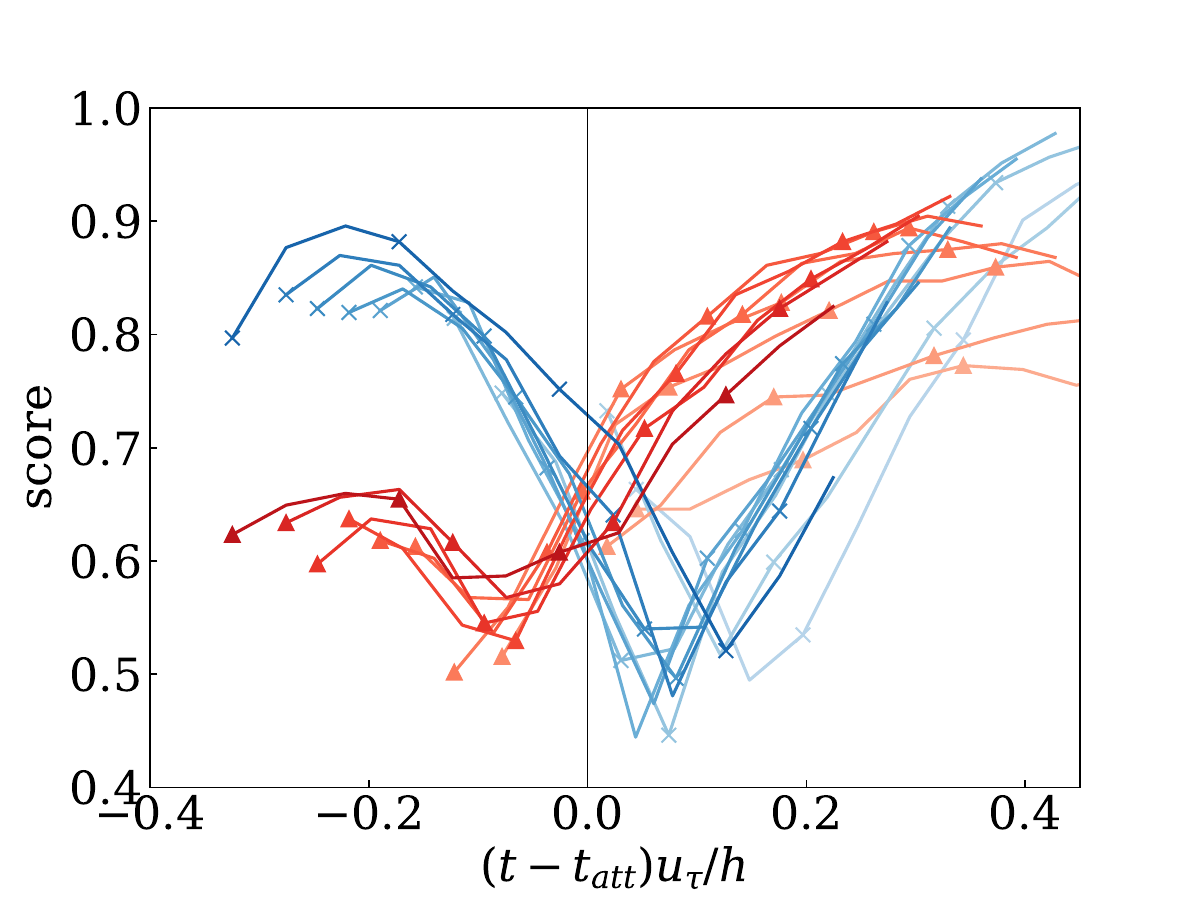}%
    \mylab{-0.33\textwidth}{0.24\textwidth}{(d)}%
}
\vspace{3mm}%
\centerline{%
    \includegraphics[width=0.4\textwidth]%
    {\figpath 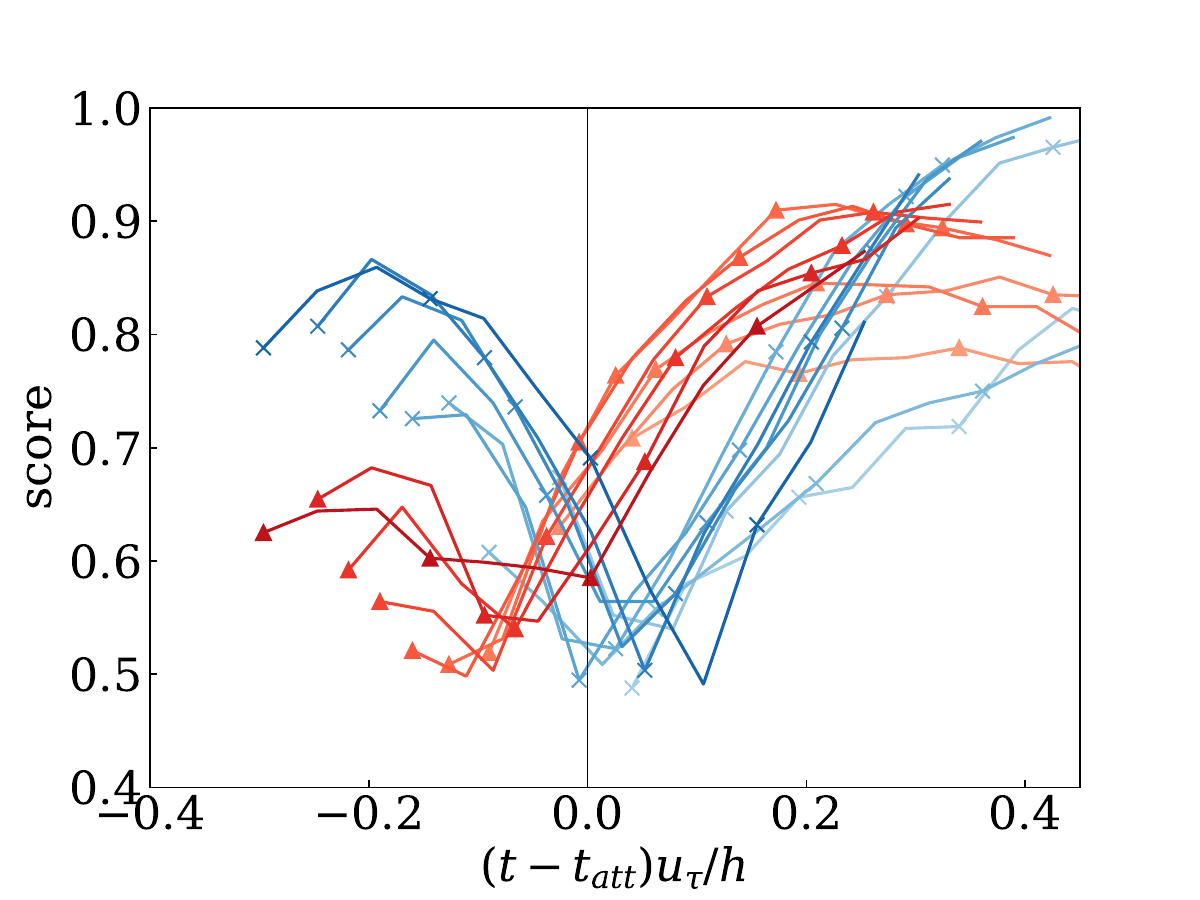}%
    \mylab{-0.33\textwidth}{0.24\textwidth}{(e)}%
\hspace{3mm}%
    \includegraphics[width=0.4\textwidth]%
    {\figpath 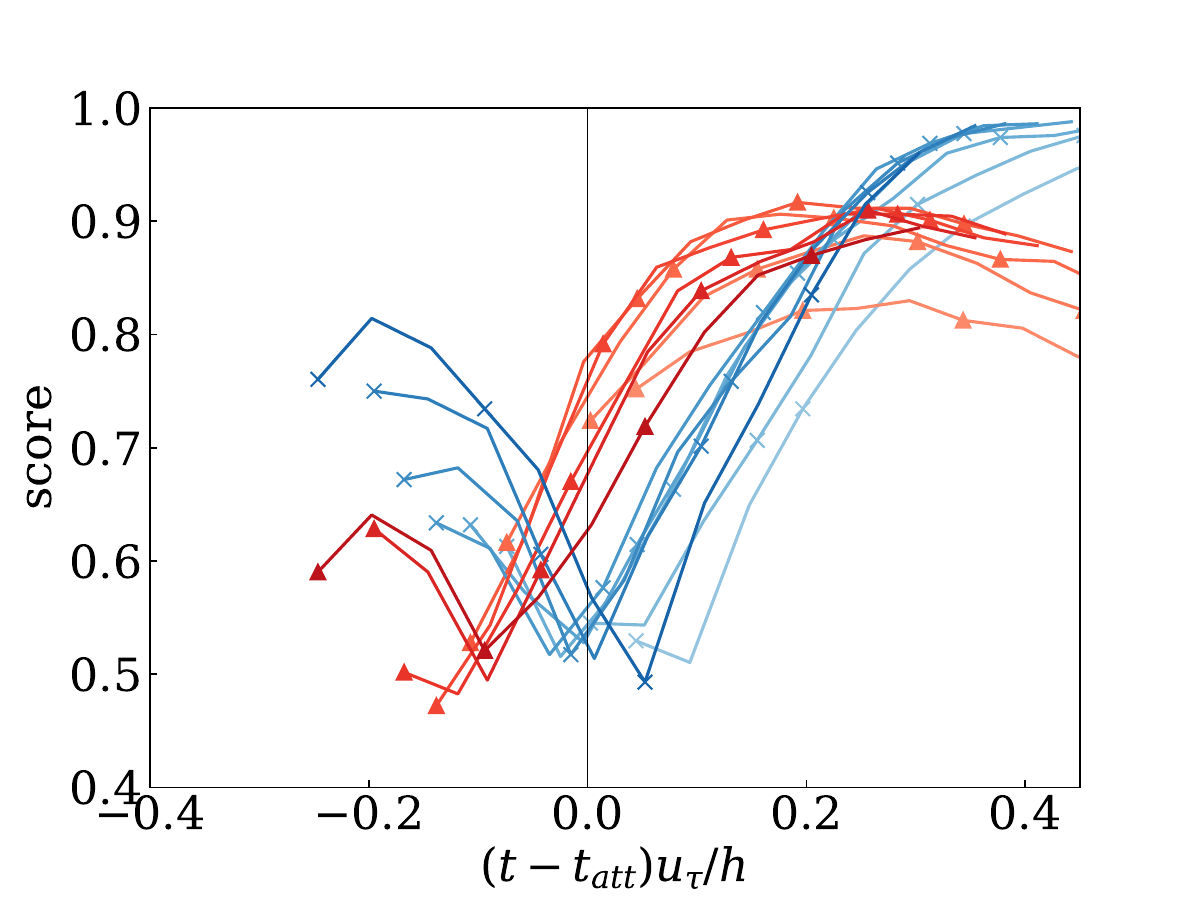}%
    \mylab{-0.33\textwidth}{0.24\textwidth}{(f)}%
}
    \caption{%
(a) Time at which the sum of the scores of the four best four observables is lowest. Dashed: $\tsig$
from figure \ref{fig_tsig}(b). Other symbols as in table \ref{tab_cell}.
(b-f) Classification score of selected observables as a function of time, computed from $\sigur$.
\fulltriangle, Small-sale quantities, average of $\erru$ and $\cave{\sijsij}$; $\times$, cell-scale
quantities, average of $\cave{u}$ and $\cave{v}$. Abscissae are offset by the attachment time
$\tatt$. Colour intensity increases with distance from the wall, from $\ycell^+=12.5$ to
$\ycell^+=275$, excluding $\ycell=0$. (b) $\lcell^+=25$. (c) $\lcell^+=50$. (d) $\lcell^+=75$. (e)
$\lcell^+=100$. (f) $\lcell^+=150$.
}
\label{fig_class_t_errur_ycell_prof}
\end{figure}

Much of this persistence can be compensated by using the relative significance $\sigur$. Figure
\ref{fig_class_yt_errur_lcell} displays the two best observables for different sizes at a fixed cell
height, and figure \ref{fig_class_yt_errur_ycell} displays results for a given size and different
heights. In both cases the best score starts being relatively high at short classification times,
decreases for intermediate ones, and increases again towards the end of the experimental run. The
evolution of the optimum diagnostic variables with the classification time can be divided in three
phases.

During the initial phase, up to the time when the scores are lowest, the best observables include
$\sijsij$, the magnitude of the initial disturbance, and the disturbance production and dissipation,
all of which are either highly correlated with the initial value of $\erru$, or are terms in its
evolution equation. This part of the table is equivalent to the observation in \S\ref{sec_perturb}
that initially stronger perturbations not only remain strong, but also grow faster than
weaker ones.

The second phase is the broad minimum of the score around $\tsig$. While the scores in this phase
are not high, the best observables change from the small-scale perturbation properties of
the initial phase to properties of the cell that do not include fluctuations, such as the cell
average of some velocity component or, equivalently, the mean shear when the cells are very close to
the wall. It is interesting that the best measure of shear near the wall is $\sfw$, which excludes
the viscous sublayer (see figure \ref{fig_class_yt_errur_ycell}c). This is consistent with the
idea that the growth of the perturbation is due to the energy production by the local shear, because the
fluctuation production term in Appendix \ref{sec_evoleq} is proportional to the shear, but also to
the Reynolds stresses, $\dif{u}_i\dif{u}_j$, which are inactive in the sublayer.

It is interesting that the longitudinal velocity derivatives, $\cave{\partial_x u}$ and
$\cave{\partial_z w}$, appear among the most diagnostic cell properties for wall-attached
perturbations in figure \ref{fig_class_yt_errur_ycell}(c). These derivatives are involved in mass
conservation, and follow naturally from the meandering of near-wall streaks, which has been
associated with streak breakdown \citep{jmoin,Waleffe95,Waleffe97} and with the generation of Orr
bursts \citep{orr07a,jim13_lin}. Although not apparent from figure  \ref{fig_class_yt_errur_ycell}(c),
it can be shown that significant cells are associated with $\p_x u<0, \p_z w>0$, with the opposite 
association for irrelevant ones. 

Finally, at longer times of the order of $t-\tsig \approx 0.25 h/\ut$ the score recovers, and the most 
diagnostic observable reverts to the small-scale quantities that dominate short times. Since we saw in
\S\ref{sec_perturb} that $\sigur$ at long times is  essentially equivalent to $\erru(0)$, this final phase is a
reflection of the behaviour at short times, and does not represent new physics.

Figure \ref{fig_class_t_errur_ycell_prof}(a) shows the dependence on $\ycell$ of the time at which
the accumulated score of the top four observables reaches its minimum. After an initial transient
that gets shorter as the cell size increases, $t_{\mbox{\scriptsize\it min}}$ grows with the distance
from the wall, and approximately tracks the optimum classification time, $\tsig$.

Figure \ref{fig_class_t_errur_ycell_prof}(b--f) summarises the evolution of the scores as functions
of time. The blue lines are averages of the scores of several fluctuation quantities, and the red
ones are averages of cell-scale properties. The figures are offset by their attachment time,
$\tatt$, which improves their collapse significantly, and reflect the decreasing influence of the
small-scale quantities as the perturbations approach the attachment time, as well as the increasing
importance of the cell-scale properties as the perturbations intensify. It
should be mentioned that offsetting $t$ with the optimum classification time, $\tsig$, instead of
with $\tatt$, also collapses most scores, as could be expected from the similarity of both times in
figure \ref{fig_tsig}. It also collapses better the case $\ycell=0$, which is not included in figure
\ref{fig_class_t_errur_ycell_prof}, and for which $\tatt$, defined at the arbitrary
distance $y^+=50$, does a poor job. In spite of this, $\tatt$ is used in figure
\ref{fig_class_t_errur_ycell_prof} because it improves the case $\lcell^+=25$, and underscores the
already mentioned connection between significance and the energy production from the near-wall
shear. It is also interesting that the score of the velocities has a secondary maximum at $t=0$,
probably due to the known correlation of small-scale vorticity with the large-scale
streamwise-velocity streaks \citep{tana04}.

\begin{figure}
    \centering
    \includegraphics[height=0.32\textwidth]{\figpath 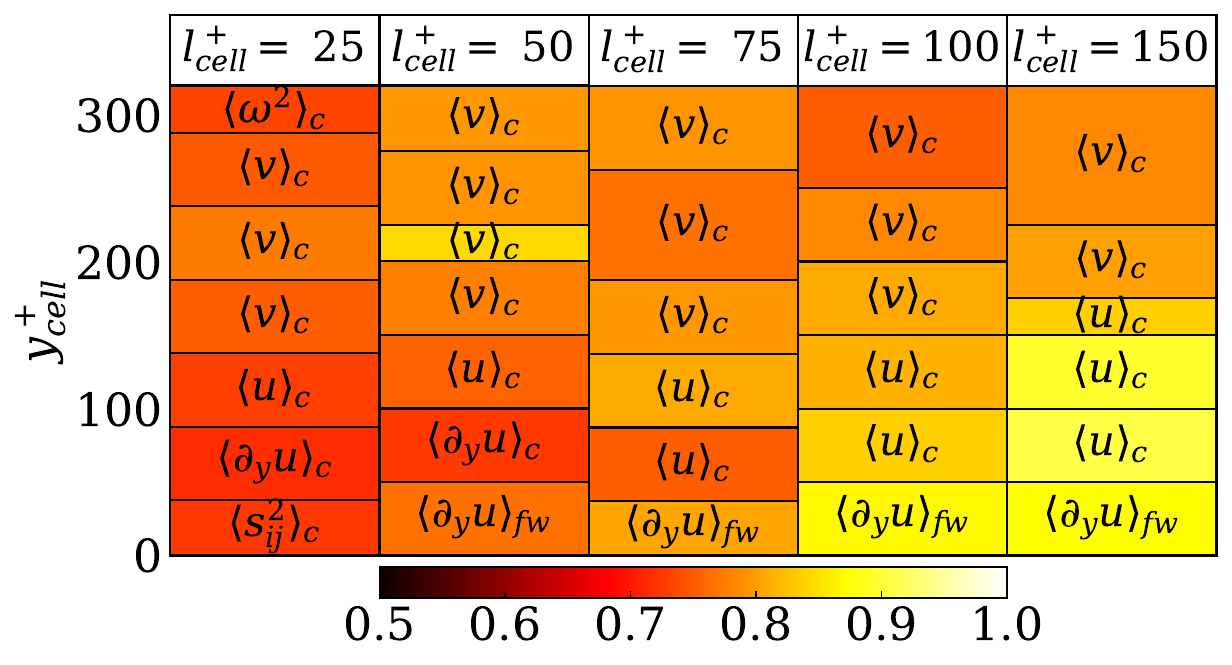}%
\caption{Best observable at $\tsig$. Color: score. The bottom of each tile in the figure aligns with
the left-hand $\ycell$ axis.
\label{fig_class_tsig}}
\end{figure}

From now on, we will mostly focus on results classified at $\tsig$, for which the optimum
observables are collected in figure \ref{fig_class_tsig}. As seen above, they are mostly 
the average cell velocities, $\cave{u}$ or $\cave{v}$. The exceptions are cells in the buffer layer
$\ycell^+\lesssim 50$, in which shear can probably be taken as a proxy for the streamwise velocity,
and cells with $\lcell^+=25$ very far from the wall. We have already mentioned that these perturbations
are probably too small to survive for the relatively long times required to reach the wall, in agreement
with the assimilation results from \cite{Wang22} mentioned in the introduction. 

%

\section{Conditional flow fields}\label{sec_template}
 
\begin{figure}
\vspace*{0mm}
\centerline{%
    \includegraphics[width=0.47\textwidth]{\figpath 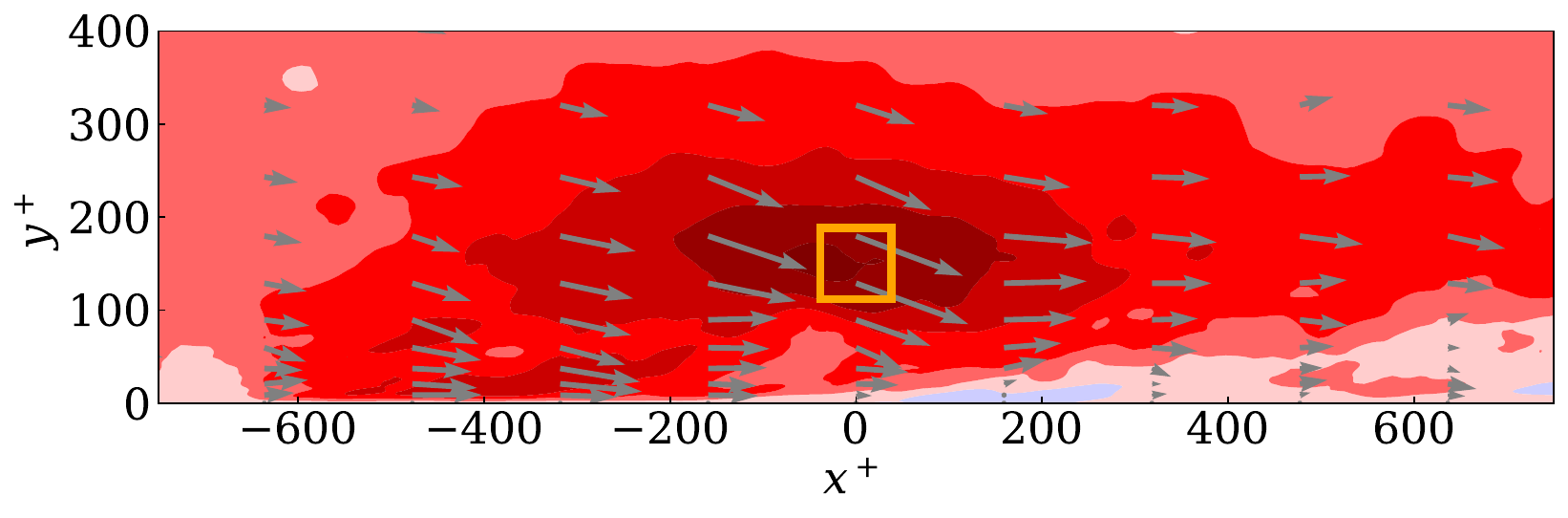}%
    \mylab{-0.48\textwidth}{0.12\textwidth}{(a)}%
\hspace{3mm}%
    \includegraphics[width=0.47\textwidth]{\figpath 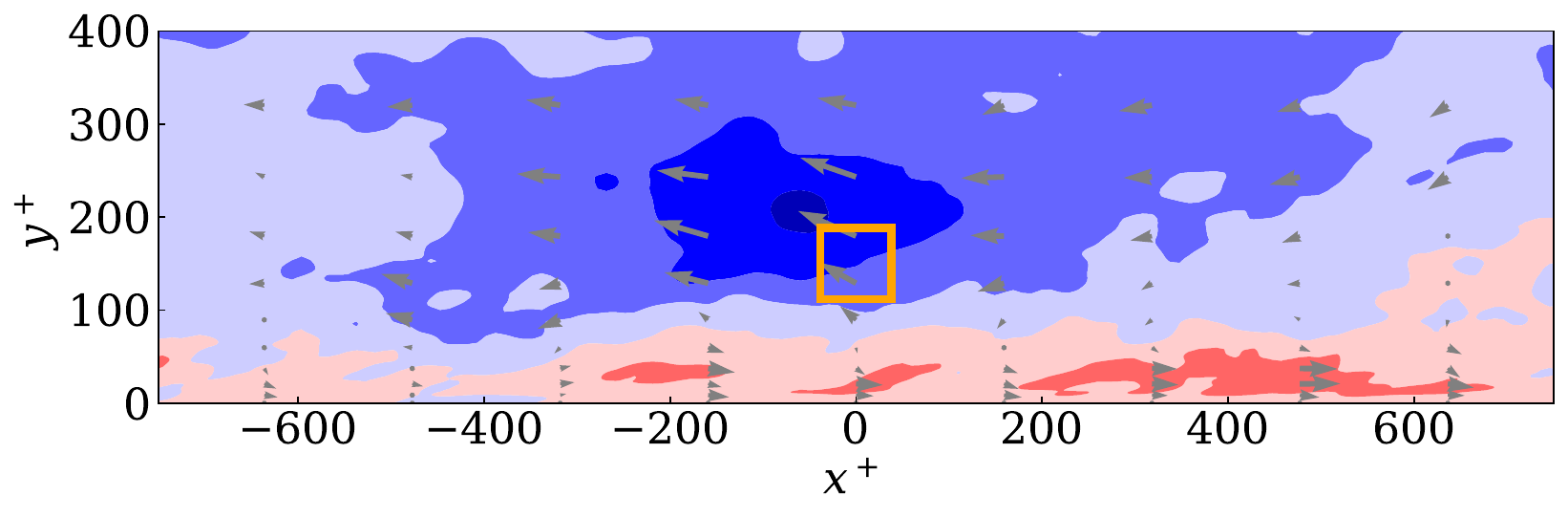}%
    \mylab{-0.48\textwidth}{0.12\textwidth}{(b)}%
}
\vspace*{0mm}
\centerline{%
    \includegraphics[width=0.47\textwidth]{\figpath 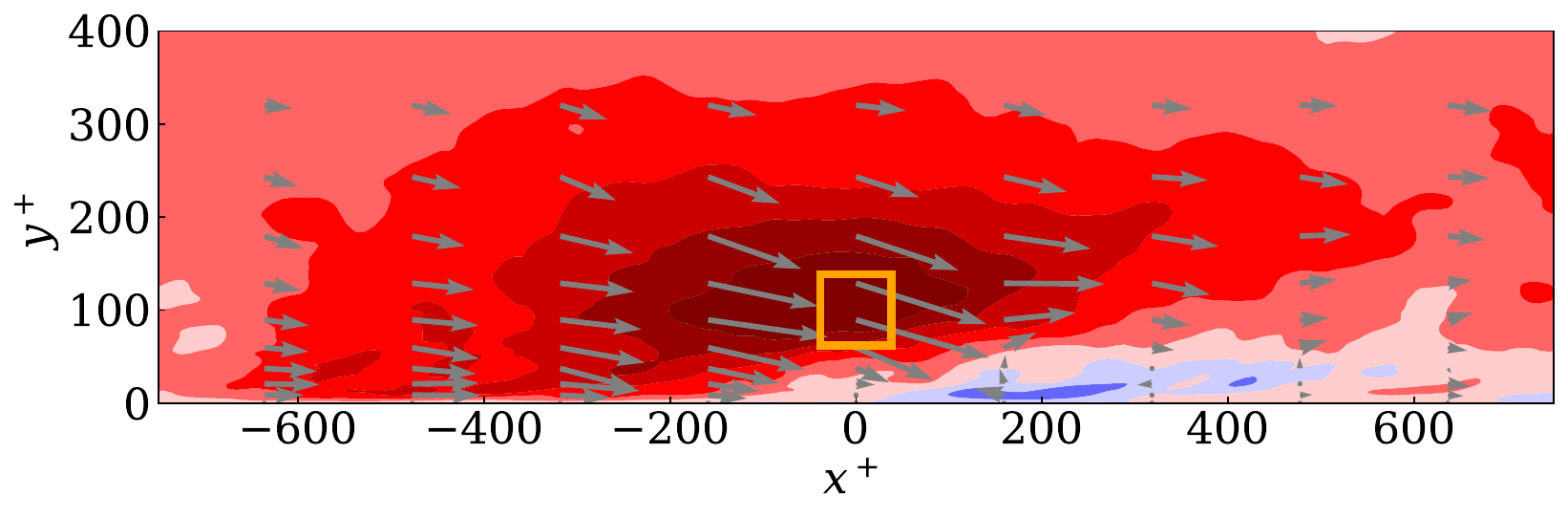}%
    \mylab{-0.48\textwidth}{0.12\textwidth}{(c)}%
\hspace{3mm}%
    \includegraphics[width=0.47\textwidth]{\figpath 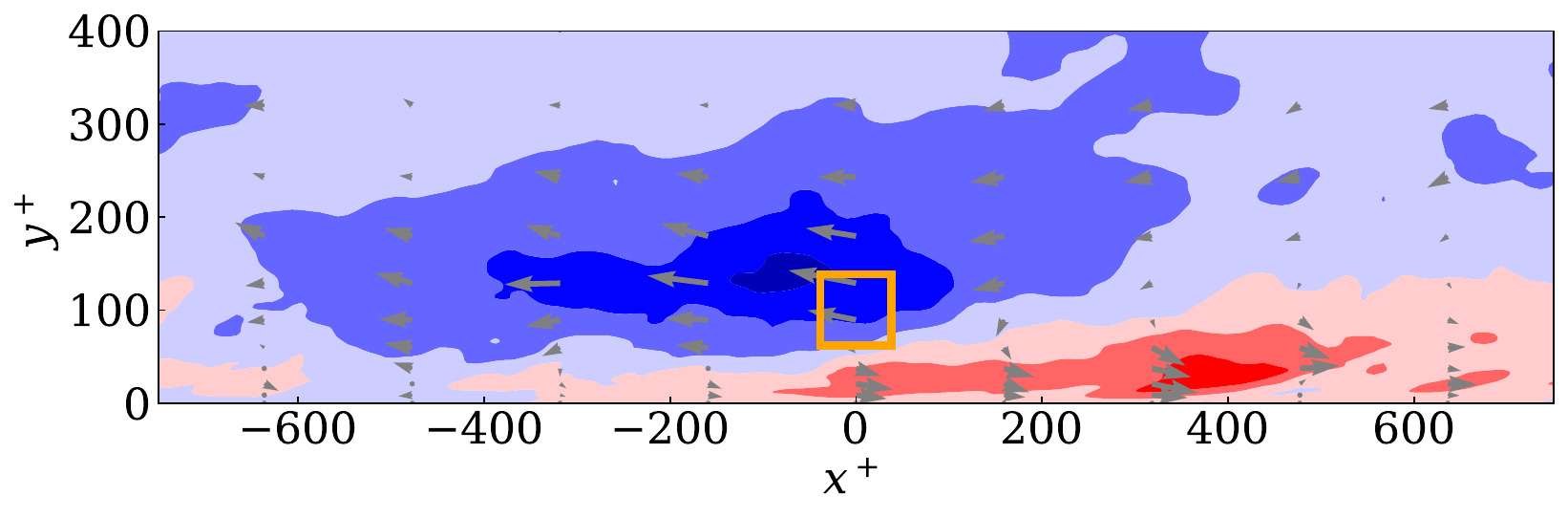}%
    \mylab{-0.48\textwidth}{0.12\textwidth}{(d)}%
}
\vspace*{0mm}
\centerline{%
    \includegraphics[width=0.47\textwidth]{\figpath 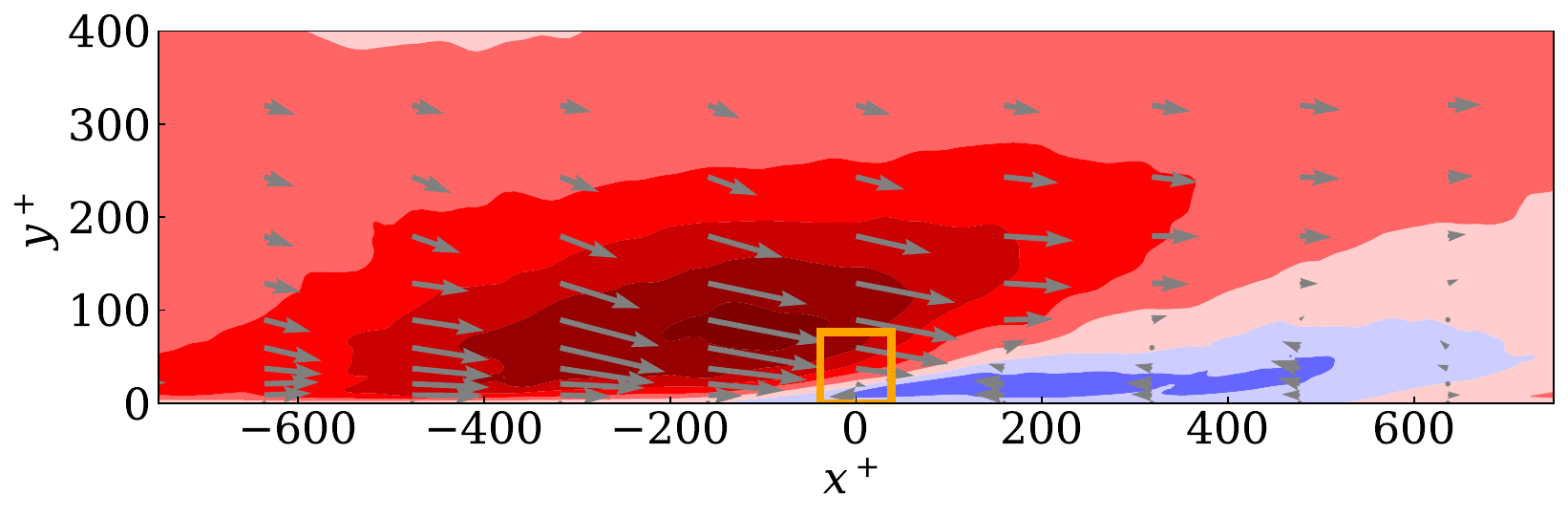}%
    \mylab{-0.48\textwidth}{0.12\textwidth}{(e)}%
\hspace{3mm}%
    \includegraphics[width=0.47\textwidth]{\figpath 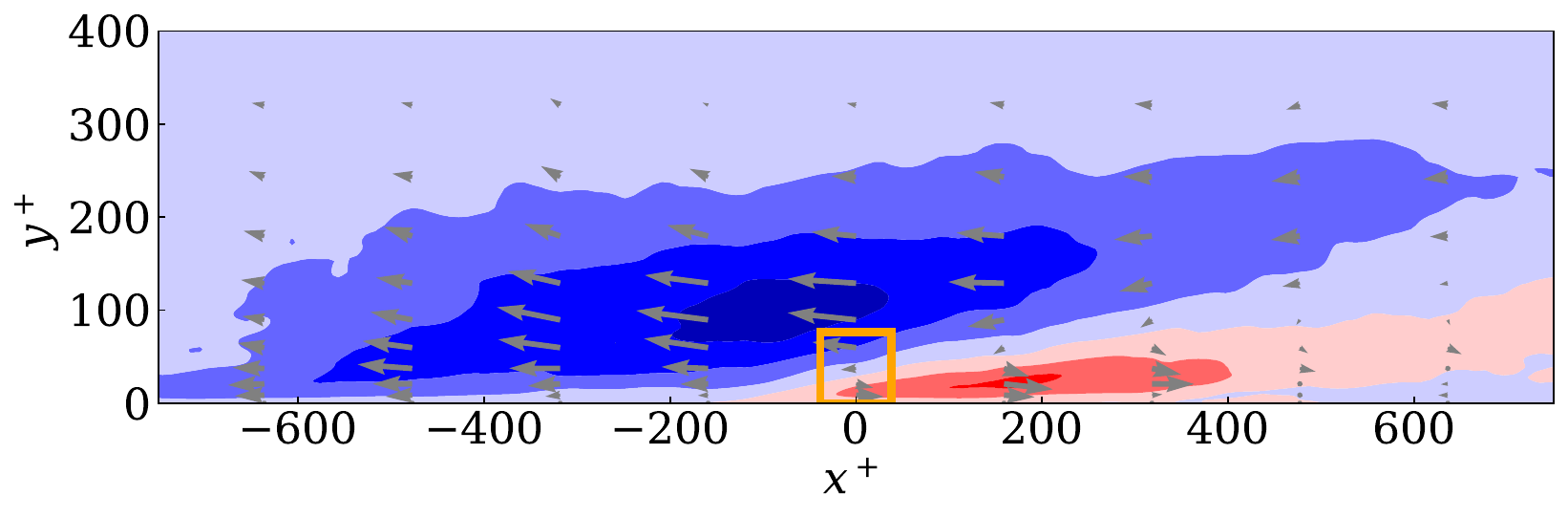}%
    \mylab{-0.48\textwidth}{0.12\textwidth}{(f)}%
}
\vspace*{-2mm}
\centerline{%
    \includegraphics[width=0.35\textwidth]{\figpath 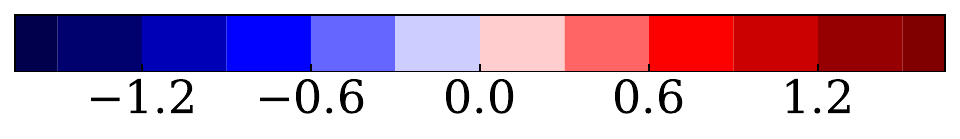}%
}%
\caption{Streamwise section at $z=z_c$ of the conditional velocity field of the reference flow at $t=0$
around the perturbation cell. The colour background is the conditional streamwise velocity. Arrows
are velocity fluctuation vectors parallel to the plane of the figure, and the light-coloured box is
the perturbation cell. $\lcell^+=75$. (a,b) $\ycell^+=113$. (c,d) $\ycell^+=62.6$ (e,f) $\ycell=0$.
(a,c,e) Significants. (b,d,f) Irrelevants.
\label{fig_temp_xy}}
\vspace{2mm}%
\centerline{%
    \includegraphics[width=0.47\textwidth]{\figpath 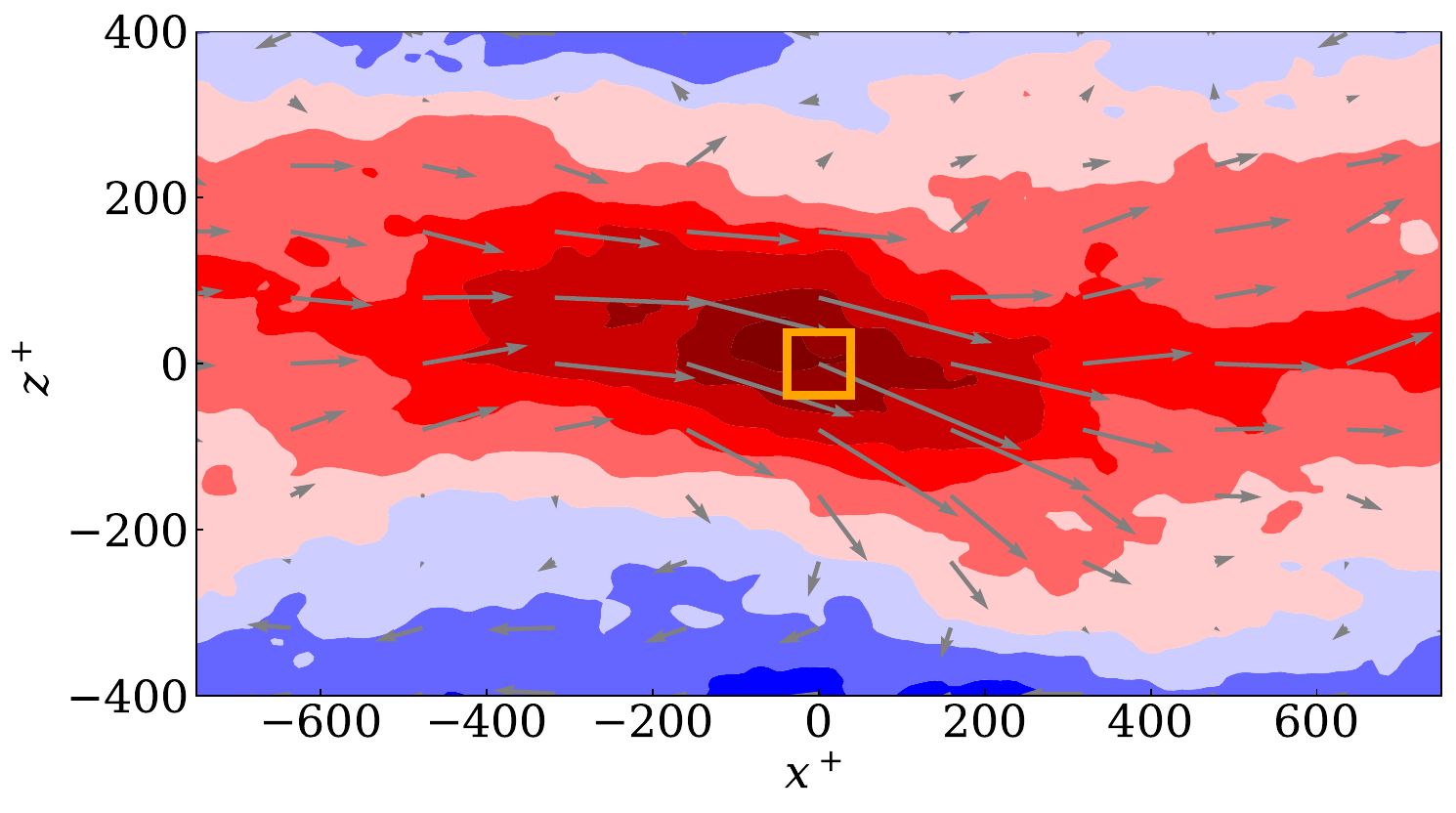}%
    \mylab{-0.47\textwidth}{0.165\textwidth}{(a)}%
\hspace{1mm}%
    \includegraphics[width=0.47\textwidth]{\figpath 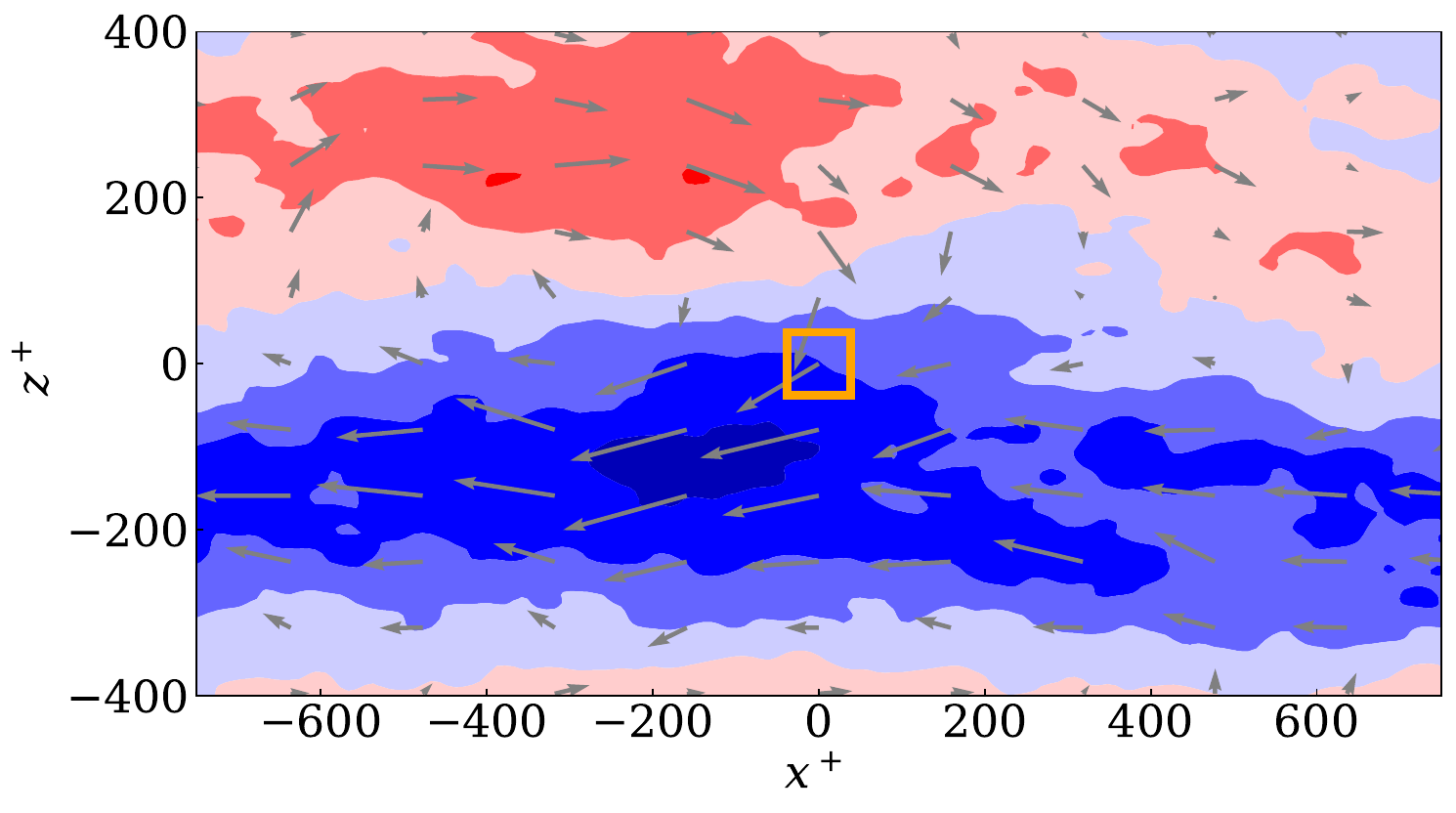}%
 \mylab{-0.47\textwidth}{0.165\textwidth}{(b)}%
}
\vspace*{3mm}
\centerline{%
    \includegraphics[width=0.47\textwidth]{\figpath 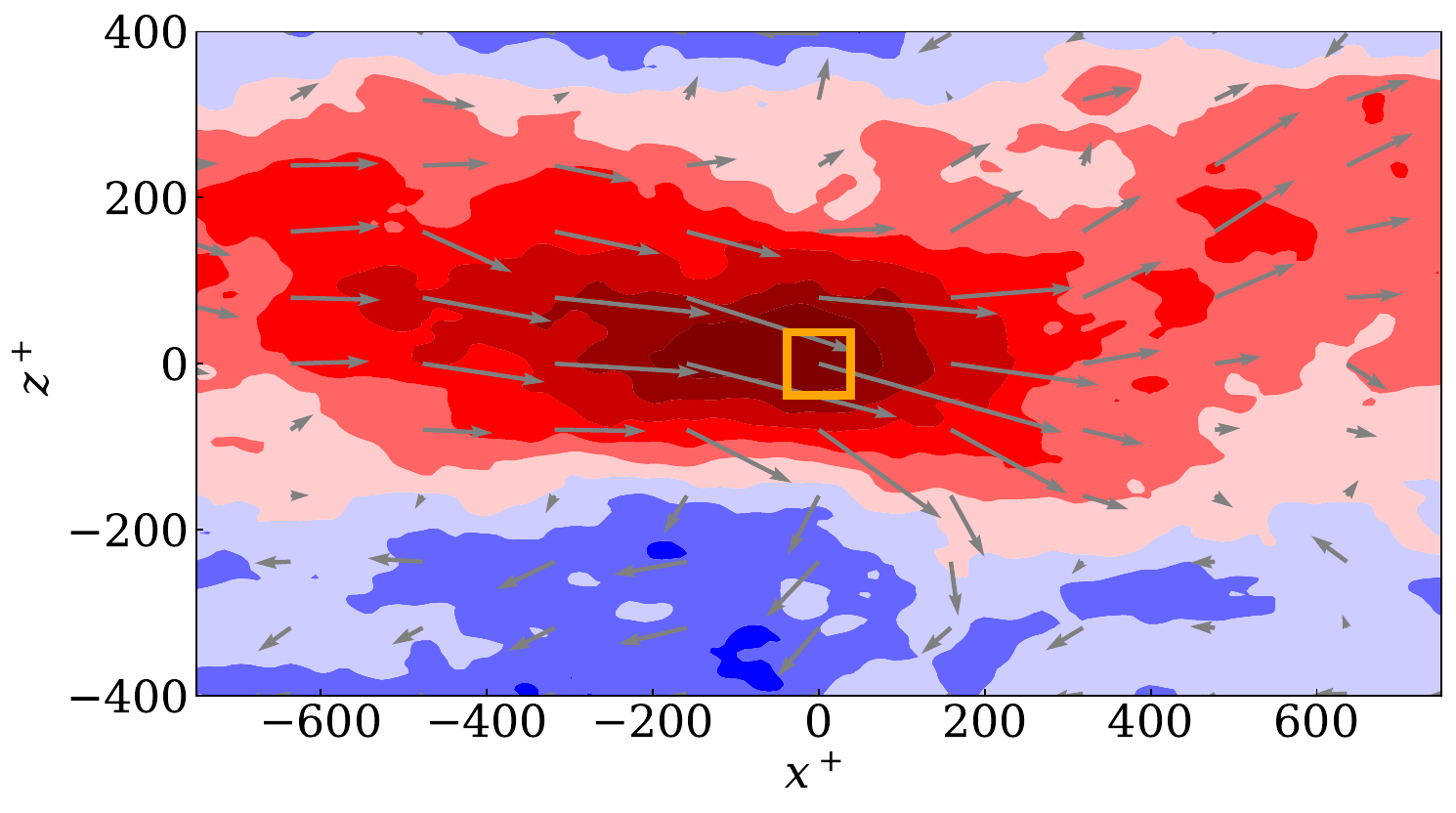}%
    \mylab{-0.47\textwidth}{0.165\textwidth}{(c)}%
\hspace{1mm}%
    \includegraphics[width=0.47\textwidth]{\figpath 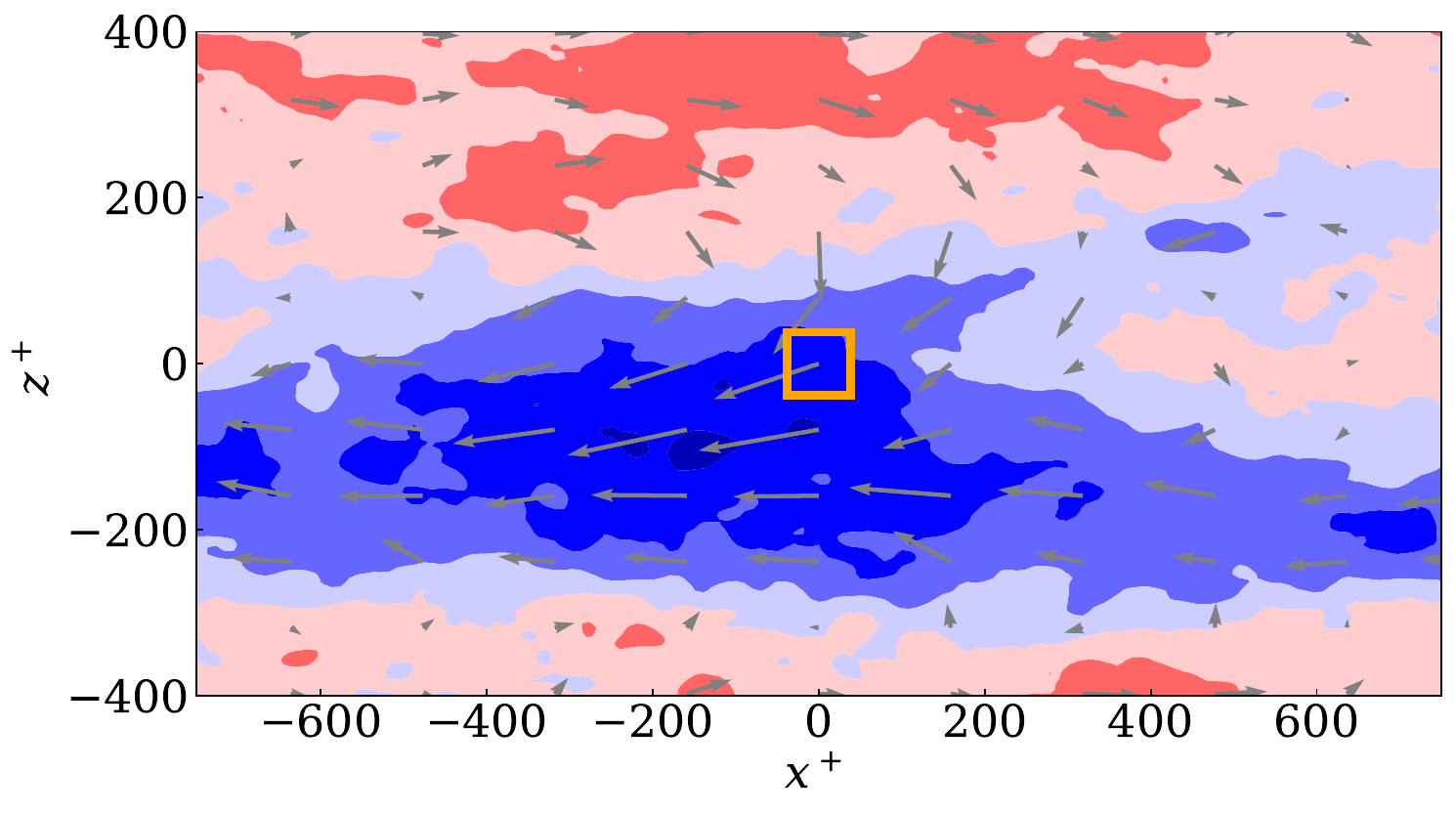}%
 \mylab{-0.47\textwidth}{0.165\textwidth}{(d)}%
}
\vspace*{3mm}
\centerline{%
    \includegraphics[width=0.47\textwidth]{\figpath 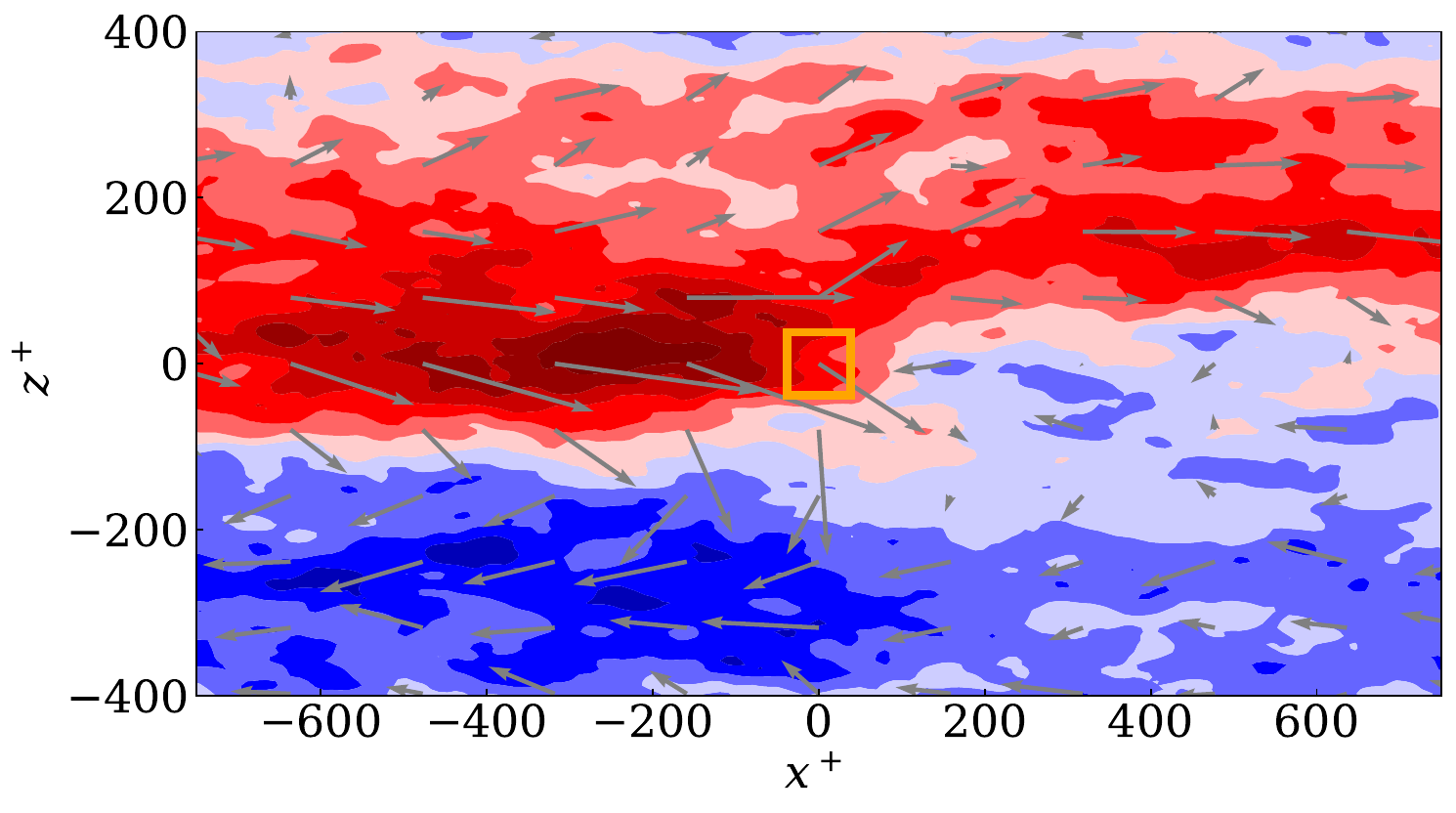}%
    \mylab{-0.47\textwidth}{0.165\textwidth}{(e)}%
\hspace{1mm}%
    \includegraphics[width=0.47\textwidth]{\figpath 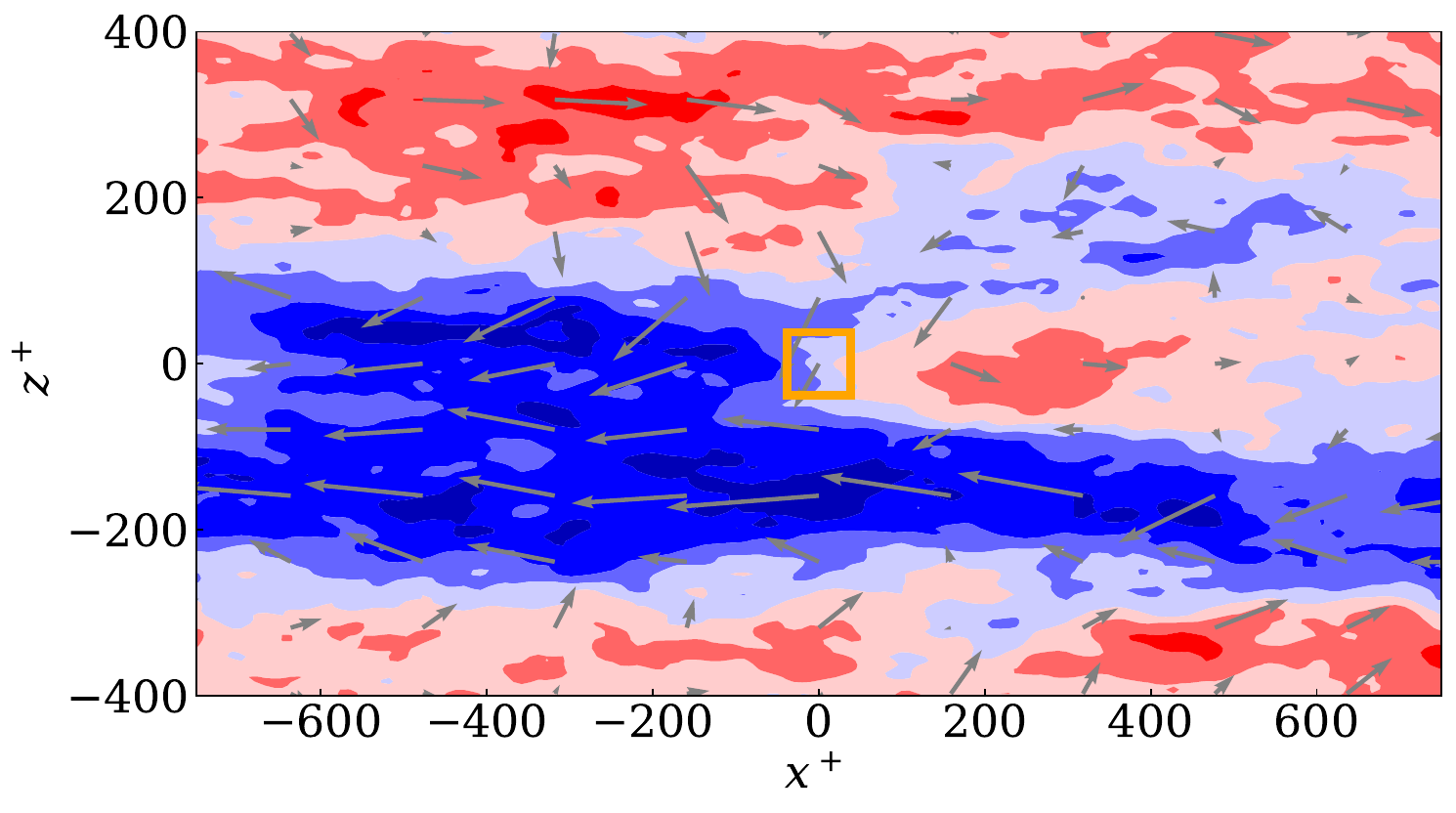}%
 \mylab{-0.47\textwidth}{0.165\textwidth}{(f)}%
}
\vspace*{-2mm}
\centerline{%
    \includegraphics[width=0.35\textwidth]{\figpath template_u_cbar.pdf}%
}%
    \caption{As in figure \ref{fig_temp_xy}, for a wall-parallel section at $y=y_c$.
\label{fig_temp_xz}}
\end{figure}

\begin{figure}
\vspace*{3mm}%
\centerline{%
    \includegraphics[width=0.47\textwidth]{\figpath 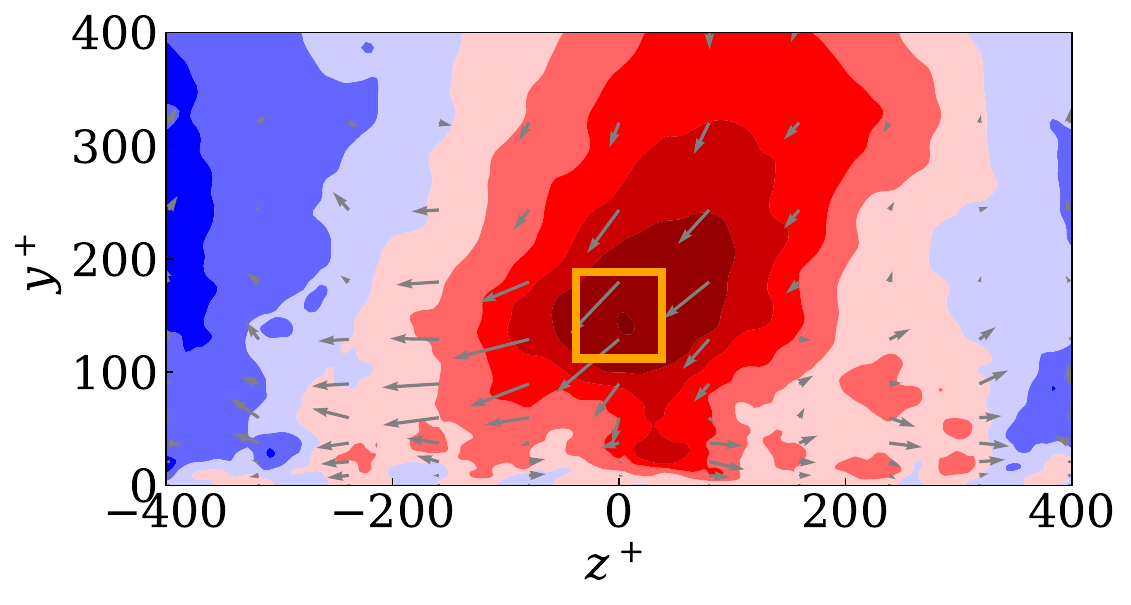}%
    \mylab{-0.215\textwidth}{0.255\textwidth}{(a)}%
\hspace{0mm}%
    \includegraphics[width=0.47\textwidth]{\figpath 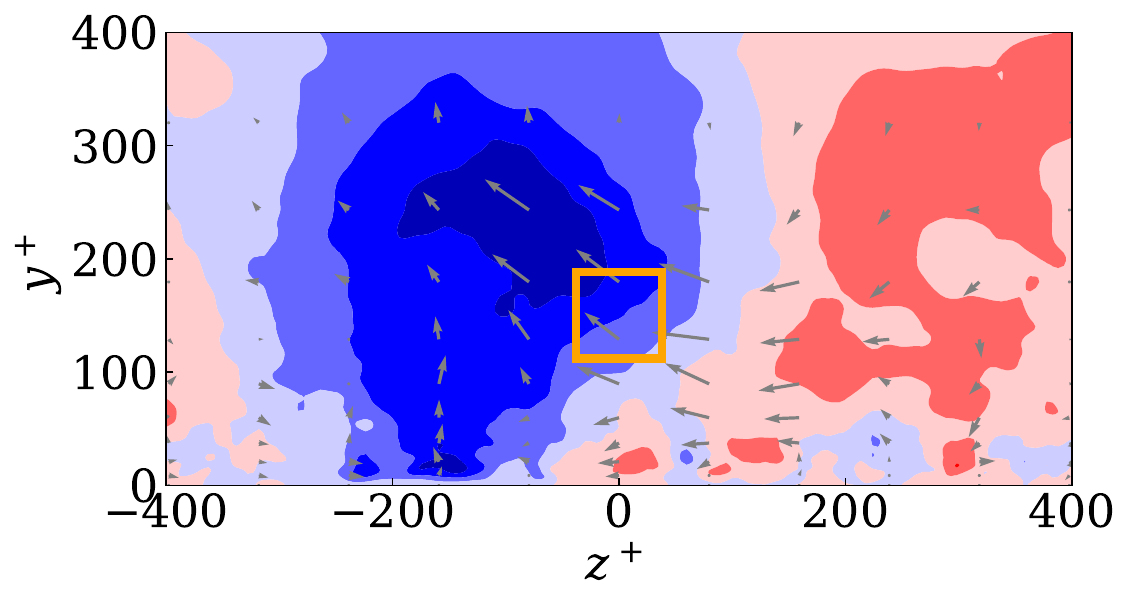}%
    \mylab{-0.215\textwidth}{0.255\textwidth}{(b)}%
}
\vspace*{-2mm}
\centerline{%
    \includegraphics[width=0.35\textwidth]{\figpath template_u_cbar.pdf}%
}%
\caption{As in figure \ref{fig_temp_xy}(a,b), for the $(z-y)$ cross-flow section at $x=x_c$.
(a) Significants. (b) Irrelevants. $\ycell^+=113, \, \lcell^+=75$.
\label{fig_temp_zy}}
\end{figure}

While we have seen that the cell-averaged velocity fluctuations are diagnostic quantities for
causality, our analysis did not include their sign. Figures \ref{fig_temp_xy} to \ref{fig_temp_zy},
which display averaged velocity fields conditioned to the position of significant or
irrelevant perturbation cells, shows that the sign is important.

Figure \ref{fig_temp_xy} displays longitudinal $(x-y)$ sections of the streamwise and wall-normal
velocities, conditioned to either significant (left column) or irrelevant cells (right column). It
is evident that the former are biased towards fourth-quadrant regions $(u>0,\,v<0)$, while the
latter are in second-quadrant regions $(u<0,\,v>0)$.

The three rows in figure \ref{fig_temp_xy} corresponds to perturbations introduced at decreasing
distances from the wall. The frames are centred at the streamwise position of the perturbation cell,
and the figure shows a fluid wedge entering the frame from its downstream right-hand edge, becoming
more prominent as the cell approaches the wall. At the same time, there is an upstream drift of the
darker core of the velocity distribution. This is clearest in the significant cases in the left-hand
column, where the incoming wedge is low-speed fluid, but it can also be traced in the irrelevant
cases in the right-hand column. The result is that the position of the causally significant cells
moves downstream towards an interface at which high-speed fluid overtakes a low-speed one.
Irrelevants are associated to an interface in which low-speed fluid is left behind by higher speed
ahead of it.

Figure \ref{fig_temp_xz} displays wall-parallel $(x-z)$ sections of the same cases as figure
\ref{fig_temp_xy}, and figure \ref{fig_temp_zy} shows cross-flow $(z-y)$ sections of figure
\ref{fig_temp_xy}(a,b). There is no orientation ambiguity in the longitudinal sections in figure
\ref{fig_temp_xy}, but figures \ref{fig_temp_xz} and \ref{fig_temp_zy} would be statistically
symmetric even if individual flow fields were not. To preserve possible systematic asymmetries, the
$z$ coordinate of all the flow fields is reflected so that the spanwise velocity averaged over a
cube of side $3\lcell$, centred at the perturbation cell and possibly truncated by the wall, is
$\bra w \ket_{3c}<0$. The orientation of the sections in figures \ref{fig_temp_xz} and
\ref{fig_temp_zy} is therefore not physically meaningful, but the asymmetry of the different frames
is consistent and complements the information in figure \ref{fig_temp_xy}.
 
Figure \ref{fig_temp_xz} shows that the velocity interfaces in figure \ref{fig_temp_xy} correspond
to kinks in the large-scale streaks that dominate the flow.

Although the frames in figures \ref{fig_temp_xy} or \ref{fig_temp_xz} represent unrelated
experiments, it is tempting to interpret them as a temporal evolution in which perturbations
introduced farther from the wall correspond to earlier times, and travel downstream and towards the
wall until they reach it at $\tatt$ (or $\tsig$). The velocity maximum in figures
\ref{fig_temp_xz}(a,c,e) shifts by $\Delta x^+\approx 200$ from the top to the bottom row of frames.
Assuming, from figure \ref{fig_tatt}, that $\ut\tatt/h \approx 0.2$, this corresponds to a velocity
difference $\Delta u^+\approx 1.5$, which is a reasonable estimate for the difference in the
advection velocity of features in a high-velocity streak with respect to the average flow velocity
\citep{kro:kas:rim:98,lozano_time}.

Interestingly, the streamwise drift of the irrelevant velocity features in the right-hand columns of
figures \ref{fig_temp_xy} and \ref{fig_temp_xz} is less clear than in the significant ones in the
left-hand columns. The environment of the irrelevants can rather be described as an interface that
gets wider as the cell approaches the wall. The flow cross-sections in figure \ref{fig_temp_zy} support this
description, and the combined evidence from the three sets of sections is consistent with a model in
which causally significant cells are associated with the front of a high-speed sweep that steepens
as it approaches the wall and overtakes a lower-speed region. Conversely, irrelevant cells are
located at the trailing end of a high-speed region that leaves behind a lower-speed flow. We may
recall at this point that the table in figure \ref{fig_class_yt_errur_ycell}(c) showed that the
wall-parallel mass-conservation derivatives are diagnostic of causality near the wall and that,
while $\p_x u<0$ signals significance, $\p_x u>0$ signals irrelevance. This asymmetry suggests that
the generation of structures strong enough to have a global effect on the flow depends on mass
conservation failures when streaks of different velocity run into each other. This is most probably
due to meandering, as in figure \ref{fig_temp_xz}(a,c,e), and the effect is strongest when this
happens within the strong shear near the wall. The trailing edge of the meander, as in figure
\ref{fig_temp_xz}(b,d,f), or the tails at which high-speed streaks pull away from low speed ones,
are passive.

The association of strong near-wall $v$ structures with the downstream end of high-speed
streaks and with the upstream end of low-speed ones was already noted by \cite{jim04b} and
\cite{JimKaw13}. Their interpretation was that $v$ creates the streaks, but the arguments above,
together with the fact that causality can be traced to flow locations far from the wall, suggest
that the sequence of events is the other way around, and that the formation of near-wall bursts
depends on continuity failures of nonuniform $u$-streaks.

\subsection{The shear time}\la{sec_shear}

\begin{figure}
\vspace*{3mm}%
\centerline{%
\includegraphics[width=0.43\textwidth]{\figpath 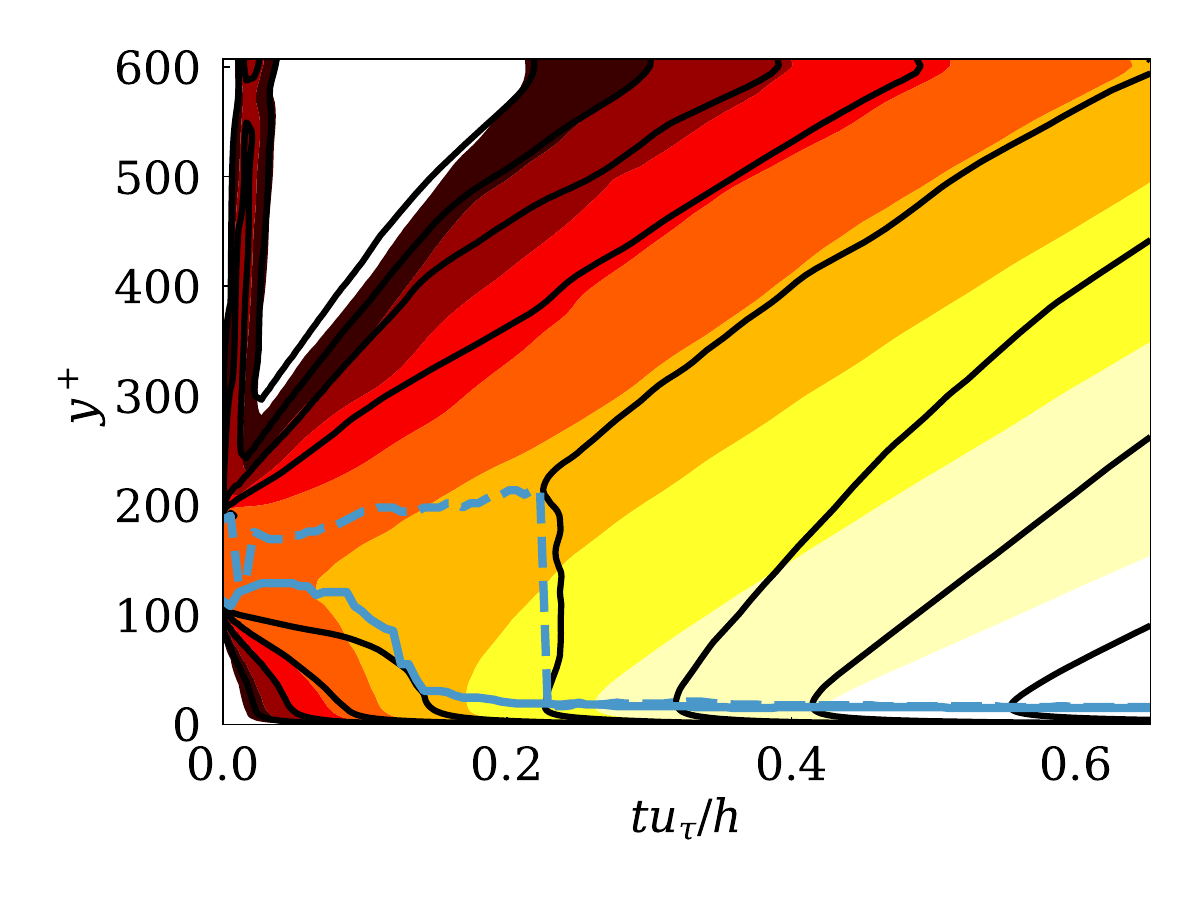}%
\mylab{-0.20\textwidth}{0.32\textwidth}{(a)}%
\hspace{3mm}%
\includegraphics[width=0.43\textwidth]{\figpath 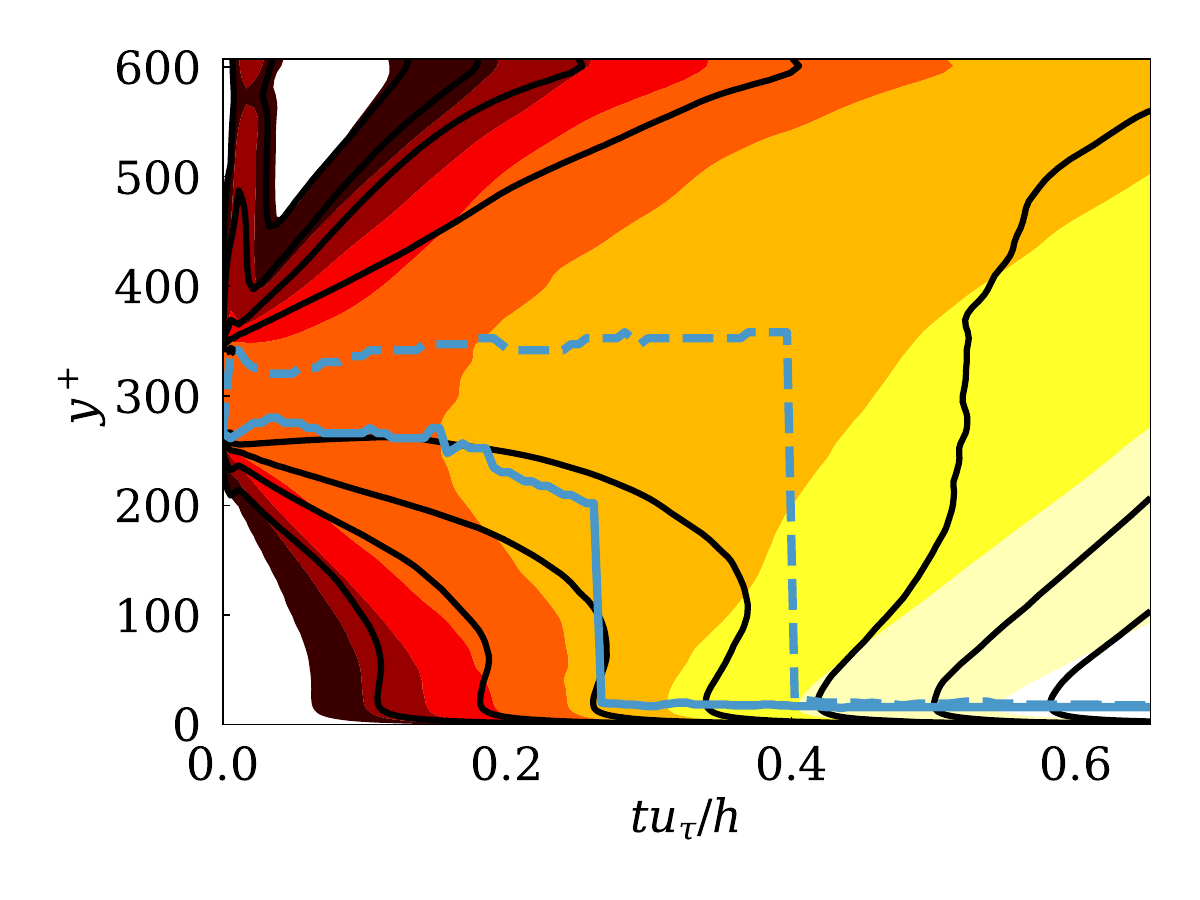}%
\mylab{-0.20\textwidth}{0.32\textwidth}{(b)}%
}
\caption{Plane-averaged perturbation magnitude \r{eq:avexz}, conditionally averaged over significant or irrelevant
perturbations and normalized with the maximum of the conditioned initial value.  
The bold cyan lines are the instantaneous position of the perturbation maximum.
Filled contours and solid line are significants; line contours and dashed line are irrelevants.
Contour levels are ${\bra\erru\ket}(y,t)/\max_y\bra\erru\ket(y,0)= 10^{-4}(\times 10)10^3$.
Classification is done at $\tsig$ using $\sigur$. $\lcell^+=75$.  (a) $\ycell^+=113$. (b) $\ycell^+=263$
\label{fig_err_t_bewo}}
\end{figure}

Figure \ref{fig_err_t_bewo} is similar to the evolution of the plane-averaged fluctuation magnitude
in figure \ref{fig_err_y_t}, but is here separated into conditionally significant cases (filled
contours) and irrelevant ones (lines). In both cases, as well as in the unconditional evolution in
figure \ref{fig_err_y_t}, the perturbation initially remains at the height at which it is
introduced, before spreading vertically. All perturbations eventually fill the channel, as expected
for a chaotic system, but the growth is faster in the significant cases. The bold cyan lines in
figure \ref{fig_err_t_bewo} are the wall-normal position of the perturbation maximum. It is
clear that the solid lines representing significants initially trend downwards and attach to the
wall faster than the dashed lines representing irrelevants, which initially trend away from the
wall or drift little. This supports the interpretation that significants are sweep-like, and the
conjecture in \S \ref{sec_signi} and \S \ref{sec_classify} that causal significance depends on the
amplification of perturbations by the strong shear near the wall.

\begin{figure}
\centerline{%
\includegraphics[width=0.4\textwidth]{\figpath 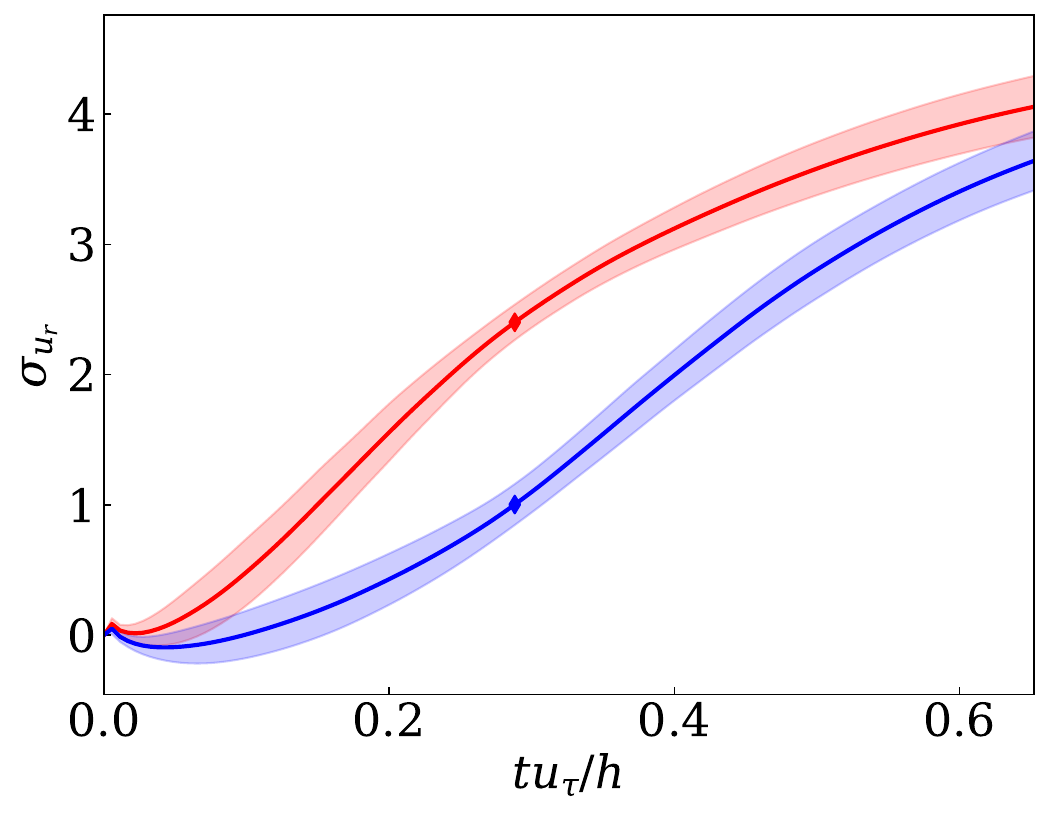}%
\mylab{-0.35\textwidth}{0.27\textwidth}{(a)}%
\hspace{3mm}%
\includegraphics[width=0.39\textwidth]{\figpath 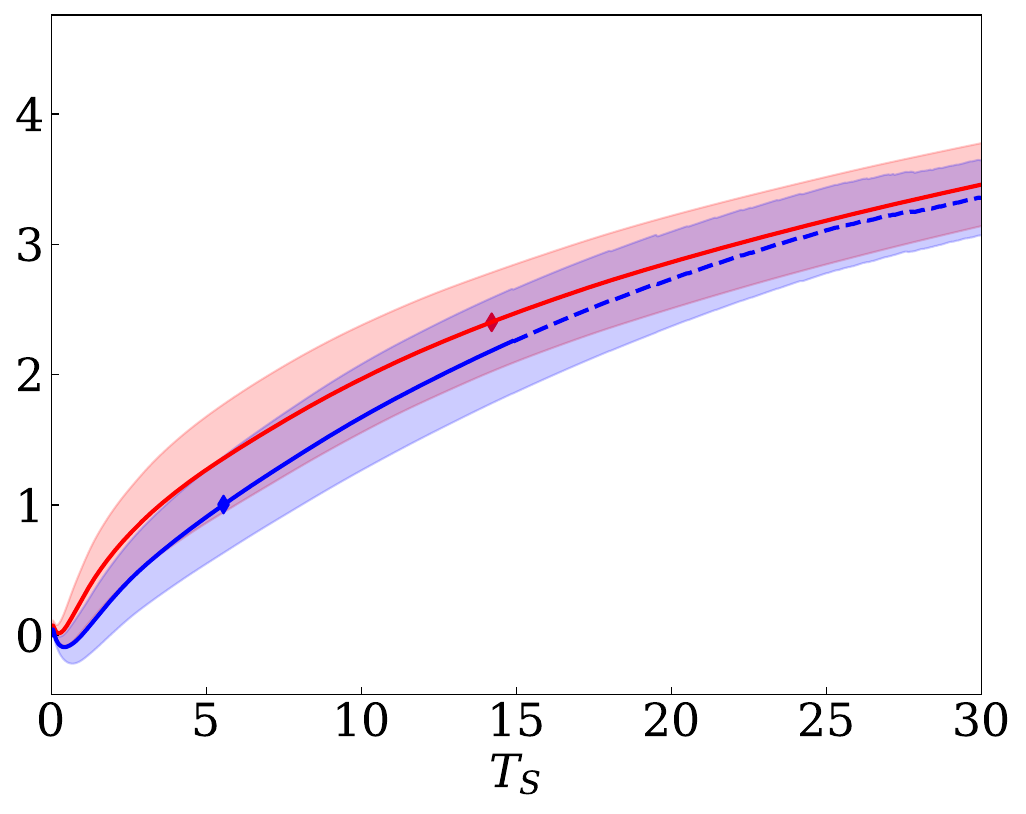}%
\mylab{-0.36\textwidth}{0.27\textwidth}{(b)}%
}
\caption{Relative perturbation growth, $\sigur(t)$, conditioned to significant (red) and
irrelevant (blue) samples, classified at $\tsig$. Solid line, mean; shading: standard deviation;
\fulldiamond, $\tsig$. (a) In eddy turnovers. (b) In local shear time. The dashed part of the
irrelevant line in (b) corresponds to times for which not all experiments are available, because some
of them end within the plot. $\lcell^+=75$, $\ycell^+=133$.
\label{fig_tshear}}
\end{figure}

This is tested directly in figure \ref{fig_tshear}. Figure \ref{fig_schem} shows that perturbations
spread with time along the three directions, and that it is difficult to define an instantaneous
location to measure the shear that they encounter, but figures \ref{fig_err_y_t} and
\ref{fig_err_t_bewo} suggests that it is possible to define an effective shear by weighting the mean
velocity profile, which depends only on $y$, with the perturbation magnitude,
\begin{equation}
S_\varepsilon (t)=\frac{\int \erru(y,t)\, \partial_y U(y) \dd y}{\int \erru(y,t) \dd y},
\la{eq_shearsig}
\end{equation}
and an effective shear time
\begin{equation}
T_S (t)=\int{S_\varepsilon \dd t}.
\la{eq_sheartime}
\end{equation}
Figure \ref{fig_tshear}(a) illustrates the development of the relative growth of causally
significant and irrelevant perturbations in terms of the global eddy-turnover-time. The significant
perturbations grow from the start, while the irrelevant ones initially decay and only later grow to
match the causal case. Most of the initial decay and of the slow growth of irrelevants can be
attributed to their failure to initially approach the wall. Figure \ref{fig_tshear}(b) plots the
same data using the local shear time computed for individual experiments. The two evolutions now
approximately coincide, supporting the importance of the local shear, and providing an explanation
for the association of sweep-like flows with causality. The effect of the negative wall-normal
velocity is to bring perturbations close to the wall. Ejection-like regions move perturbations away
from the wall to layers where the shear is low, and they become causally relevant only after they
eventually diffuse into the near-wall layer.

The collapse with the shear time only applies to significant and irrelevant perturbations at the same
distance from the wall. Perturbations introduced at different distances behave differently, at least 
at the relatively low Reynolds number of our experiments for which self-similar behaviours 
with respect to $y$ are necessarily limited.

\subsection{Geometry of causal events}\la{sec_Qs}

While we have seen that significant and irrelevant cells are associated with sweep- and
ejection-like regions of the flow, it remains unclear whether the quadrants discussed in the
previous section are the same as the intense events traditionally associated with sweeps and
ejections in wall turbulence \citep{lu:wil:73}. This is the purpose of the experiments in the second
column of table \ref{tab_cell}, which cover selected wall-parallel planes with a dense grid of
perturbation experiments.

\begin{figure}
\centerline{%
    \includegraphics[height=0.32\textwidth]{\figpath 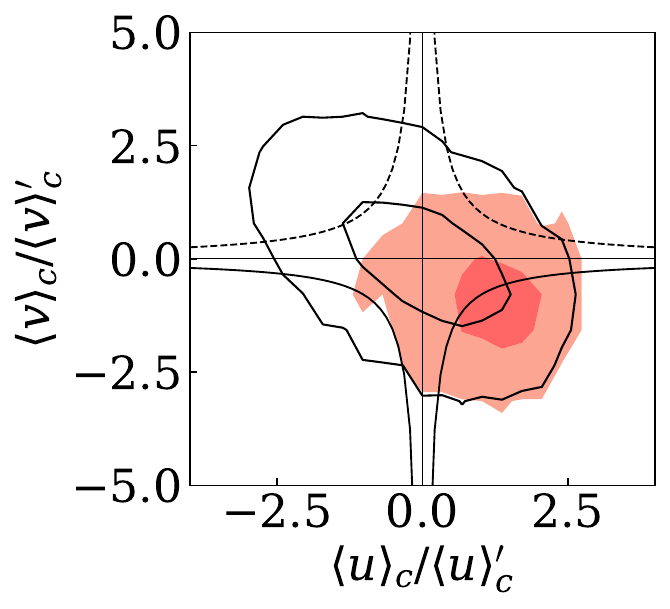}%
\mylab{-0.24\textwidth}{.27\textwidth}{(a)}%
    \hspace{3mm}%
    \includegraphics[height=0.32\textwidth]{\figpath 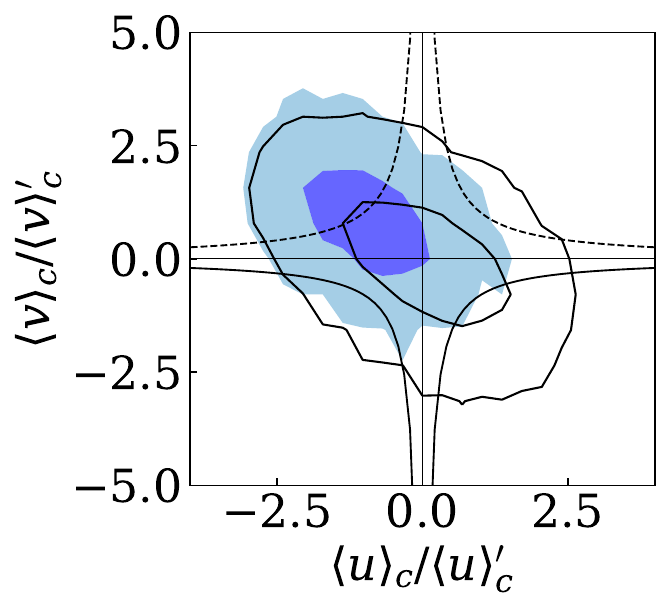}%
\mylab{-0.24\textwidth}{.27\textwidth}{(b)}%
}%
\caption{Joint probability density function of the cell-averaged velocities. Line contours are
unconditional. Filled ones are conditioned to: (a) significant cells; (b) irrelevants. Both
contain 60\% and 99\% of the data. $\ycell^+=113$, $\lcell^+=75$.
The solid hyperbolae are the $H_-$ threshold that isolates the 10\% most intense velocity quadrants with
$\cave{v}<0$, as in \r{eq_uvhole}, and the dashed ones are $H_+$ for $\cave{v}>0$. 
\label{fig_jpdf_uv}}
\end{figure}

\begin{table}
\begin{center}
\def~{\hphantom{0}}
    \begin{tabular}{>{\hspace{\colwd}}c>{\hspace{\colwd}}c>{\hspace{\colwd}}c>{\hspace{\colwd}}c>{\hspace{\colwd}}c>{\hspace{\colwd}}c>{\hspace{\colwd}}cl>{\hspace{\colwd}}c<{\hspace{\colwd}}}
$\ycell^+$ & $H_-$ & $H_+$    & $S_1(\%)$  & $S_2(\%)$ & $S_3(\%)$ &  $S_4(\%)$ \\[3pt]
  1 & 0.74 & 0.75 & 4.17 & 45.94 & 3.14& 46.75 \\
 38 & 0.77 & 0.87 & 2.58 & 47.17 & 1.81 & 48.44 \\
113 & 0.80 & 1.03 & 1.44 & 48.08 & 2.00 & 48.47\\
188 & 0.76 & 1.04 & 1.39 & 48.17 & 2.72 & 47.72\\
263 & 0.75 & 0.99 & 0.92 & 48.25 & 4.14 & 46.69\\
-- & 1.75 & 1.75 & 4.40 & 61.54 & 6.59 & 27.47 \\
    \end{tabular}
\caption{Parameters of intense quadrant structures for cell-averaged velocities, compiled over
wall-parallel planes for $\lcell^+=75$. The thresholds $H_-$ and $H_+$ are as in figure
\ref{fig_jpdf_uv}, and the $S_j$ are the fraction of the intense area associated to each quadrant.
The bottom row are volume fractions for point-wise quadrant structures in the $Re_\tau=935$
channel of \cite{lozano-Q}, in which the combined intense quadrants fill 9\% of the channel volume.
}
    \label{table_uvster}
    \end{center}
\end{table}

Figure \ref{fig_jpdf_uv} displays the resulting quadrant plot, drawn for cell-average quantities.
The line contours are the unconditional joint probability density function of $\cave{u}$ and
$\cave{v}$, while the coloured ones are conditioned to either significant cells in figure
\ref{fig_jpdf_uv}(a), or irrelevants in figure \ref{fig_jpdf_uv}(b). The hyperbolic lines in the
figure are intensity limits for sweeps and ejections, defined as
\beq
|\cave{u}(\vec{x})\cave{v}(\vec{x})|\ge H_\pm \cave{u}'(y)\cave{v}'(y).
\label{eq_uvhole}
\eeq
In the classical quadrant plot for point velocities, the threshold $H$ only depends on $y$
\citep{lozano-Q}, and results in different volume fractions for sweeps and for ejections (see the
last line in table \ref{table_uvster}). To facilitate comparison with the our choice of a common
area fraction for significant and irrelevant cells, figure \ref{fig_jpdf_uv} uses two different
thresholds: $H_-$ for sweep-like structures with $\cave{v}<0$, and $H_+$ for ejection-like ones with
$\cave{v}>0$. They are adjusted so that the total areas for sweeps and for ejections are the ones used for the
significance analysis, $\fsig=10\%$. They are given in table \ref{tab_cell} for the five
experimental wall distances, and are lower than the $H\approx 1.75$ used in \cite{lozano-Q} and in
other studies.  Correspondingly, they select a larger area fraction, 20\% in total, rather than the approximately 9\%
volume fraction in \cite{lozano-Q}. However, table \ref{table_uvster} shows that the distribution of
quadrants in these intense regions is not very different from the classical values, once the
relative fractions of sweep- and ejection- like structures are taken into account. As in the case of
point velocities, most strong structures are either pure sweeps, $\cave{u}>0,\, \cave{v}<0$, or pure
ejections, $\cave{u}<0,\, \cave{v}>0$, and there are comparatively few intense $Q_1$ or $Q_3$. The
filled contours in figure \ref{fig_jpdf_uv}(a) are cell-averaged velocities of the significant
cells, and those in figure \ref{fig_jpdf_uv}(b) are irrelevant ones. It is evident that significants
tend to be in strong $Q_2$ sweeps, while irrelevants are in strong $Q_4$ ejections.

\begin{figure}
\centerline{%
    \includegraphics[height=0.33\textwidth]{\figpath 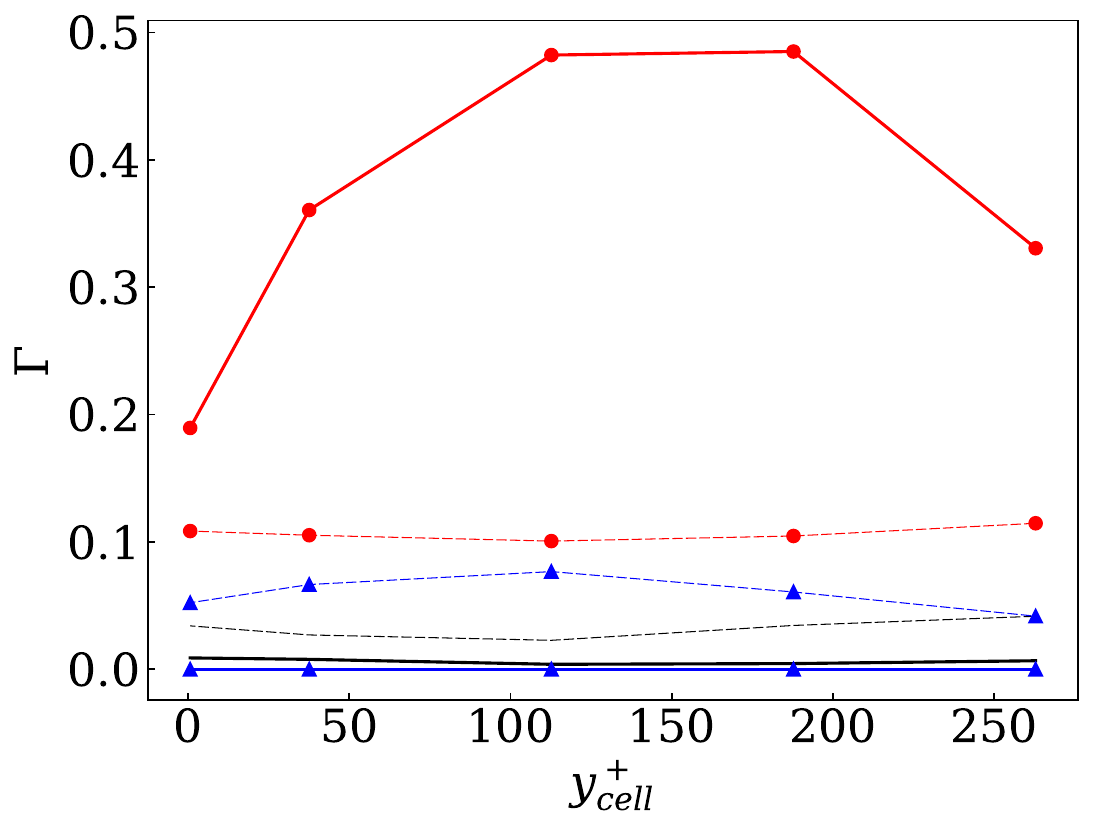}%
\mylab{-0.37\textwidth}{.29\textwidth}{(a)}%
\hspace{3mm}%
    \includegraphics[height=0.33\textwidth]{\figpath 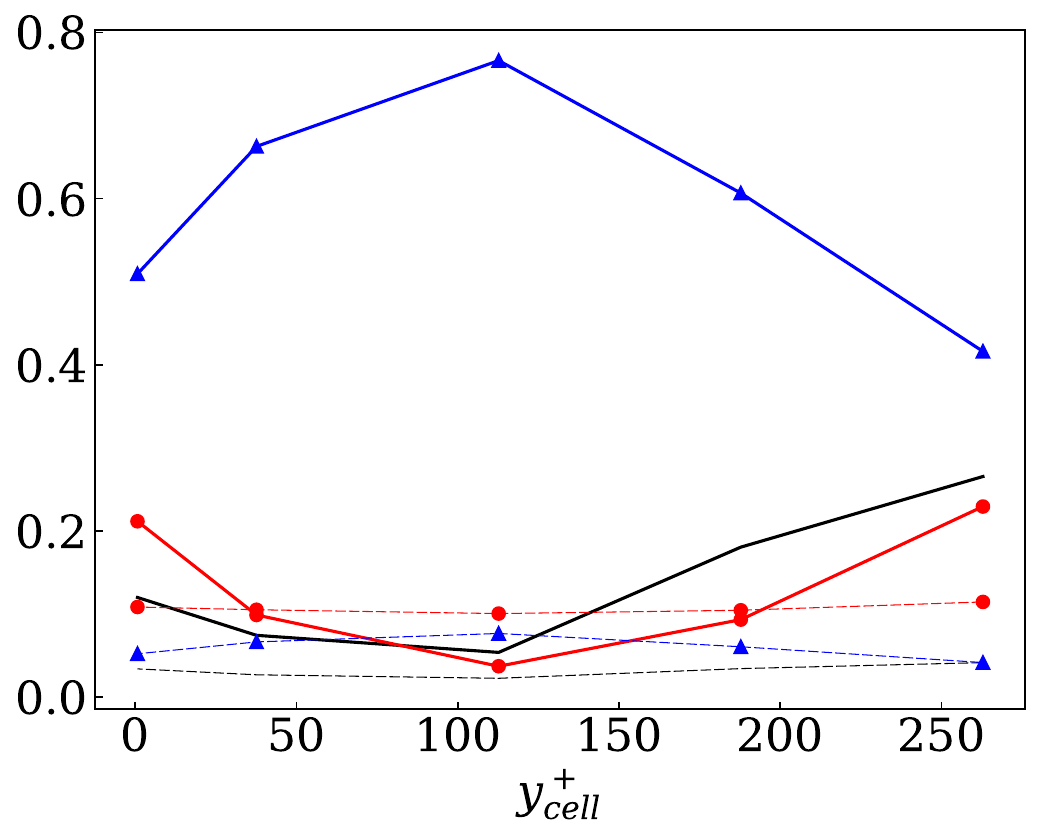}%
\mylab{-0.36\textwidth}{.29\textwidth}{(b)}%
}
\caption{Area fraction of the significance structures intersected by intense quadrants. $\lcell^+=75$.
Red, $Q_4$; blue, $Q_2$; black, $Q_1\cup Q_3$. Solid lines are
conditioned to: (a) significants, (b) irrelevants. Dashed ones are unconditional. 
\label{fig_crossrate}}
\end{figure}

This association is quantified in figure \ref{fig_crossrate}. The area fraction of the intersection
between two classes, $A$ and $B$, is defined as $\Gamma(A,B) = 2 S(A \cap B)/(S(A)+S(B))$, where $S$
denote the area covered by each class. Figure \ref{fig_crossrate}(a) shows the intersection of
significant structures with intense quadrants, $Q_j$. The dashed lines are area fractions of
the intersection of randomly $Q_j$ structures with the same area as the significance structures. Figure
\ref{fig_crossrate}(b) repeats the analysis for irrelevants. It is again clear that significants
predominantly overlap $Q_4$, and irrelevants overlap $Q_2$, with a maximum at $y^+\approx
100-150$.
 
\begin{figure}
\vspace*{3mm}%
\centerline{%
\includegraphics[height=0.31\textwidth]{\figpath 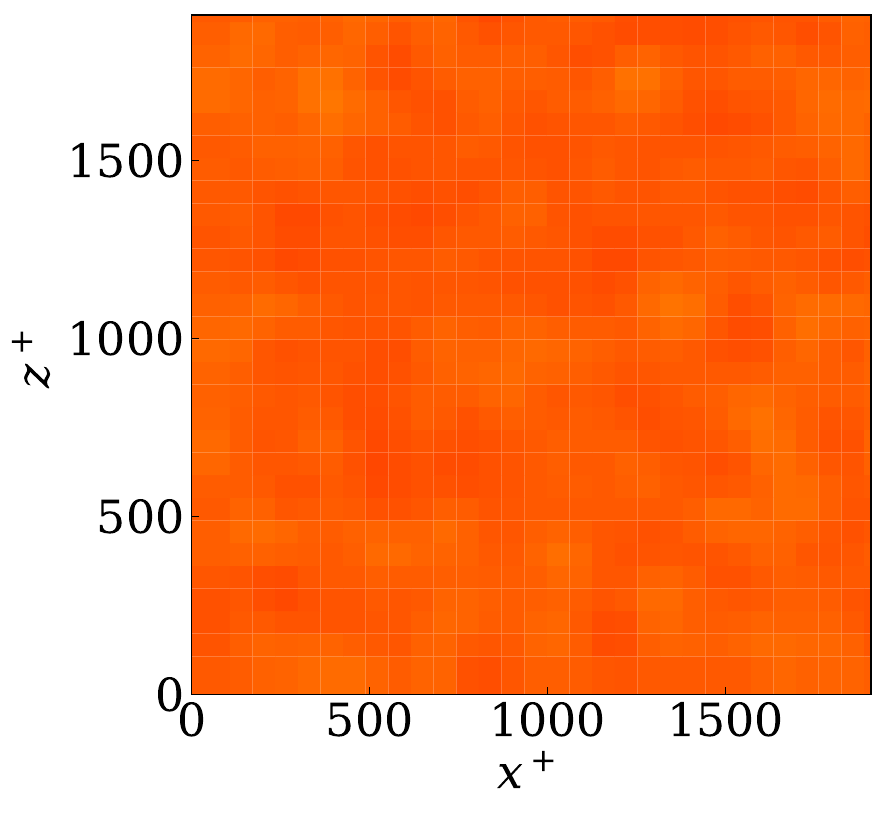}%
\mylab{-0.15\textwidth}{.31\textwidth}{(a)}%
\hspace{1mm}%
\includegraphics[height=0.31\textwidth]{\figpath 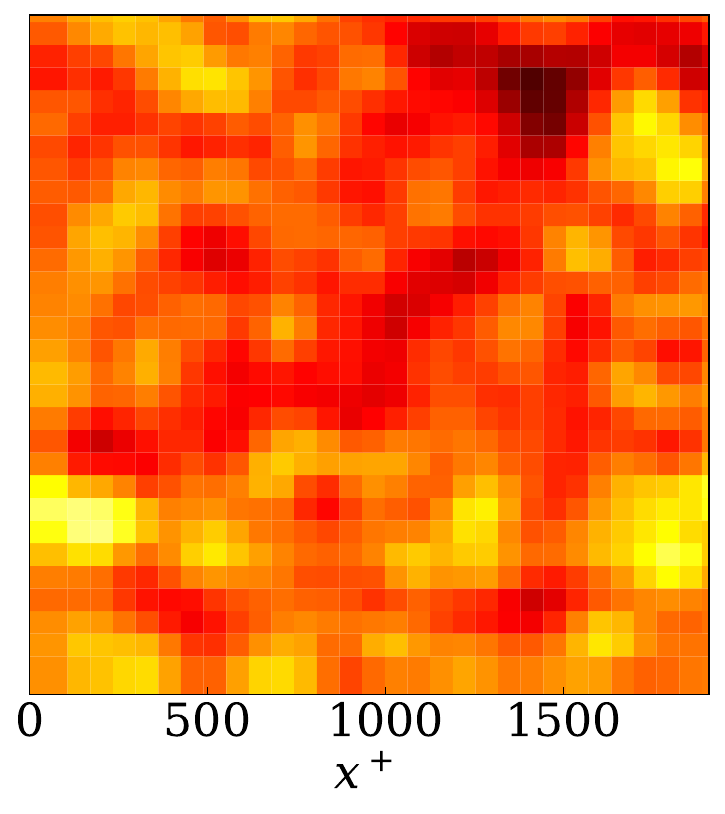}
\mylab{-0.15\textwidth}{.31\textwidth}{(b)}%
\hspace{1mm}%
\includegraphics[height=0.31\textwidth]{\figpath 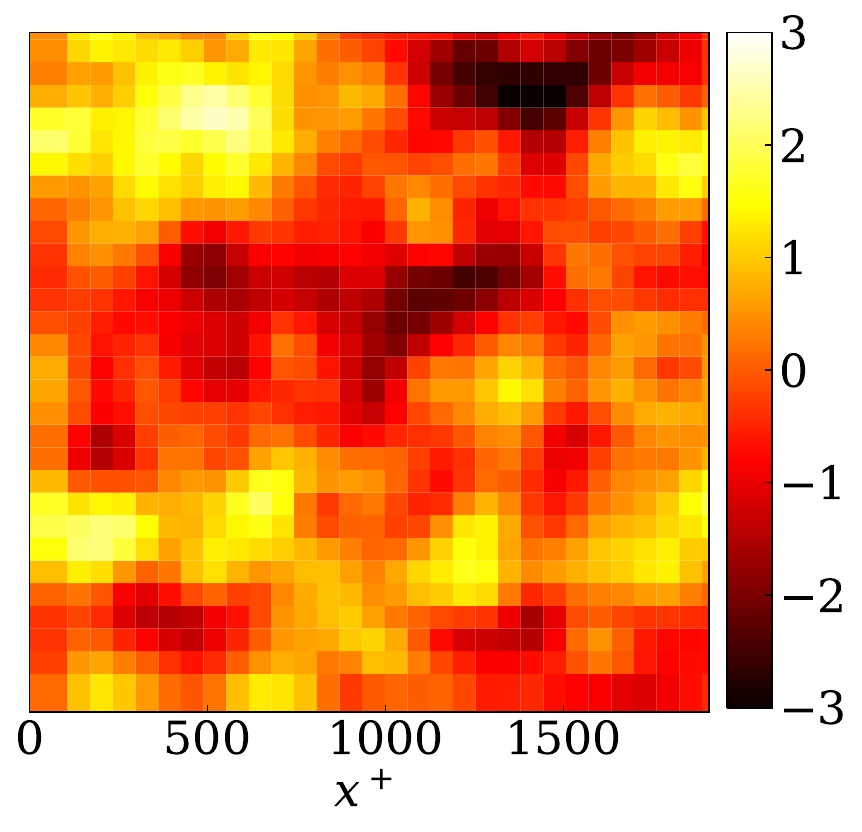}%
\mylab{-0.19\textwidth}{.31\textwidth}{(c)}%
}
\caption{Heat maps of the relative significance $\sigur$ at various evaluation times. Note
that the grid shows the position of the initial cells, whereas the colour indicates their relative significance
evaluated some time in the future. $\lcell^+=75$, $\ycell^+=113$. (a) Evaluation time, $0.01
h/u_\tau$. (b) $0.14 h/u_\tau$. (c) $0.28 h/u_\tau \approx \tsig$.
\label{fig_schem_grid}}
\end{figure}

The question of whether significant cells are organised into structures similar to those of intense
quadrants is addressed in figure \ref{fig_schem_grid}. Each panel displays the same plane away from
the wall. Cells are shown at their position at $t=0$, but labelled by their relative significance
evaluated some time after they are perturbed. The evaluation time increases from left to right. The
heat map in figure \ref{fig_schem_grid}(a) is featureless, reflecting the difficulty discussed in \S
\ref{sec_signi} of predicting the future significance of a cell from its growth at short times. The
organisation increases in figure \ref{fig_schem_grid}(b), and is best developed in figure
\ref{fig_schem_grid}(c), where significance is evaluated at the optimum classification time,
$\tsig$. The size and organisation of the significance in the last figure are very similar to those
in the velocity maps in figure \ref{fig_temp_xz}(a,b), with structures of $O(h)$ organised into
longer streaky structures.

This suggests the possibility that irrelevants and significants are essentially the same as sweeps
and ejections. To test this hypothesis, both sets of cells are collected into individual connected
objects for which the product $\cave{u}\cave{v}$, or $\sigur$, are above or below the threshold required to
isolate the $\fsig\%$ area fraction of their wall-parallel plane, as in \cite{lozano-Q}. It is known
that the wall-attached sweeps and ejections of the point-wise velocity form spanwise pairs
\citep{lozano-Q}. A similar analysis is done here for the cell-averaged quadrants and for the
significance structures. The centroid of all the structures and their pairwise distance is computed
first, and two structures, such as $A_i$ and $B_j$ are defined as a pair if $B_j$ is the closest
object to $A_i$ and $A_i$ is the closest object to $B_j$. It can be shown that the probability that
significant and irrelevant structures are part of a close pair is similar to that of the $Q_2$ and
$Q_4$, and much larger than for randomly located objects.

\begin{figure}
\vspace*{3mm}%
    \centerline{%
    \includegraphics[height=0.28\textwidth]{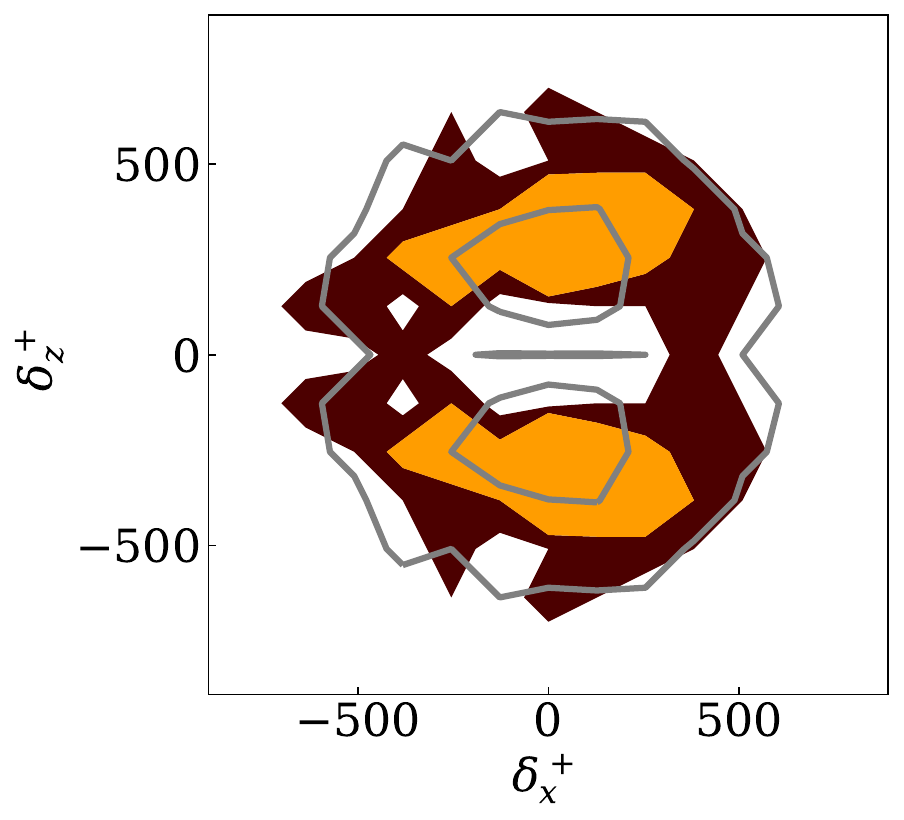}%
    \mylab{-0.14\textwidth}{.29\textwidth}{(a)}%
    \hspace{2mm}%
    \includegraphics[height=0.28\textwidth]{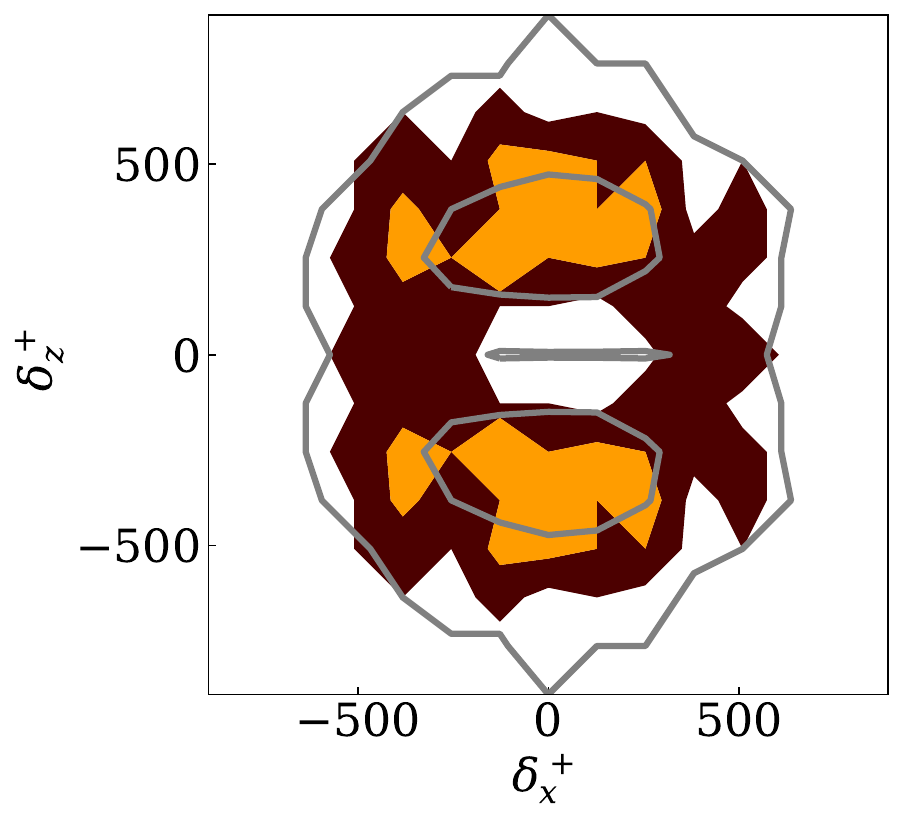}%
    \mylab{-0.14\textwidth}{.29\textwidth}{(b)}%
    \hspace{2mm}%
    \includegraphics[height=0.28\textwidth]{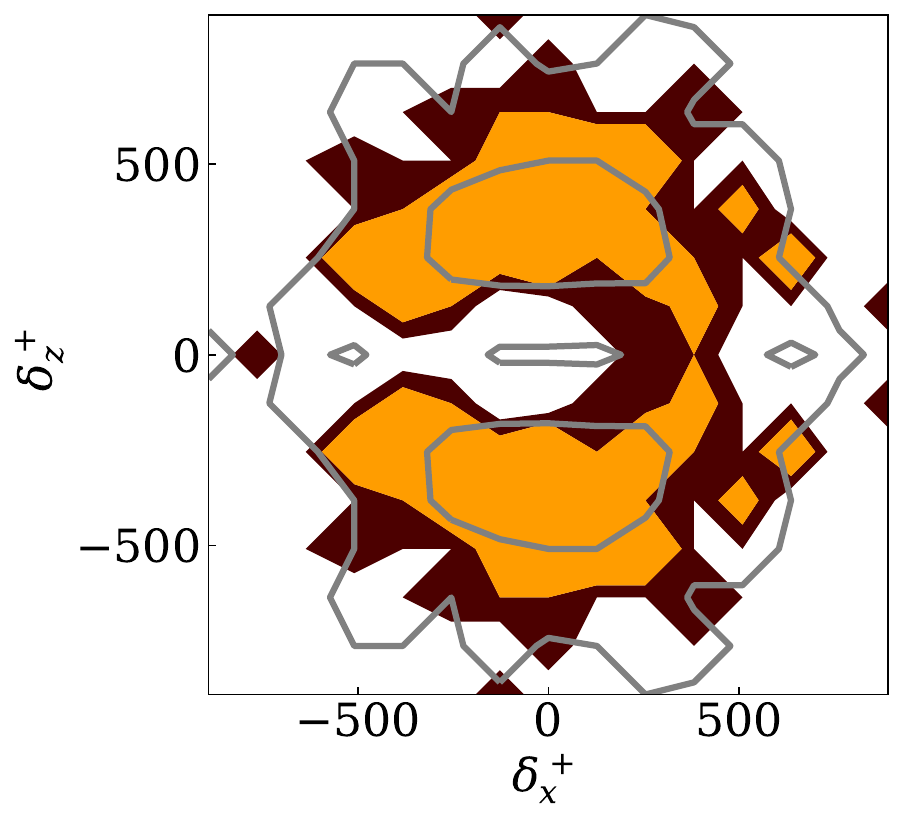}%
    \mylab{-0.14\textwidth}{.29\textwidth}{(c)}%
    }%
    
\caption{Joint probability density function of the relative position of structures in close pairs.
Filled contours are irrelevant--significant pairs, and lines are $Q_2$--$Q_4$. 
Contours contain 60\% and 99\% of the data, and $z$-symmetry is enforced.
$\lcell^+=75$. (a) $\ycell^+=0$. (b) $\ycell^+=113$. (c) $\ycell^+=263$.
 \label{fig_pairjpdf}}
\end{figure}

\begin{figure}
    \centerline{%
    \includegraphics[height=0.30\textwidth]{\figpath 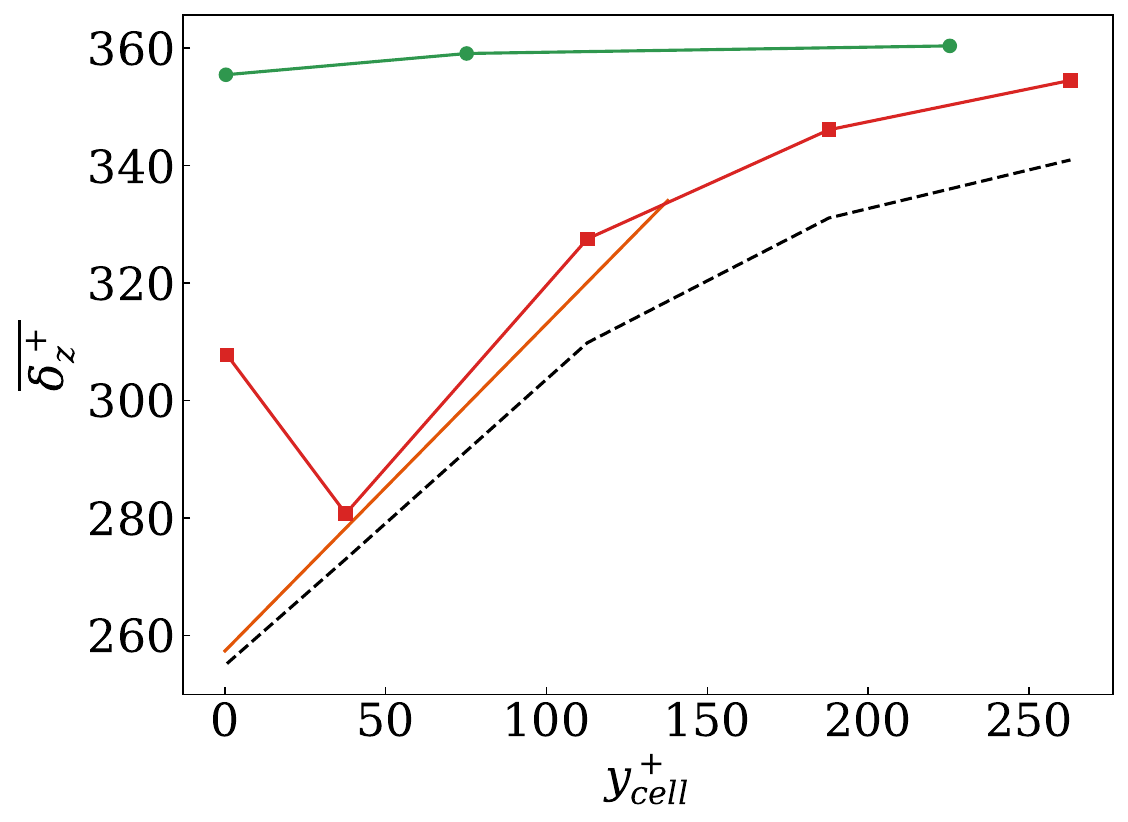}%
}%
\caption{Mean spanwise distance among nearest significance or quadrant structures. Lines with
symbols are significant--irrelevant pairs, with symbols denoting $\lcell$, as in table
\ref{tab_cell}. The dashed line is $Q_2$--$Q_4$.
\label{fig:meanz}}
\end{figure}

Figure \ref{fig_pairjpdf} presents joint probability density functions of the position of the
nearest structures of similar kinds. Filled contours depict irrelevants around significants, and
lines indicate ejections around sweeps. Both use spanwise symmetry to enhance statistical
convergence, and it is clear that the pairs of the two types of structures have a similar
organisation. Figure \ref{fig:meanz} shows the average spanwise width of the pairs as a function of
wall distance and cell size. It is known from \cite{lozano-Q} that this width scales with $y$ for
attached $Q_2$--$Q_4$ pairs far from the wall. Our Reynolds number is too low for this
self-similarity to hold, but figure \ref{fig:meanz} shows that, at least for the two smallest cell
sizes, the width of the significance pairs grows with the distance from the wall, and approximately
follows that of the quadrants. It is difficult to make an exact correspondence between
single planes and slabs of relatively large cells, and the green line in the figure is probably a reflection
of this difficulty. Its cell size, $\lcell^+=150$, is of the same order as the distance from the
wall. Similarly, the apparently large discrepancy of the $\lcell^+=75$ red line with the quadrant
pairs at $\ycell=0$ is put in perspective by figure \ref{fig_pairjpdf}(a), which represents the
same data.

\begin{figure}
    \centerline{%
    \includegraphics[height=0.28\textwidth]{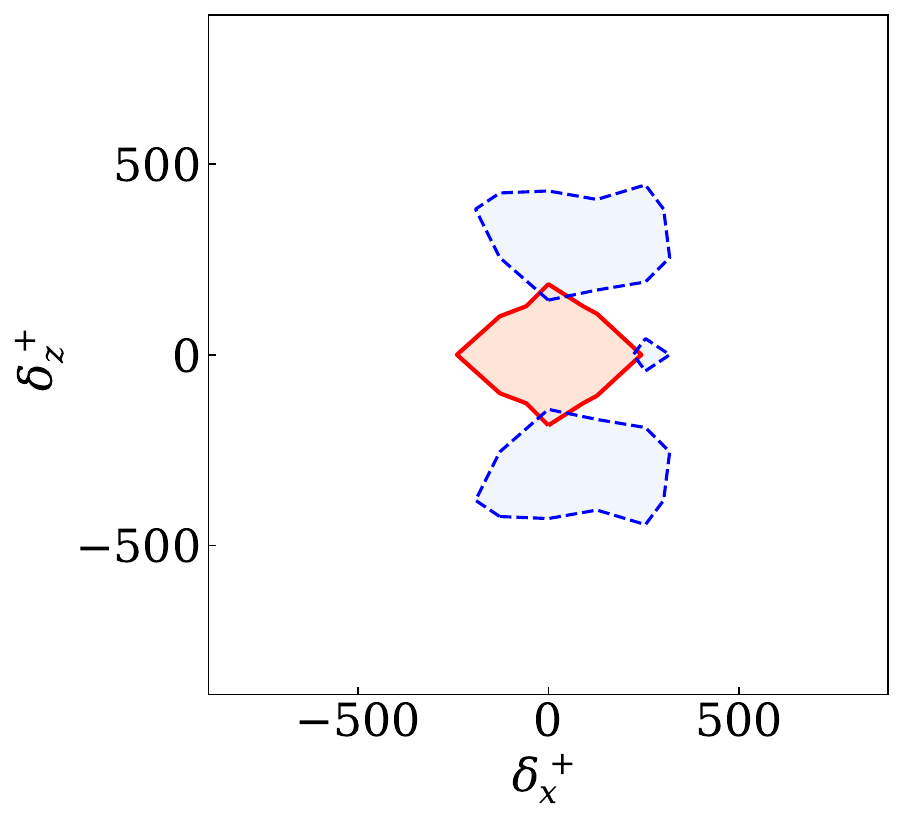}%
    \mylab{-0.22\textwidth}{.24\textwidth}{(a)}%
    \hspace{2mm}%
    \includegraphics[height=0.28\textwidth]{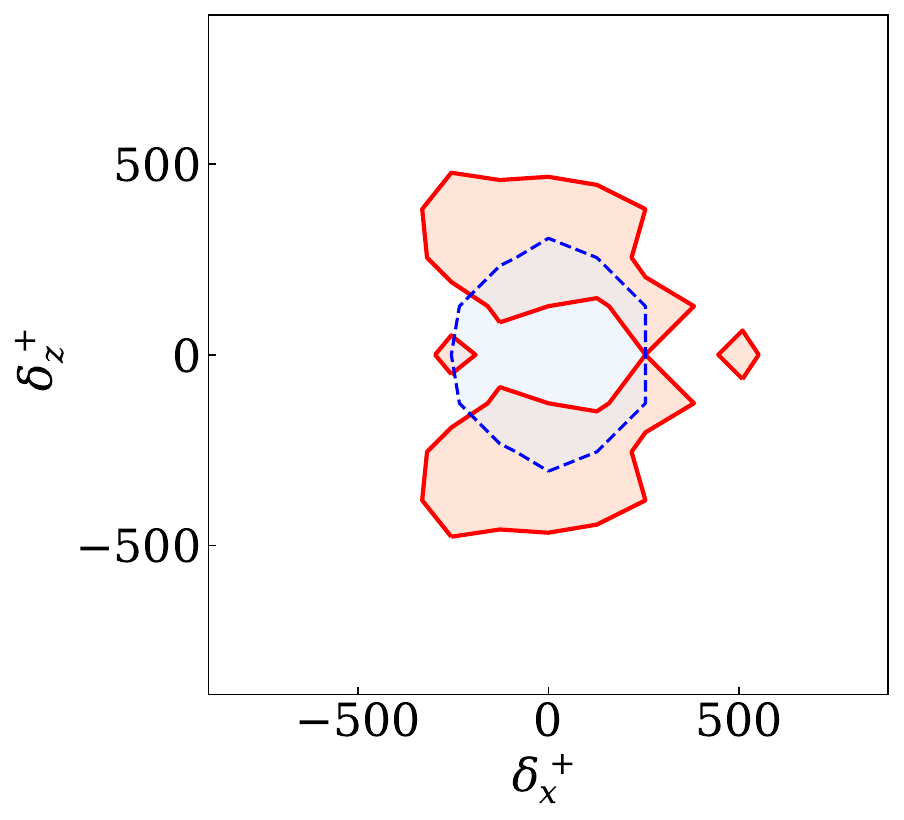}%
    \mylab{-0.22\textwidth}{.24\textwidth}{(b)}%
    }
\caption{Joint probability density function of the relative position of closest structures of different types.
(a) Quadrants around significants. (b) Quadrants around irrelevants. Red, nearest $Q_4$; blue,
nearest $Q_2$. Contours contain 60\% of data. $\ycell^+=113$. $\lcell^+=75$.
\label{fig_pairjpdf_sigQ}}
\end{figure}
%

Finally, figure \ref{fig_pairjpdf_sigQ} shows the relative position of intense quadrants with respect to
significance structures. Figure \ref{fig_pairjpdf_sigQ}(a) is centred on significants,
and shows that the closest sweep coincides with the significant structure, while the closest ejection
avoids it. Figure \ref{fig_pairjpdf_sigQ}(b), which is centred on irrelevants, shows that the
opposite is true for them.

In summary, the results in this section show that highly significant and irrelevant structures
respectively coincide with intense ejections and sweeps, at least statistically. They are organised
in a similar way, and they most probably refer to the same structures, although it should be
emphasised that this does not imply that all significance structures are intense quadrants, or vice
versa.

\section{Discussion and conclusions}\la{sec_conc}
 
We have analysed the causal relevance of flow conditions in wall-bounded turbulence, using ensembles
of interventional experiments in which the effect of locally perturbing the flow in a small cell is
monitored at some future time. We have shown that the evolution of the kinetic energy of the
perturbation velocity is mostly determined by its initial intensity, but that, when the effect is
characterised by the relative amplification of the perturbation energy, causality depends on the
cell size, on the flow condition within the cell, and on its distance from the wall. It is then
possible to enquire which properties of the flow at the time of the initial perturbation determine
causality and, in this way, use the causality experiments as probes of the flow dynamics, rather than simply
as a reflection of the dynamics of the perturbations \citep{JJ20,encinar23}.

We have shown that there is an optimum time at which causality can be measured most effectively,
because the influence of different cell conditions is most pronounced. This time is proportional to
the distance from the wall of the original intervention, and we have related it to the mean shear
that the perturbation experiences as it evolves. When time is normalised with this shear, the
evolution of perturbation applied to causally significant and to causally irrelevant cells collapses
reasonably well (figure \ref{fig_tshear}).

For perturbations away from the wall, the variables that predominantly determine causality are the
streamwise and wall-normal velocities within the cell, $\cave{u}$ and $\cave{v}$, with the latter
becoming more influential farther from the wall. For wall-attached perturbations, the dominant
variable is the local cell-averaged wall shear. Positive $\cave{u}$, negative $\cave{v}$ and high
wall shear are associated with high causality, and negative $\cave{u}$, positive $\cave{v}$ and low
shear are associated with irrelevance. This, together with the shear scaling mentioned above,
suggests that wall-detached significant cells are predominantly associated with sweeps that carry
the perturbation towards the stronger shear near the wall, whereas irrelevant ones are associated with
ejections that carry it towards the weaker shear in the outer layers. This is confirmed by the
conditional flow fields in figures \ref{fig_temp_xy}--\ref{fig_temp_zy}, and by the quadrant
analysis in figures \ref{fig_jpdf_uv}--\ref{fig_crossrate}.

We have also shown that causally significant and irrelevant cells are themselves organised into
structures that share many characteristics of classical sweeps and ejections. For example,
the latter are known to be organised in spanwise pairs, and we have shown that the same is true of
causally significant and irrelevant structures. The dimensions of the two types of pairs are similar,
and their relative positions are consistent with the identification of sweeps with significants and
of ejections with irrelevants (figure \ref{fig_pairjpdf_sigQ}).

However, as already noted at the end of \S \ref{sec_Qs}, not all sweeps and ejections are
causally significant or irrelevant. Figures \ref{fig_temp_xy}--\ref{fig_temp_xz} show
that, as the perturbation experiments are performed closer to the wall, significant cells move
towards the downstream end of the sweep, while irrelevant ones drift towards an interface between
the ejection and a sweep downstream and underneath it. At the wall, this is consistent with a
causally significant configuration in which a high-speed streak overtakes a low-speed one, and with
a causally irrelevant situation in which the two streaks pull apart. In fact, $\cave{\p_z w}$ and
$\cave{\p_x u}$ are among the leading indicators of causality for some wall-attached perturbations
(figure \ref{fig_class_yt_errur_ycell}).

This raises the question of how two structures that form a close pair can lead to different
outcomes. A similar question was raised by \cite{lozano_time} when they found that the vertical
advection velocity of the sweep and ejection components of attached pairs are $-u_\tau$ and
$+u_\tau$, respectively. During the lifetime of the pair, this leads to relative vertical
displacements of the order of the height of the pair, and to its dissolution. The answer offered at
the time was that this was the mechanism that limits the lifetime of the pair, and a similar answer
may apply here, since one of the results of this paper is that causality can only be traced to the
moment that significant perturbations reach the wall. In fact, sweep--ejection pairs are known to be
located at the interface between high- and low-velocity streaks \cite{lozano-Q}, and it is easy to see that their
effect on the streamwise velocity would be to deform the streaks into meandering. Although the statistical
confirmation of this model is beyond the scope of the data used in the present paper, figure
\ref{fig_temp_xz} strongly suggests that the association of causality at the wall with the
streamwise variation of $\cave{u}$ refers to the leading and trailing edges of a streak meander. The
causal significance of the off-wall sweeps would then reduce to their role in modifying the active
downstream edge of the meander.

From an application perspective, the results of this study suggest new possibilities for turbulence
control. Currently, most active control research centres on wall-attached devices, but some of
the results above suggest that, if the focus is on altering the overall state
of the flow, it may be more efficient to act on motions detached from the wall and moving towards
it. While the technical challenges associated with manipulating the flow in the far field are
clear, this approach is intriguing because it is not currently receiving much attention, although one
cannot avoid to be reminded of LEBUs \citep{Lebus:2018}.

As for further work, several avenues should be explored, although most would require substantially
more data and processing than the ones used here. For example, statistical confirmation of the
continuity model outlined above should probably be done on larger computational boxes than the
present one, to avoid artefacts on the behaviour of streaks. The same can be said about different
Reynolds numbers. Similarly, the characterisation of causality by the integrated perturbation energy
over the whole domain could probably be gainfully substituted by more specific measures, such as the
energy of a particular layer, near or far from the wall, or by more practical ones, such as the skin
friction. This would make the results more relevant to control, and probably illuminate flow
interactions that are relevant to the physical understanding of the flow (e.g. small near-wall
scales with large far-wall ones). Unfortunately, studies like the present one involve large amounts
of data that cannot be fully stored for re-processing. Much of the analysis is done `on the fly'
while the experiments are being carried out, and applying a new processing strategy involves a new
set of simulations. On the other hand, prospective studies like the present one are crucial to the
design of any such future extension.


\backsection[Funding]{This work was supported by the European Research Council under the Caust grant
ERC-AdG-101018287.  }

\backsection[Declaration of interests]{The authors report no conflict of interest.}

\backsection[Author ORCIDs]{K. Osawa, https://orcid.org/0000-0001-6535-6241; 
J. Jim\'enez, https://orcid.org/0000-0003-0755-843X}

\appendix
\section{The perturbation evolution equation}\label{sec_evoleq}

Consider a generic quantity $T$ with source $S$, advected by a velocity field $u_i$, 
\begin{equation}
    \partial_t T = - u_j \partial_j T + S,
    \la{eq:Teq}
\end{equation}
where repeated indices imply summation, and two independent experiments `$a$'  and `$b$', denoted by 
superscripts of the respective fields. Consider now the evolution equation for
the difference between the experiments. Define
\begin{align}
    \dif{g} & =g^a-g^b, \\
    \ave{g} &=(g^a+g^b)/2.
\end{align}
for any $g$. Particularising \r{eq:Teq} for $T^a$ and $T^b$, and subtracting one from the other,
\beq
    \partial_t \dif{T} = - u_j^a \partial_j T^a + u_j^b \partial_j T^b + \dif{S},
\eeq
which, since
\begin{align}
     \dif{u}_j\partial_j\ave{T} + \ave{u}_j\partial_j\dif{T} 
     =&\,
     (u_j^a-u_j^b)\partial_j(T^a+T^b)/2 
     (u_j^a+u_j^b)\partial_j(T^a-T^b)/2 \notag\\
    =&\,
      u_j^a \partial_j T^a/2 - u_j^b \partial_j T^b/2 
    + u_j^a \partial_j T^b/2 - u_j^b \partial_j T^a/2 \notag\\ 
    &+ u_j^a \partial_j T^a/2 - u_j^b \partial_j T^b/2 
    - u_j^a \partial_j T^b/2 + u_j^b \partial_j T^a/2 \notag\\
    =&\, u_j^a \partial_j T^a -u_j^b \partial_j T^b,
\end{align}
can be written as
\beq
    \partial_t \dif{T} = -\dif{u}_j\partial_j\ave{T}  -\ave{u}_j\partial_j\dif{T} + \dif{S}.
    \la{eq:difT}
\eeq
Multiplying \r{eq:difT} by $\dif{T}$,
\begin{align}
    \partial_t \dif{T}^2 &= 
     -2\dif{u}_j\dif{T}\partial_j\ave{T} -2\ave{u}_j\dif{T}\partial_j\dif{T}
     +2\dif{T}\dif{S} \notag\\
    &=
    -2\dif{u}_j\dif{T}\partial_j\ave{T}
    -\ave{u}_j\partial_j\dif{T}^2
    + 2\dif{T}\dif{S}.
    \la{eq:dTsq}
\end{align}

The evolution of the velocity perturbation magnitude follows from substituting $u_i$ for $T$ in
\r{eq:dTsq}. The source of the evolution equation for $u_i$ is $S_i=-\partial_i p + \nu \partial_j^2
u_i$, so that $ \dif{S}_i=-\partial_i \dif{p} + \nu \partial_j^2 \dif{u}_i$, and
\begin{align}
    \partial_t \dif{u}_i^2 &= 
    -2\dif{u}_i\dif{u}_j\partial_j\ave{u}_i
    -\ave{u}_j\partial_j\dif{u}_i^2
    -2\dif{u}_i\partial_i \dif{p}
    +2\nu\dif{u}_i\partial_j^2 \dif{u}_i \notag\\
    &=
    -2\dif{u}_i\dif{u}_j\partial_j\ave{u}_i
    -\ave{u}_j\partial_j\dif{u}_i^2
    -2\partial_i (\dif{u}_i\dif{p})
    +2\nu\partial_j(\dif{u}_i\partial_j \dif{u}_i)
    -2\nu(\partial_j \dif{u}_i)^2 \notag\\
    &=
    -2\dif{u}_i\dif{u}_j\partial_j\ave{u}_i
    -\partial_j \{
        \ave{u}_j\dif{u}_i^2
        +2\dif{u}_i\dif{p} \delta_{ij}
        -\nu\partial_j \dif{u}_i^2
    \}
    -2\nu (\partial_j \dif{u}_i)^2 \notag\\
    &=\Perr+\Cerr+\Derr,
    \la{eq:balance}
\end{align}
where $\delta_{ij}$ is Kronecker's delta. 
The first and last term in \r{eq:balance} represent the production and dissipation of the perturbation
energy, respectively. Note that velocity gradient in the production term is the average of two
independent fields, and does not necessarily agree with the usual ensemble averaged gradient.
The terms in curly brackets are fluxes that do not contribute to $\erru$ when \r{eq:balance} is
integrated over the whole computational box. From left to right, they represents convection,
pressure-strain and viscous diffusion, respectively. It is noteworthy that swapping $a$ and $b$ do
not change the \r{eq:balance}, in agreement with the symmetric way in which they are defined. In
fact, \r{eq:balance} is similar to the evolution equation for the structure function between the
velocities at two neighbouring points, with the  difference that there are no interactions here
between the fields $a$ and $b$.
 
%

\bibliographystyle{jfm}
\bibliography{jjploffv3}

\end{document}